\def\junk#1{}
\documentclass[journal]{IEEEtran}
\newcommand{\dc}[1]{\textcolor{black}{#1}}

\newcommand{\dcR}[1]{\textcolor{black}{#1}}

\ifCLASSOPTIONcompsoc
\else
\fi

\ifCLASSINFOpdf
 \usepackage{times}
  \usepackage{caption,color}
  \usepackage{subcaption}
  \usepackage[pdftex]{graphicx}
\else
\fi

\usepackage[cmex10]{amsmath}
\usepackage{amsthm}
\usepackage{amssymb}

\usepackage[linesnumbered,ruled,vlined]{algorithm2e}
\usepackage{multirow}
\usepackage{epstopdf}
\usepackage{soul}
\usepackage{url}
\newtheorem{definition}{Definition}

\newtheorem{example}{Example}
\newtheorem{problem}{\textbf{Problem}}

\newcommand{\superscript}[1]{\ensuremath{^{\textrm{#1}}}}

\begin{document}
%
\title{A Constrained Shortest Path Scheme for \\ Virtual Network Service Management 
}
\author{
Dmitrii Chemodanov\superscript{*}, Flavio Esposito\superscript{$\dagger$*}, Prasad Calyam\superscript{*}, Andrei Sukhov\superscript{$\ddagger$}\\
\small{\superscript{*}University of Missouri-Columbia, USA; \superscript{$\dagger$}Saint Louis University, USA; \superscript{$\ddagger$}Samara University, Russia;} \\ \small{Email: \textit{dycbt4@mail.missouri.edu, espositof@slu.edu, calyamp@missouri.edu, amskh@yandex.ru}}\\
\vspace{3mm}
\large
\textit{\textbf{Extended Technical Report}}
\vspace{-11mm}
}

\maketitle



\begin{abstract}
Virtual network services that span multiple data centers are important to support emerging data-intensive applications in fields such as bioinformatics and retail analytics. Successful virtual network service composition and maintenance requires flexible and scalable `constrained shortest path management' both in the management plane for virtual network embedding (VNE) or network function virtualization service chaining (NFV-SC), as well as in the data plane for traffic engineering (TE). In this paper, we show analytically and empirically that leveraging constrained shortest paths within recent VNE, NFV-SC and TE algorithms can lead to network utilization gains (of up to 50\%) and higher energy efficiency. The management of complex VNE, NFV-SC and TE algorithms can be, however, intractable for large scale substrate networks due to the NP-hardness of the constrained shortest path problem. To address such scalability challenges, we propose a novel, exact constrained shortest path algorithm viz., `Neighborhoods Method' (NM). Our NM uses novel search space reduction techniques and has a theoretical quadratic speed-up making it practically faster (by an order of magnitude) than recent branch-and-bound exhaustive search solutions. Finally, we detail our NM-based SDN controller implementation in a real-world testbed to further validate practical NM benefits for virtual network services.


\end{abstract}

\begin{IEEEkeywords}
Constrained Shortest Path, Virtual Network Embedding, NFV Service Chaining, Traffic Engineering.
\end{IEEEkeywords}

\maketitle

\IEEEdisplaynotcompsoctitleabstractindextext

%
\IEEEpeerreviewmaketitle

\vspace{-1mm}
\section{Introduction}
\label{Introduction}

\IEEEPARstart{T}{he} advent of network virtualization has enabled new business models allowing infrastructure providers to share or lease their physical networks that span multiple data centers. 
Network virtualization is being increasingly adopted to support data-intensive applications within enterprises ($e.g.$, retail analytics) and academia ($e.g.$, bioinformatics and high energy physics)
over wide-area 
federated infrastructures such as the VMware Cloud~\cite{vm_cloud} and the Global Environment for Network Innovations (GENI)~\cite{geni}. 
A major challenge for infrastructure providers is to offer virtual network services that meet the application Service Level Objective (SLO) demands on shared (constrained) physical networks.
Examples of 
SLO demands can refer to technical constraints such as bandwidth, high reliability, or low latency. 
%


In order to compose and maintain virtual network services, infrastructure providers need to run management protocols (within the `management plane') such as $e.g.$, Virtual Network Embedding (VNE) to satisfy  users' virtual network requests. VNE is the NP-hard graph matching problem of mapping a constrained virtual network on top of a shared physical infrastructure~\cite{Esposito:2013:SES, VNE-survey}. 
Another management protocol example pertains to Network Function Virtualization service chaining (NFV-SC)~\cite{catena,  nfv_survey1}, in which a network manager is required to  \dcR{place Virtual Network Functions ($e.g.$, firewalls, load balancers, etc) and setup a path across corresponding} software-defined middleboxes to guarantee different  high-level traffic constraints $i.e.$, policies. 
While virtual network services are operating, network managers need to also deploy, within the `data plane', traffic engineering (TE) techniques to maintain profiles or improve network utilization~\cite{TESurvey}.

%
Successful virtual network service composition and maintenance in both management plane 
and data plane mechanisms requires \textit{scalable} and \textit{flexible} `constrained shortest path management' for the following reasons. The constrained shortest path problem is \dcR{the NP-hard problem of finding the shortest path that satisfies an arbitrary number of end-to-end (or path) constraints}~\cite{edfs}, and its existing exact solutions are commonly based on branch-and-bound exhaustive search algorithms~\cite{edfs,pareto_bfs,exact_qos} or integer programming. 
These techniques have exponential complexity, and hence, limit \textit{scalability} of the constrained shortest path management. 
Such scalability limitations are exacerbated by  the complexity of VNE, NFV-SC and TE algorithms  as well as the potentially large scale of substrate networks. Although some heuristics or approximation algorithms can find constrained shortest paths in polynomial time (at expense of 
optimality~\cite{QoSRoutingSurvey}), these algorithms support only a specific number of constraints, or a specific cost to optimize 
~\cite{net_utilization,yuli,jaffe}. Thereby, they limit the \textit{flexibility} of the constrained shortest path management.

By leveraging constrained shortest paths, we show how we can enhance network utilization and energy efficiency of VNE, NFV-SC and TE services 
by both analytical and empirical means. 
Reasons for the observed benefits are as follows: Firstly, by using a constrained shortest path, we \dc{can} minimize a provider's cost associated with flows allocation subject to the application SLO constraints \dc{in TE~\cite{swan, b4}}. Such costs potentially hinder long-term infrastructure providers' revenue maximization; this can be understood using a simple example: 
a path comprising of two physical links to maintain a single flow has a higher allocation cost than an alternative solution that uses only one physical link.
For example,  if the flow SLO demand is 10 Mbps, a path composed by a single physical link would need to provision only 10 Mbps, while a path composed by two links would require 20 Mbps.
%
%
Secondly, a more scalable and flexible constrained shortest path approach can be also beneficial for VNE and NFV-SC~\cite{vne_path,vne_cost, vne_opt} when virtual links are subject to an arbitrary number of constraints such as bandwidth, latency and loss rate. 
Other types of constraints can also be imposed by optimization methods, such as the column generation approach, typically used to speed-up integer programming~\cite{edfs, vne_path, vne_cost}.
This is because in the aforementioned cases, constrained shortest paths can be part of the optimal solution, $i.e.$, such paths can best improve the objective value under  arbitrary constraints.

\noindent
{\bf Our Contributions.}  
%
%
To achieve the constrained shortest path benefits and address their management flexibility and scalability challenges in virtual network services,
we propose a novel and exact constrained shortest path algorithm $viz.$, `Neighborhoods Method' (NM). 
Our NM is based on a novel double-pass search space reduction technique that synergizes dynamic programming with a branch-and-bound exhaustive-search. 
%
%
In addition, we propose a novel infeasibility pruning technique  $viz.$, {\it ``Look Back''}, which benefits from NM's double pass design to further 
ease the constrained shortest paths finding in practice. 
Our NM is solving an NP-hard problem, so it is exponential in its general form.
%
However, our computational complexity analysis shows that NM has a quadratically lower complexity upper-bound (halved exponent) than alternative methods.
Moreover, when synergistically used with existing search space reduction 
techniques~\cite{edfs, pareto_bfs,exact_qos}, our scalability evaluation results 
indicate how NM is faster by an order of magnitude than recent branch-and-bound exhaustive search methods, 
and hence, \textit{scales} better.
%
%
%

%
Furthermore, NM is \textit{flexible} due to its adaptable performance and applicability to different constrained shortest path scenarios with an arbitrary set of (SLO) constraints and an arbitrary cost function.
For example, when we allocate traffic flow requests with a single path and multiple link constraints, NM can find some constrained shortest path variants in polynomial time. Thus, it can substitute diverse existing path finder algorithms such as the extended version of Dijkstra~\cite{wacr} and the iterative version of Bellman-Ford~\cite{net_utilization}. 
%
In its general form, NM can be also used to speed-up finding of all loop-free, $k$-constrained shortest or Pareto-optimal paths~\cite{pareto_bfs} from the source to the destination. 
Thus, NM is also applicable to diverse virtual network services including those with splittable and unsplitable flows. 
%
We demonstrate such flexibility with an extensive numerical simulation campaign, testing NM over diverse network topologies for both 
 online VNE/NFV-SC with unsplittable flows and for TE with splittable flows. 
%
%
%
We found that the number of embedded VN requests and energy efficiency (and thereby the providers' revenue) can increase when the constrained shortest path management is used for tested VNE/NFV-SC solution\dc{s}: either one-shot centralized ($i.e.$, with joined node and link embedding)~\cite{vne_path,vne_cost,vne_opt} or  two-stages distributed ($i.e.$, with separate node and link embedding)~\cite{nodeembedding}. 
%
%
%
%
When using the constrained shortest path management within either linear programming~\cite{b4_infocom} or greedy-based TE solutions~\cite{swan,b4}, our simulation results indicate gains of up to $50\%$ in network utilization 
and lower energy consumption in some cases. 

Finally, we implement an open-source Software-Defined Networking (SDN) based NM controller that is available at~\cite{nm_repo}. Our GENI~\cite{geni} evaluation experiments with our implementation prototype confirm our analytical and empirical findings in real-world settings and show no constrained shortest path overhead \dc{(sought by NM)} on the virtual network service management and data plane mechanisms \dc{at large scale}. 

\noindent
{\bf Paper organization.} 
In Section~\ref{problem}, we formally state the constrained shortest path problem using optimization theory. 
Section~\ref{related_problems} shows importance of the constrained shortest path management in VNE/NFV-SC and TE. 
In Section~\ref{nm}, we present details of our NM approach. The complexity improvements of our NM w.r.t. recent branch-and-bound exhaustive search algorithms 
are presented in Section~\ref{complexity_section}. Section~\ref{sdn} describes our NM prototype implementation. Section~\ref{implementation} describes our evaluation methodology, performance metrics and results.
\junk{
In particular, we first evaluate our NM approach  showing how it is beneficial in terms of physical network utilization, by applying NM in combination with existing management plane (VNE) solutions.\footnote{We preserved the node embedding algorithm and replace the virtual link embedding phase with our NM.} 
 We then show how NM is beneficial within data plane (SLO-adhering) solutions such as on-demand traffic steering. In both cases, we test NM's scalability with respect to its related solutions. Finally, we evaluate our NM prototype in GENI~\cite{geni} testbed and reproduce some of the simulation results. 
 }
 Section~\ref{conclusion} concludes the paper.
 
\section{The Constrained Shortest Path Problem}
\label{problem}
The constrained shortest path problem is the NP-hard problem of finding the shortest path subject to an arbitrary set of hop-to-hop and end-to-end constraints. 
In this section, we define this  problem using optimization theory. In the subsequent section, we motivate the importance of its \textit{flexible} and \textit{scalable} management in diverse virtual network services. 

%
%

\subsection{Problem Overview}
The problem of providing a (shortest) path with multiple  (SLO) constraints is NP-hard~\cite{edfs}, and its 
complete survey 
can be found in~\cite{QoSRoutingSurvey}. Herein, we mention a few representative solutions that help us present our novel contributions.
Most heuristics group multiple metrics into a single function reducing the problem to a single constrained routing problem~\cite{ksts}, and then solve the routing optimization separately,  $e.g.$, using Lagrangian relaxation~\cite{jsmr}. 
%
The exact pseudo-polynomial algorithm proposed by Jaffe $et$ $al.$~\cite{jaffe} offers a distributed path finder solution limited to a two-path constraints problem.
%
%
%
%
Wang $et$ $al.$~\cite{wacr} use extended version of Dijkstra algorithm (EDijkstra), where all links with infeasible hop-to-hop constraints are excluded. 
EDijkstra runs in polynomial time but may omit any (path hop count) minimization, desirable in network virtualization to optimize the physical network utilization.
%
To minimize the path hop count under a single path constraint ($e.g.$, delay) an iterative modification of Bellman-Ford (IBF) algorithm was proposed in~\cite{net_utilization}. 
%
%
Our approach is not limited to a single path constraint and can be adapted to subsume both EDijkstra and IBF as we discuss in Section~\ref{complexity_section}. 

The authors in~\cite{exact_qos} propose an exact algorithm for the constrained shortest path problem, and apply several search space reduction techniques such as dominated paths (pruning by dominance and bound) and the look-ahead (pruning by infeasibility) notion for the exponential complexity exhaustive search, utilizing the $k$-shortest path algorithm. 
We also apply a similar technique to reduce the constrained path search space, though without any look-ahead, since it is  computationally expensive. Instead, our design uses a more efficient ``Look-Back" pruning technique (see Section~\ref{space_reduction_sec}).
%
In~\cite{widyono}, the authors propose an Exhaustive Breath-First Search (EBFS) based approach to solve the constrained path finder problem, focussing on delay. 
Another more recent work~\cite{pareto_bfs} also uses EBFS with a dominant path space reduction technique to find multi-criteria Pareto-optimal paths. 
Alternatively, an exhaustive Depth-First Search (EDFS) can be used as a branch-and-bound algorithm. For example, the authors in~\cite{edfs} proposed the ``pulse" algorithm that uses EDFS with dominated paths and look-ahead search space reduction techniques. 
Both of EBFS and EDFS algorithms have exponential worst case time complexity. 
%
%
Our solution however quadratically reduces the worst case complexity of these algorithms (see Section~\ref{complexity_section}). 
%
%



\vspace{-2mm}
\subsection{Constrained Shortest Path Problem}
Let $l$ be the number of hop-to-hop or {\it link constraints} for min/max network metrics, $e.g.$, bandwidth, and  $p$ be the number of end-to-end {\it path constraints} for additive/multiplicative network metrics, $e.g.$, delay or loss. Moreover, we denote with $l \oplus 1$ paths with multiple links and a single path constraints, and with and $l \oplus p$ paths with multiple links and multiple path constraints. 
Given the above notation, we define the constrained shortest path problem as follows:
%
%
%
\begin{problem}[constrained shortest path]
\label{csp_def}
Given a physical network $G = (V, E)$, 
let us  denote with $D$ the vector of flow demands to be transferred and let $u_{ij}$ denote a capacity of the directed edge $e_{ij}$;
let $f_{ij}$ be a variable $f_{ij} \in \{0, D\}$ denoting an amount of flow on the edge $e_{ij}$, and let $c_{ij}$ denote a cost of transferring a unit of flow through such edge; 
finally, let $\bar{l}$ and $\bar{p}$ denote vectors of link (hop-to-hop) and path (end-to-end) constraints, excluding capacity constraints,
where $\bar{l}$ corresponds to min/max edge $e_{ij}$ weights $w^{\bar{l}}_{ij}$ ($i.e.$, $\ge$ or $\le$, respectively), 
and $\bar{p}$ corresponds to additive edge $e_{ij}$ weights~\footnote{Note that multiplicative constraints (e.g., packet loss) can be converted to additive by composing them with a logarithmic function 
to avoid nonlinearity.
} $w^{\bar{p}}_{ij}$; 
the problem of finding constrained shortest path between source $v_s$ and destination $v_t$ vertices can be formulated as follows:\\
\begin{equation}
\text{minimize} \underset{e_{ij} \in E}{\sum}c_{ij}f_{ij} 
\label{csp_objective_eq}
\end{equation}
\textit{subject to}\\
\textbf{Flow Conservation Constraints}\\
\begin{equation}
\underset{v_{j} \in V}{\sum}f_{ij} - \underset{v_{k} \in V}{\sum}f_{ki}=
\begin{cases} 
D, & i=s \\
0, & i\ne s \text{ or } t \\
-D, & i=t 
\end{cases}
, \forall v_i \in V
\end{equation}
\textbf{Capacity Constraints}\\
\begin{equation}
f_{ij} \le u_{ij}, \forall e_{ij} \in E
\end{equation}
\textbf{Other Link Constraints}\\
\begin{equation}
\frac{1}{D}w^l_{ij}f_{ij} \leq l, \forall e_{ij} \in E, l \in \bar{l}  
\label{link_con_eq}
\end{equation}
\textbf{Path Constraints}\\
\begin{equation}
\frac{1}{D}\underset{e_{ij} \in E}{\sum}w^p_{ij}f_{ij} \leq p, \forall p \in \bar{p}
\label{path_con_eq}
\end{equation}
\textbf{Existential Constraints}\\
\begin{equation}
f_{ij} \in \{0, D\}, \forall e_{ij} \in E, D > 0.
\label{domain_con_eq}
\end{equation}
\end{problem}
Finding a \textit{shortest path} (without constraints) has a polynomial time complexity: 
consider Equations~\ref{link_con_eq} and~\ref{path_con_eq}: in absence of any link or path constraints, the constraint matrix of the above optimization problem is unimodular~\cite{opt_book}. 
%
This condition allows us to solve the optimization problem using linear programming. 
Such time complexity bound does not necessarily hold in presence of at least a single link or path constraint ($\bar{l} \neq 0$ or  $\bar{p} \neq 0$). In that case, we have to solve the above optimization problem using integer programming~\cite{opt_book}. 
%

%
In the next section we show how finding a flexible and scalable solution to Problem~\ref{csp_def} benefits a wide range of path finder subproblems to manage several virtual network services. 

\section{Constrained Shortest Path for Virtual Network Service Management}
\label{related_problems}
%
Using optimization theory, in this section we motivate the need for a flexible and scalable constrained shortest path to manage virtual network services such as
%
%
Traffic Engineering (TE), Virtual Network Embedding (VNE) and NFV Service Chaining (NFV-SC). \dcR{We also show how such a constrained shortest path scheme does not introduce any additional interoperability issues for aforementioned virtual network services with respect to traditional shortest path schemes.}
%

\vspace{-3mm}
\subsection{Finding Virtual Paths in Resource Constrained Scenarios} 
We begin by considering a variant of the constrained shortest path  that operates on a  resource constrained scenario, $e.g.$, a natural or man made disaster scenario where connectivity is scarce. In those cases, one aim is to minimize the overall physical resource consumption of virtual paths.
To this end, we define the {\it resource optimal constrained path} by modifying the objective of Problem~\ref{csp_def} as follows:
\begin{problem}[resource optimal constrained path]
\label{rocp_def}
The resource optimal constrained path is a path that satisfies an arbitrary set of link/path constraints using minimal amount of physical bandwidth:
\begin{equation}
\textit{minimize} \underset{e_{ij} \in E}{\sum}f_{ij} 
\label{rocp_objective_eq}
\end{equation}
where $f_{ij}$,$e_{ij}$ and $E$ are as defined in Problem~$1$.
\end{problem}
Note how by defining an equal weight $c$ to all edges we seek the minimum hop path that satisfies an arbitrary set of link/path constraints ($e.g.$, imposed by SLO).  

\subsection{Traffic Engineering}
Traffic engineering (TE) techniques today can be roughly divided into two groups: oblivious $i.e.$, no a-priori knowledge of the SLO demands~\cite{coyotte}, and demands-aware, when such knowledge is available~\cite{swan,b4}. 
Moreover, the later has a superior performance ($e.g.$, can better utilize substrate network resources) than the former~\cite{coyotte} at the expense of having a centralized forwarding (or routing) control 
~\cite{swan,b4,fibbing}.

%

We broadly classify demands-aware traffic engineering solutions (see $e.g.$~\cite{swan, b4}) with the following network utility maximization problem: 
\begin{equation}
\textit{maximize} [\underset{f_i \in F}\min fairness_i (f_i)]
\label{te_objective_eq}
\end{equation} 
where $F$ denotes a set of all demands (or commodities); in~\cite{b4}, for example, such commodities are \{$src$, $dst$, $SLO$\} tuples;
$f_i \in [0,D_{i}]$ is continuous variable that denotes the total amount of flow for commodity $i$ with bandwidth demand $D_i$; and  $fairness_i(f)$ 
is a linear piecewise-defined function whose definition is based on path service's demands SLO constraints. 
For a complete problem formulation we refer to~\cite{b4_infocom, arc_max_min}.


\noindent
{\bf Constrained shortest path relevance.}
There are two standard ways of formulating the TE optimization problem shown in Equation~\ref{te_objective_eq} --- the arc-based~\cite{arc_max_min} and path-based~\cite{b4_infocom} 
 formulations. 
In practice, due hardware granularity limitations, the arc-based linear programming solution can be infeasible to implement, 
or require use of NP-hard integer programming which can be intractable even for moderate size networks. 
The shortest path algorithms such as Dijkstra ($e.g.$, within k-shortest path algorithms~\cite{eppstein}) are currently used to find a set of paths as an input for the path-based linear programming formulations or for their simplified greedy solutions~\cite{swan, b4}. 
The hope is also to map the highest priority flows to the minimum latency paths first. 
However, if we are aware of the services' SLO demands ($e.g.$, bandwidth, latency, loss rate, etc.), 
we can use them as constraints to optimize physical network utilization (and hence the flow fairness) as described in Problem~\ref{rocp_def}. 
%
%

\begin{figure}[t!]
\centering
\includegraphics[width=0.75\linewidth]{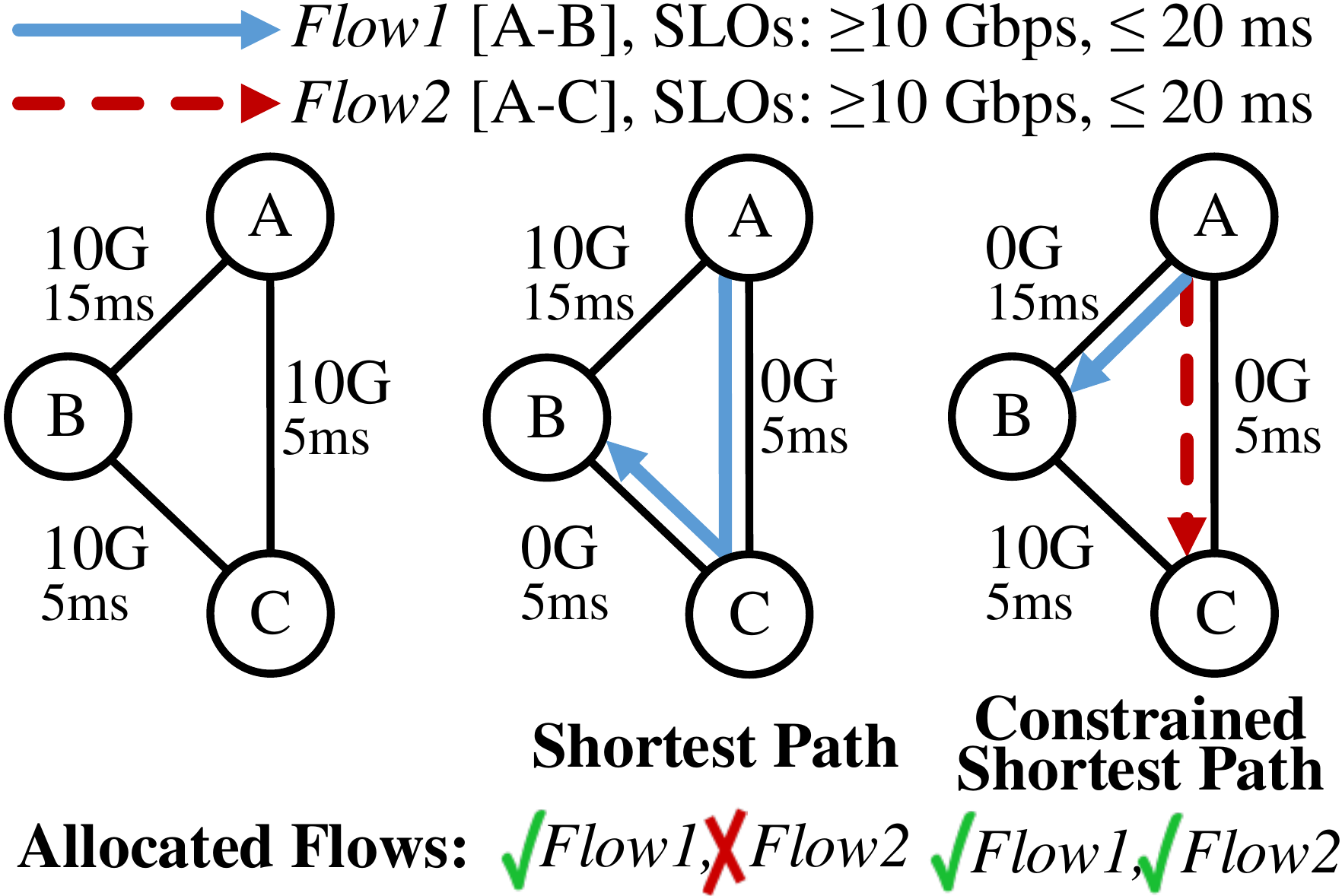}
\caption{\footnotesize{Applying constrained shortest paths to allocate traffic flows improves network utilization: the widely applied shortest paths, $e.g.$, to allocate traffic flows, in this case $Flow1$, can hinder allocation of other flows; allocating $Flow1$ on a constrained shortest path as in Problem~\ref{rocp_def} instead permits the allocation of $Flow2$ as well. 
}}
\label{constrained_shortest_path}
\vspace{-4mm}
\end{figure}
\noindent
{\bf Motivating example.}  Consider Figure~\ref{constrained_shortest_path}: two flows  $Flow1$ and $Flow2$ with  bandwidth and latency constraints are to be allocated on a physical network of Figure~\ref{constrained_shortest_path}-left. Attempting an allocation of $Flow1$ by merely considering a shortest path algorithm Figure~\ref{constrained_shortest_path}-center --- as current TE solutions do, --- the hosting physical path would have to use up to two physical links ($A \rightarrow C \rightarrow B$), thus preventing the allocation of the subsequent requested flow $Flow2$.
If instead a constrained shortest path is used to allocate $Flow1$ and $Flow2$ as shown in Figure~\ref{constrained_shortest_path}-right, then path $A \rightarrow B$ could be found, leaving capacity for the allocation of $Flow2$ (resulting in its higher fairness) as well. 

On the contrary, when shortest path algorithms are used to find min-hop paths, such algorithms can lead to an infeasible solution even when a feasible one exists,
%
leading to TE performance degradation. 
For example, consider the mapping of $Flow1$ with a latency constraint of 10 ms (tighter than 20 in Figure~\ref{constrained_shortest_path}). The min hop (shortest) path $A \rightarrow B$ violates the $10$ms latency constraint, preventing the allocation of $Flow1$, whereas the constrained shortest path algorithms would find the feasible and optimal  $A \rightarrow C \rightarrow B$ path even though more hops are needed to satisfy such tighter SLO demand.
We dissect such network utilization gains when constrained shortest paths are applied for TE in Section~\ref{implementation}. 

\subsection{Embedding Virtual Networks and Service Chains}
\label{csp_vne_sc_sec}

Embedding a virtual network (service chain) requires a constrained virtual service request  to be mapped on top of a physical network hosted by a single infrastructure provider, or by a federation of providers. 
To solve this (NP-hard~\cite{np_hard_vne}) graph matching problem, earlier work proposed centralized (see $e.g.$~\cite{vne_path, vne_cost, vne_opt, yyrc, zham,joaz,lika}) and distributed~\cite{nodeembedding,polyvine,HS} algorithms. Such solutions either separate the node embedding from the link embedding phase~\cite{nodeembedding,yyrc,polyvine,HS}, or simultaneously apply the two phases~\cite{vne_path, vne_cost, vne_opt, lika}.
When link embedding is considered separately, paths can be found dynamically or precomputed for each virtual link using Dijkstra or $k$-shortest path algorithms~\cite{nodeembedding, yyrc, HS}. 
When instead  node and link embedding are considered \dc{jointly}, recent work has shown how embedding problems can be formulated as known multi-commodity flow problems~\cite{vne_path, vne_opt, vne_cost}. 
On the one hand, all these problems show how there is a need to minimize providers' management costs when allocating the virtual network or the service chain~\cite{vne_cost} as shown in the following objective: 
%
\begin{equation}
\textit{minimize} \underset{e_{st} \in E_V}{\sum}\underset{e_{ij} \in E_S}{\sum}c_{ij}f^{st}_{ij} +  \underset{v_{s} \in V_V}{\sum}\underset{v_{i} \in V_S}{\sum}c_ix^s_i
\label{vne_cost_objective_eq}
\end{equation}
where $V_S$ and $E_S$ ($V_V$ and $E_V$) denote sets of physical (virtual) vertices and edges, respectively; $c_{ij}$ and $c_i$ denotes unitary bandwidth and CPU cost on physical edge $e_{ij}$, respectively; 
$f^{st}_{ij} \in [0,D^{st}]$ is continuous variable that models the amount of flow  from $v_s$ to $v_t$ virtual vertices transferred through the physical edge $e_{ij}$ when a flow with demand $D^{st}$ is splittable, or, in its binary form, $f^{st}_{ij} \in \{0,D^{st}\}$ for unsplittable flows.
Finally, $x^{s}_{i} \in \{0,D^{s}\}$ is the binary variable that denotes whether the virtual vertex $v_s$ with computation demand $D^s$ is assigned to the physical vertex $v_i$ or not.
For a complete problem formulation we refer readers to~\cite{vne_path, vne_cost, vne_opt}.
On the other hand, when virtual network (or service chain) requests are unknown in advance, the cost minimization strategy presented in Equation~\ref{vne_cost_objective_eq} can cause physical network partitioning, that in turn can lead to lower physical network utilization~\cite{vne_path, vne_opt}.
To maximize the long term provider's revenue $i.e.$ to allocate more virtual network requests, often a load balancing objective is sought \dc{~\cite{vne_path, vne_opt}:} 
%
\begin{equation}
\textit{minimize} \underset{e_{st} \in E_V}{\sum}\underset{e_{ij} \in E_S}{\sum}\frac{c_{ij}}{u_{ij}}f^{st}_{ij} +  \underset{v_{s} \in V_V}{\sum}\underset{v_{i} \in V_S}{\sum}\frac{c_i}{u_i}x^s_i
\label{vne_lb_objective_eq}
\end{equation}
where $u_{ij}$ and $u_i$ denote available capacities of physical edge $e_{ij}$ and vertex $v_i$, respectively.

\noindent
{\bf Constrained shortest path relevance.}
To cope with integer programming intractabilities for solving path-based multi-commodity flow problems, the well-known {\it column generation approach} can be applied as in~\cite{vne_path, vne_cost}.
The column generation approach iteratively adds only those paths ($i.e.$, flow variables or columns) to the problem formulation that improve objective.
%
%
In~\cite{vne_path, vne_cost} the Dijkstra shortest path algorithm is used to find the best shadow price paths (columns) for each virtual link to include them in the formulation. Although such approach best improve the objective value of $e.g.$, Equation~\ref{vne_lb_objective_eq}, \dc{it can be suboptimal when virtual links have additional constraints such as latency or loss rate. In this scenario, a path between any two physical nodes (found by solving Problem~\ref{csp_def}) can be a part of the optimal solution, even if it has a worse objective value than the shortest path, but satisfies all virtual link constraints. } 
\dc{Generally, for any two stage or one-shot VNE algorithm, constrained shortest paths can be used to best improve the objective value while satisfying an arbitrary number of virtual link constraints.}
%
%
In Section~\ref{implementation}, we confirm such intuition empirically, by studying the allocation ratio improvements (a proxy for provider's revenue) when our method is used to find paths for  embedding services.
\junk{
\begin{definition}[optimal path]
   The \textit{optimal} path for virtual path embedding is the shortest path (in terms of number of hops) that satisfies all (SLO or QoS) constraints ($i.e.$, constrained shortest path) hence minimizing the network over provisioning. 
\end{definition}
We remark that finding a constrained shortest path is an NP-hard problem~\cite{ktmk}. 
}
%
%
%

\junk{
\begin{figure}[t!]
\centering
\begin{subfigure}[t]{0.24\textwidth}
\centering
\includegraphics[width=1\linewidth]{img/pdf/ibf_example1.pdf}
\caption{}
\label{ibf_example1}
\end{subfigure}
\begin{subfigure}[t]{0.24\textwidth}
\centering
\includegraphics[width=1\linewidth]{img/pdf/ibf_example2.pdf}
\caption{}
\label{ibf_example2}
\vspace{-5mm}
\end{subfigure}
\caption{\footnotesize{Illustrative example of an IBF run under an arbitrary distance constraint $\le 5$: (a) IBF successively relaxes vertices (finds the shortest or least cost paths) from the source X while advancing towards the destination Y; (b) upon the first and consequent destination Y relaxations, IBF can also relax other predecessor ($\pi$) vertices of the \textit{optimal} path, $e.g.$, vertex B, thus providing $X \rightarrow A \rightarrow B \rightarrow Y$ suboptimal solution with distance 4, instead of providing the \textit{optimal} $X \rightarrow B \rightarrow Y$ path with distance 5.}}
\label{ibf_example}
\vspace{-5mm}
\end{figure}
\subsection{Motivation} 
%
%
Our notion of network over provisioning can be understood through a simple example where two virtual links ($i.e.$, $VL1$ and $VL2$) with specified bandwidth and latency SLOs are embedded on top of a physical network as shown in Figure~\ref{constrained_shortest_path}.
In this case, the shortest path embedding uses three physical nodes for maintaining $VL1$ (point-to-point path) preventing further embedding of $VL2$ due to the lack of available bandwidth, and, thus, hinders infrastructure providers' revenue maximization. Switching to the constrained shortest path embedding provides a solution alternative for $VL1$ that needs only two physical nodes leaving a room for $VL2$ embedding. 
This example demonstrates two main ideas: (i) a 100\% link utilization does not imply a 100\% network utilization~\cite{b4}, $e.g.$, in both cases 2 links are utilized on 100\%, however, in the constrained shortest path case the overall virtual link throughput is twice as higher; moreover, (ii) minimum hop count paths are desirable for a statistically higher overall network utilization~\cite{net_utilization1}.
\noindent
\textbf{Existing solutions.}
Unless P=NP, only suboptimal~\cite{ktmk,yuli,jsmr} or exponential~\cite{widyono, yyrc, exact_qos, pareto_bfs} solutions of $l \oplus p$ case can exist. 
Although some search space reduction methods such as dominant paths or look-ahead concept exists~\cite{exact_qos, pareto_bfs}, in this paper we show how the process of finding an \textit{optimal} solution in $l \oplus p$ case can be further simplified reducing the time complexity (although not the complexity class).
%
Exact algorithms such as extended Dijkstra or iterative version of Bellman-Ford (IBF) can solve $l \oplus 1$ case in polynomial time (see Section~\ref{complexity_section}) at expense of the guaranteed convergence to the \textit{optimal} path.  
%
For example, to our knowledge, the best algorithm for $l \oplus 1$ case with paths' hop count minimization - IBF~\cite{net_utilization} iteratively relaxes graph vertices $i.e.$, finds the shortest or least cost paths, starting from the source and advances towards the destination (see Figure~\ref{ibf_example1}).
This means that, upon the first destination vertex relaxation, a path with the minimum number of hops should be found. However, upon the first and consequent destination vertex relaxations, other predecessor vertices on the \textit{optimal} path may be further relaxed too. Thus, the path found may result in suboptimal hop count (see Figure~\ref{ibf_example2}). 
}

\section{Neighborhoods Method}

\label{nm}
\theoremstyle{plain}
\newtheorem{th_nm}{Theorem}
\newtheorem{col_nm}{Corollary}

To solve the NP-hard constrained shortest path 
problem with an arbitrary set of ($e.g.$, SLO) constraints in practical time, we propose the novel Neighborhoods Method (NM).

\noindent
{\bf Why the name ``Neighborhoods Method"?} Most path seeking algorithms require at least two inputs for each node: $(i)$ knowledge of neighbors, and $(ii)$ awareness of all adjacent link costs, often dictated by policies, or SLO constraints. Such constraints are then used by the path seeking algorithm to compute the lowest-cost paths. The Dijkstra algorithm $e.g.$, recursively finds the shortest path traversing the source neighbors and the neighbors of their neighbors. This recursive notion leads to our definition of ``neighborhoods'' in NM, $i.e$,  a set of nodes that can be reached from the source node with the same number of hops, where each ``neighborhood'' (being a set) contains unique elements. 
Based on Bellman's ``Principle of Optimality"~\cite{bellman}, such node repetitive structures can be used for label-correcting dynamic programming that we apply to reduce a number of exhaustive search path candidates. 
%

\begin{figure}[t!]
\centering
\includegraphics[width=1\linewidth]{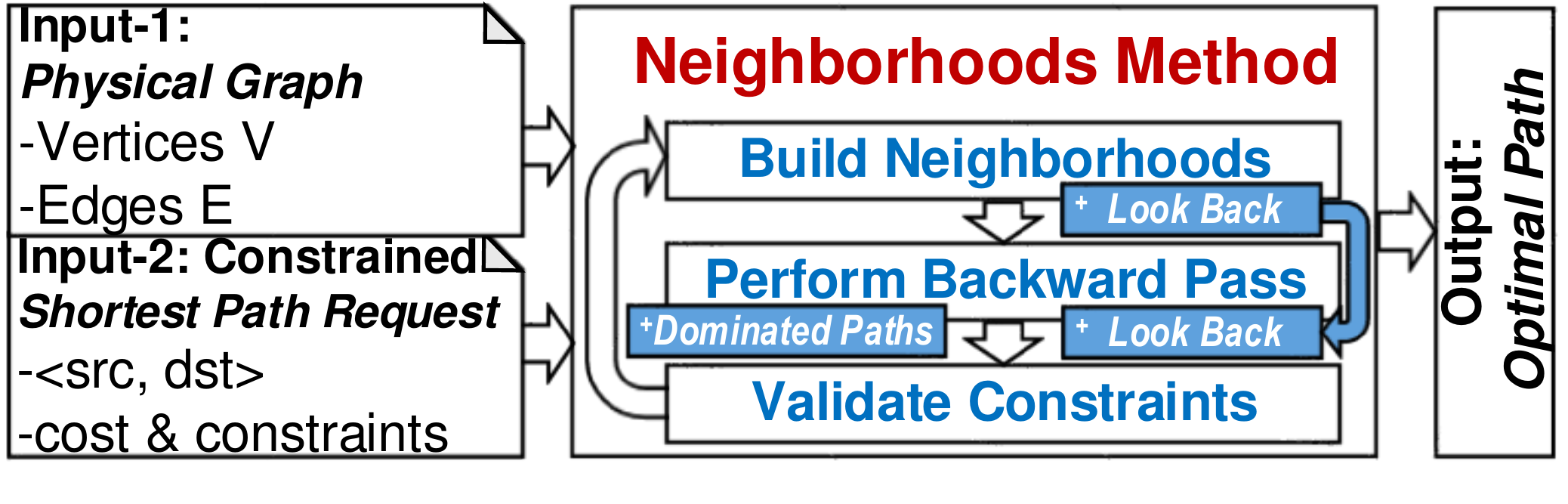}
\vspace{-2mm}
\caption{\footnotesize{Neighborhood Method Workflow in $l \oplus p$ case: The $k^{th}$-hop Neighborhood  Build (forward pass) first (using label-correcting), and the Backward Pass later find all simple paths of $k$ length (with any instance of branch-and-bound exhaustive search); in the third step, we check candidates feasibility; we then repeat recursively the backward pass with the $k+1$ neighborhoods, until the \textit{optimal} (constrained shortest) path is found or all candidates have been eliminated.
To improve the time to solution, we couple NM with a dominated path and a look-back search space reduction technique.
}}
\label{general_diagram}
\vspace{-6mm}
\end{figure}

\subsection{The NM General Case ($l\oplus p$ case)} 
As the exact constrained shortest path algorithm, 
the worst case time complexity of NM  when accepting an arbitrary set of hop-to-hop and end-to-end constraints ($l\oplus p$)  is exponential. Our NM complexity analysis shows that the time complexity exponent is, however, halved with respect to common branch-and-bound path finders methods based on exhaustive search~\cite{edfs, pareto_bfs, exact_qos} (see Section~\ref{complexity_section}).  
%
%
Note also that in addition to the constrained shortest path, NM can simplify the process of finding all simple, $k$-constrained shortest or Pareto-optimal paths~\cite{pareto_bfs} from the source to the destination. 
%
%
\junk{
Note also how the exponential nature of NM is less of a problem if we are not interested in the optimal path, as existing heuristics such as single mixed metric~\cite{ktmk} or Lagrangian relaxation~\cite{jsmr} can reduce the $l\oplus p$ constraints to a $l\oplus1$ case. 
\textcolor{red}{This is confusing, because in table 1 on page 8 we say ``guaranteed optimal path" even for $l \oplus 1$. So why do we say here, ``if we are not interested in the optimal path"? I did not understand what you wanted to say here, except that we can convert $l\oplus p$ constraints to a $l\oplus1$.  Is it because the optimal path with l+1 may not be the optimal path with $l \oplus p$?  }
}

%
The general workflow of our NM algorithm is shown in Figure$~\ref{general_diagram}$: 
%
 NM is executed in three phases: $(i)$ a \textit{forward pass} or neighborhoods building (by using label-correcting),  
 $(ii)$ a \textit{backward pass} (with any instance of branch-and-bound exhaustive search), and a final $(iii)$ \textit{constraints validation} phase.
During the forward pass, NM builds the neighborhoods to estimate the path length. The backward pass is used to find end-to-end paths with a given length (hop count). The final constraints validation phase is used to keep the best path candidate and decide whether or not the path search should be extended to longer path candidates involving more neighbors. 

\begin{figure}[t!]
\centering
\begin{subfigure}[t]{0.12\textwidth}
\centering
\includegraphics[width=1\linewidth]{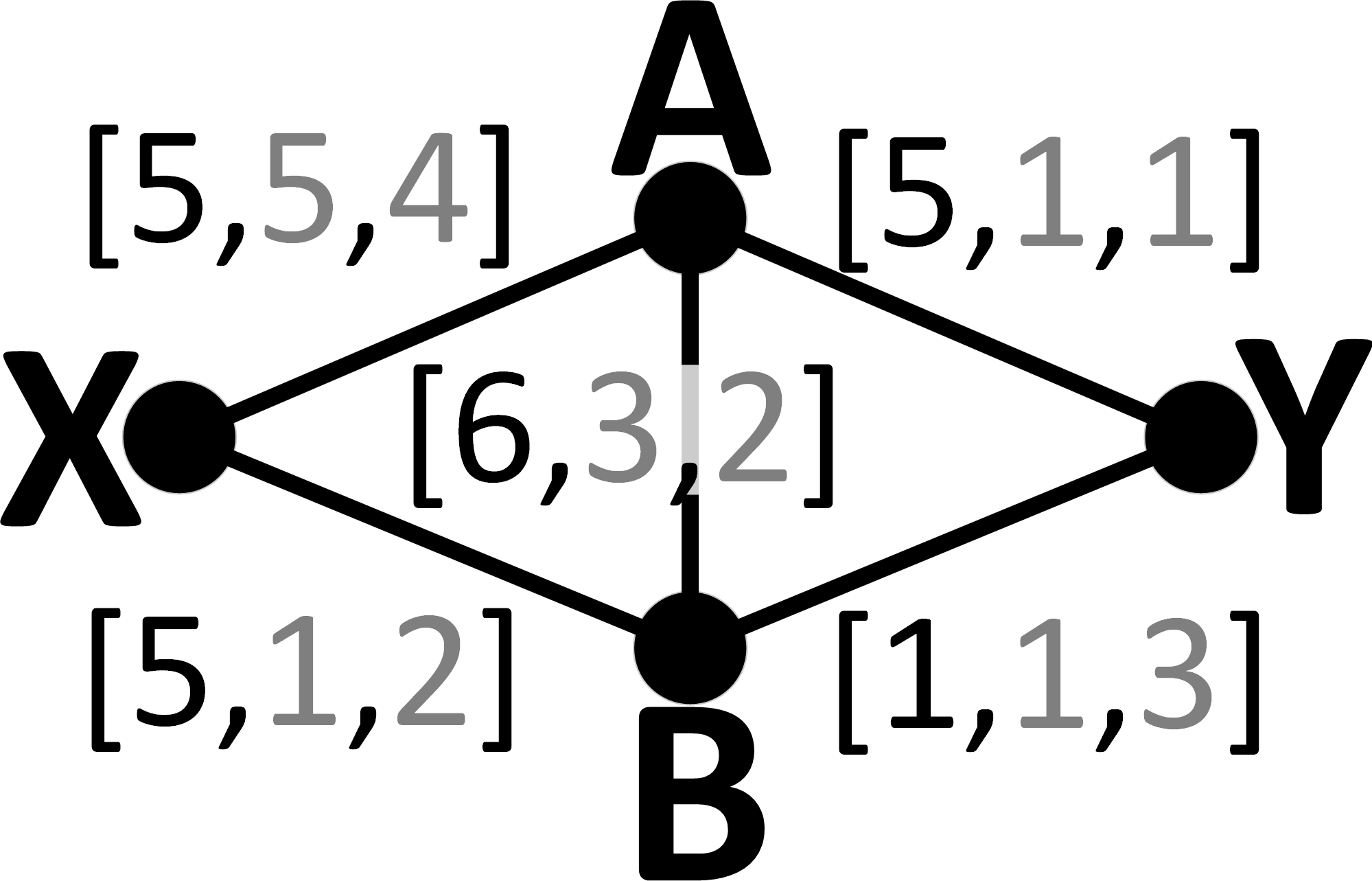}
\caption{}
\label{example_n}
\end{subfigure}
~
\begin{subfigure}[t]{0.13\textwidth}
\centering
\includegraphics[width=0.7\linewidth]{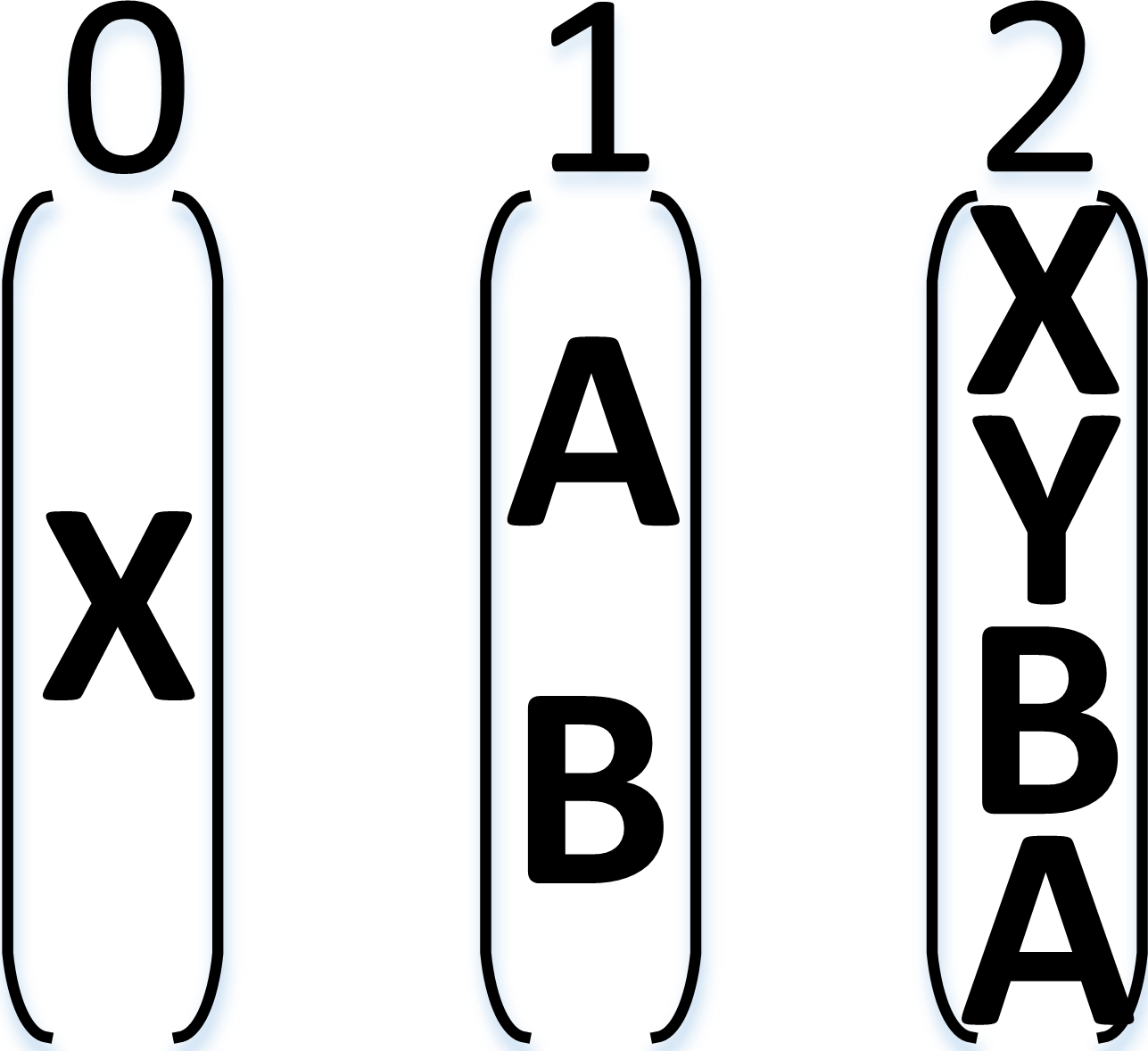}
\caption{}
\label{example_f1}
\end{subfigure}
~
\begin{subfigure}[t]{0.13\textwidth}
\centering
\includegraphics[width=0.7\linewidth]{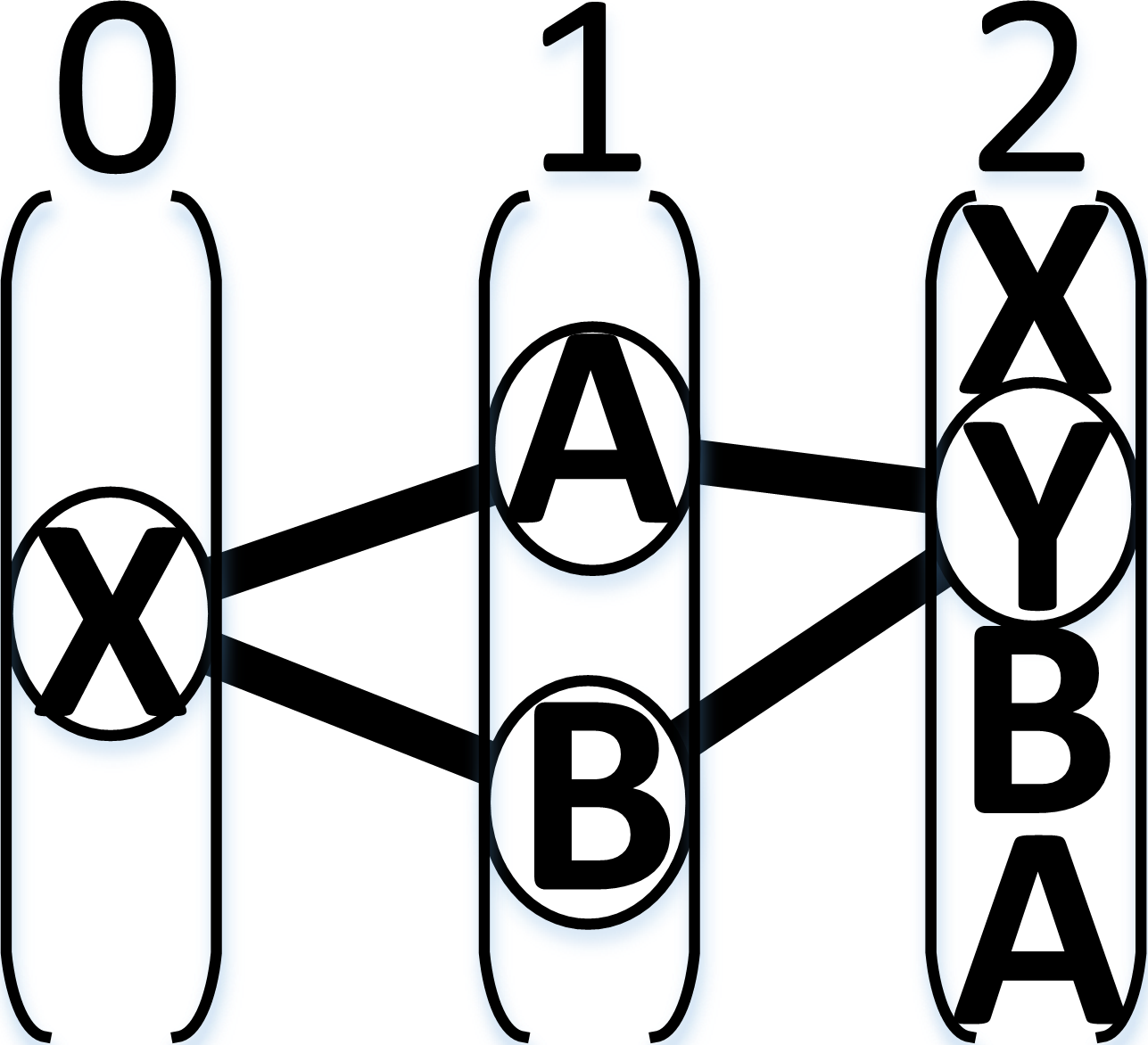}
\caption{}
\label{example_b1}
\end{subfigure}

\vspace{1mm}

\begin{subfigure}[t]{0.15\textwidth}
\centering
\includegraphics[width=0.85\linewidth]{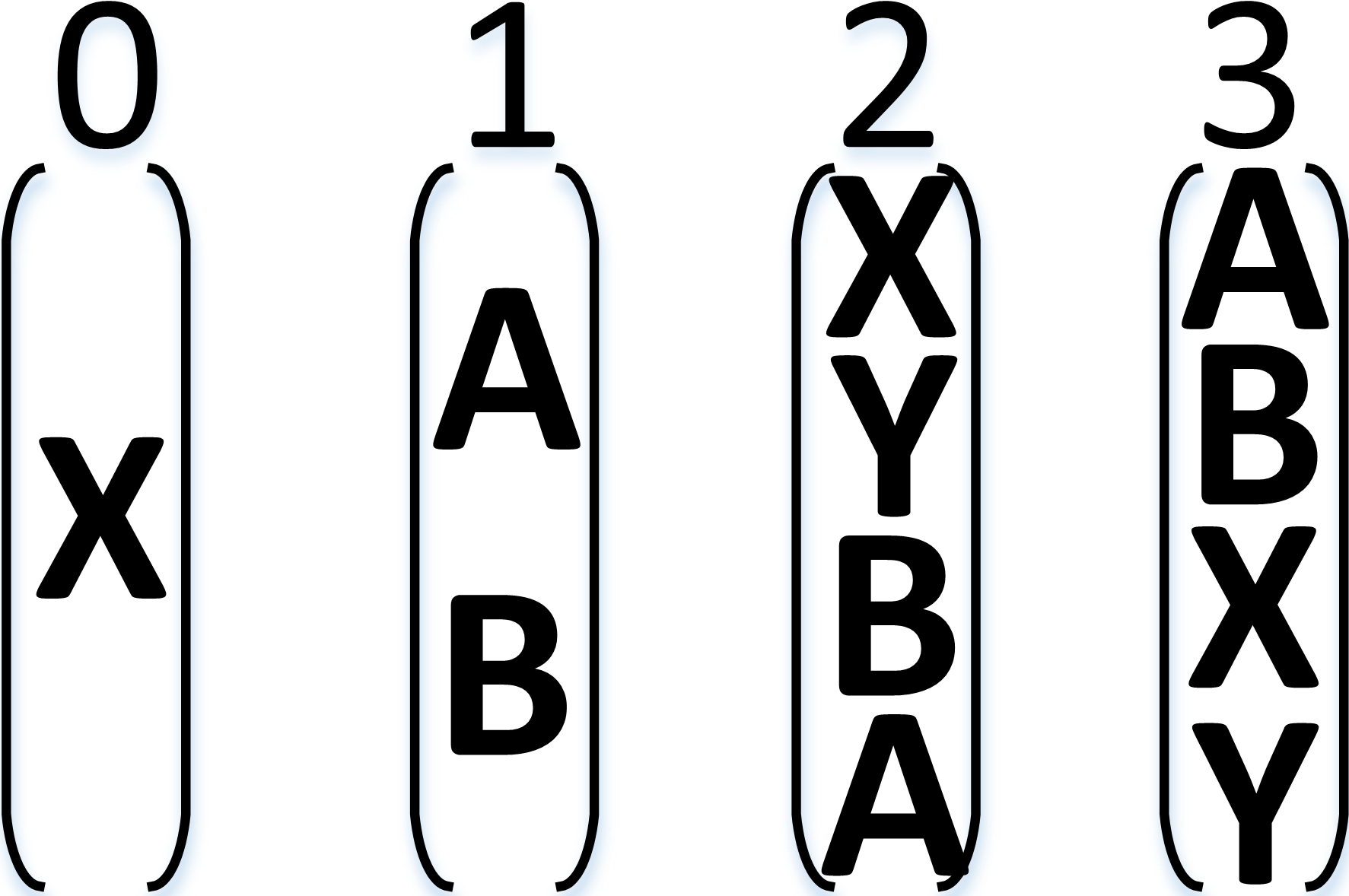}
\caption{}
\label{example_f2}
\end{subfigure}
~
\begin{subfigure}[t]{0.15\textwidth}
\centering
\includegraphics[width=0.85\linewidth]{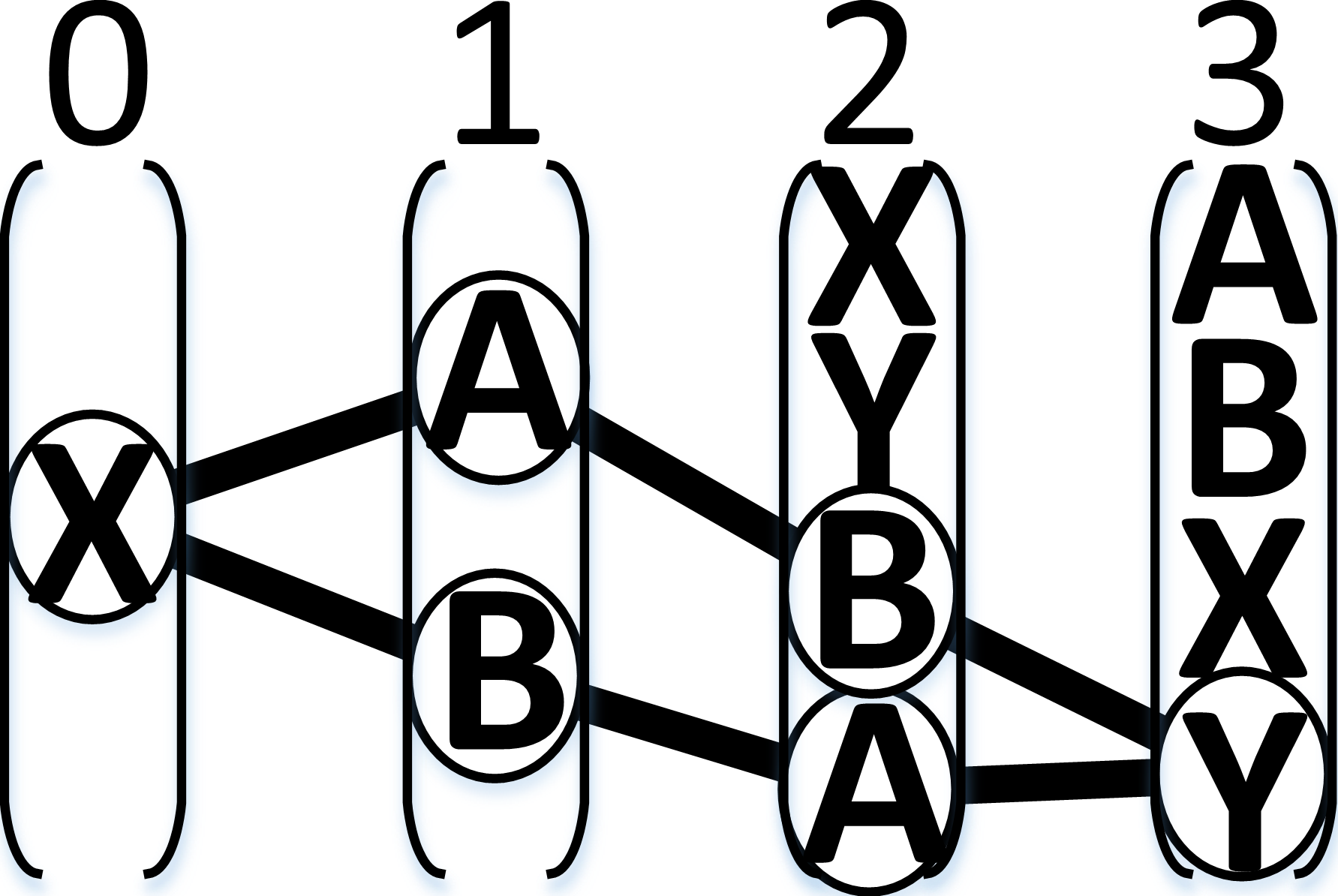}
\caption{}
\label{example_b2}
\end{subfigure}
\caption{\footnotesize{Running example of NM in $l \oplus p$ case: (a) An example network configuration with [bandwidth, delay, cost] link metrics; (b) NM forward pass phase estimates the min hop count distance to Y during the first iteration, and (c) backward pass identifies all the shortest path candidates, which in this case do not contain the \textit{optimal} path; (d) the forward pass adds more neighborhoods to find a path with longer distance to  Y. (e) The Backward pass identifies all paths of a given new length; in this case the path contains the \textit{optimal} $X\rightarrow B\rightarrow A \rightarrow Y$ solution.}}
\label{example_case3}
\vspace{-6mm}
\end{figure}

\begin{example}\label{example0}
Consider a network consisting of 4 nodes $X$, $Y$, $A$ and $B$, as shown in Figure$~\ref{example_n}$. 
On the link $(X,A)$, we denote with the first value $5$ the link constraint (in this case bandwidth), and with the second $5$ we refer to the first path constraint (in this case end-to-end delay); with the third value $4$ we refer to the second path constraint (in this case an arbitrary cost). In its general case, NM  finds a path from $X$ to $Y$ satisfying the three constraints $bw \geq 5$, $delay \leq 5$  and $cost \leq 5$ as follows: 
in the first phase, NM builds all neighborhoods starting from the source node $X$ until the destination node $Y$ is reached (Figure~\ref{example_f1}). Upon reaching $Y$, NM begins the backward pass to find the full set of min hop paths (Figure~\ref{example_b1}). If the set contains the \textit{optimal} path (in this case for Problem~\ref{rocp_def}), 
NM terminates with a solution. If no such path is found, NM builds an additional neighborhood as shown in Figure~\ref{example_f2} and performs another backward pass to check for optimality among paths that are one hop longer than at the previous iteration (Figure~\ref{example_b2}). NM iterates until either the solution is found, or the maximum path length is violated. In this example, NM returns the constrained shortest path $X\rightarrow B\rightarrow A\rightarrow Y$.
\end{example}

\noindent
{\bf Forward pass for $l\oplus p$.}
Given an arbitrary graph $G(V,E)$, where $|V|$ and $|E|$ represent the number of vertices and edges, respectively, in this phase NM successively builds neighborhoods $<NH>$ from the source $X$ to the destination $Y$. Algorithm~\ref{buildNb} describes the forward pass of NM. 
%
%
To build a new neighborhood $NH$, we add therein neighbors (adjacent vertices) of each vertex $u$ in the current neighborhood $cNH$ (line 6). For example, the first $NH$ includes nearest neighbors of the source $X$ (which is in the zero $NH$), and the second $NH$ contains nearest neighbors of vertices in the first $NH$, and so on. The first phase ends as soon as the destination $Y$ appears in $cNH$ (line 4), or $<NH>$ size is more or equal to $|V|$ (line 13).
 
\begin{algorithm}[h]
\scriptsize
\KwIn{$X$:= src, $Y$:= dest}
\KwOut{The list of neighborhoods $<NH>$ from $X$ to $Y$}

\SetAlgoLined
\Begin{
$cNH \longleftarrow X$\\
$<NH> \longleftarrow <NH> \cup cNH$\\
\While{$Y \notin cNH$} 
{
$NH \longleftarrow \emptyset$\\
\ForEach{Vertex $u \in cNH$}{$NH \longleftarrow NH \cup adjacent(u)$ } 
\eIf{$<NH>.size < |V|$}
{
$<NH> \leftarrow <NH> \cup NH$\\
$cNH \leftarrow NH$\\
}
{Return $Y$ is unreachable. }
}
}
\caption{Build Neighborhoods ($l\oplus p$ case)}
\label{buildNb}
\end{algorithm}

\noindent
{\bf Backward pass for $l \oplus p$.}
The best exhaustive search strategy may depend on the network topology, constraint and cost functions, and we leave it as a policy for NM. In this paper, we use EBFS for the backward pass of NM detailed in Algorithm~\ref{backwardAlg}. 
Note however, that this phase can use any exhaustive search ($e.g.$, EBFS~\cite{pareto_bfs}, EDFS~\cite{edfs} or exhaustive $k$-shortest path~\cite{exact_qos}) with the only difference being that \textit{we do not process all neighbors (adjacent vertices) of each vertex $u$ but only those which are within previous} $NH$ (line 10). 
%
%
The first step is to find the intersection between neighbors of the destination $Y$ with its previous $NH$ (line 5). This intersection is not an empty set, it contains at least one vertex $v$. For all obtained vertices we again build the intersection of their neighbors with the penultimate $NH$ (line 16). The second phase ends as soon as we hit the zero $NH$ (line 6), and as a result we obtain the collection of all paths with a length of $<NH>$ size between the source $X$ and the destination $Y$.


\begin{algorithm}[h]
\scriptsize
\KwIn{The list of neighborhoods $<NH>$  from $X$ to $Y$}
\KwOut{All paths $<path>$ from $X$ to $Y$ of $<NH>.size$ length} 

\SetAlgoLined
\Begin{
$path \longleftarrow Y$\\
$<path> \longleftarrow <path> \cup path$\\
$k \longleftarrow 1$\\
$NH \longleftarrow <NH>[size-k]$\\
\tcc{\textbf{EBFS:}}
\While{$NH$ $\neq <NH>[0]$}
{
$<tempPath> \longleftarrow \emptyset$\\
\ForEach{$path \in <path>$}
{
Vertex $u \longleftarrow path[1]$\\
\ForEach{Neighbor $v \in adjacent(u) \cap NH$}{$<tempPath>\longleftarrow v \cup path$}
}
$<path> \longleftarrow <tempPath>$ \\
$k \longleftarrow k+1$\\
$NH \longleftarrow NH[size-k]$\\
}
} 
\caption{Perform Backward Pass ($l\oplus p$ case)}
\label{backwardAlg}
\end{algorithm}

\noindent
{\bf Constraints validation for $l\oplus p$.}
In this last phase,  we check for candidates optimality, $i.e.$, we check whether or not a candidate path satisfies all $l \oplus p$ constraints, and keep the best candidate. 
At each consequent iteration, we first build an additional ($N+1$) neighborhood, repeat the backward pass and subsequently obtain all paths of length $N+1$, and then we check their feasibility and update the best known path candidate, if needed. 
Similarly to IBF~\cite{net_utilization}, we keep iterating while the candidate path length is less than the number of vertices $|V|$ and then return the optimal solution.


We close this subsection with three important remarks:
{\bf (1)  NM can be used to find $k$-constrained shortest paths}: the backward path at each iteration returns all possible path candidates of the same hop count. 
To find $k$-constrained shortest paths we need to keep not a single best (shortest) path candidate, but a set of $k$ best path candidates at each iteration and update this set if needed. Clearly, as in the worst case NM traverses all possible path candidates, its upper bound complexity does not change (see Section~\ref{complexity_section}).
{\bf (2) NM can be applied to both directed and undirected graphs making it an interesting solution even for NFV chain instantiations}. 
When traversing directed graphs, NM simply uses vertices' outgoing neighbors during the forward pass and incoming neighbors during the backward pass.
%
\noindent
{\bf (3) A distinctive feature of our NM is the construction of the intersection of two neighborhoods}. This leads to a quadratic reduction of path candidates for exhaustive search algorithms (see Section~\ref{complexity_section}).

\subsection{NM Search Space Optimizations}
\label{space_reduction_sec}
In this subsection we show how NM can be coupled with existing search space reduction techniques, $i.e.$, dominant paths or look ahead~\cite{edfs, pareto_bfs, exact_qos}, to speed up its time to solution. 
By leveraging the NM's double pass, we also propose a variant of the look ahead technique, $viz.$, ``Look Back"  without complexity overhead. We first describe the {\it dominated paths} method.
Observe that the dominant paths technique is not applicable in case we wish to use NM as $k$-constrained shortest paths, as it removes suboptimal candidates which can be among $k$ paths.

\noindent
{\bf Dominated paths (pruning by dominance or bound).}
The basic idea behind dominated paths
states that an algorithm can avoid evaluating candidate paths when their (multidimensional) distance is longer than other candidates distance, since they cannot be a part of the shortest solution~\cite{pareto_bfs, exact_qos}.
Consider for example Figure$~\ref{example_n}$: note how path $X \rightarrow A \rightarrow Y$ with distance vector $(6,5)$ is dominated by $X \rightarrow B \rightarrow Y$ path with  distance vector $(2,5)$ and hence it can be excluded to avoid unnecessary time-consuming processing. 
Path dominance is formally defined as follows:
\begin{definition}[dominant path]
We say that path $P_1$ dominates path $P_2$ if and only if:
\begin{equation*}
\begin{split}
\exists \; i \in 1,\ldots,p,c: d_i(P_1) < d_i(P_2) \land d_j(P_1)\le d_j(P_2),\\ \forall j\ne i \in 1,\ldots,p,c
\end{split}
\label{dominant_def}
\end{equation*}
\noindent
where $d_i$ is a distance corresponding to $p_i$ path constraint or to $c$ path cost.\footnote{Note how, if only the cost distance $d_c(P)$ is used in Equation~\ref{dominant_def}, pruning by path dominance is the well-known pruning by bound technique~\cite{edfs}.}
\end{definition}
%
\noindent
In its general form, we  can reduce NM's search space by removing the dominated paths. This is accomplished by keeping track of only non-dominated paths during its backward pass (Algorithm~\ref{backwardAlg}, line $11$). In particular, a path is added to a vertex $v$ if it is not dominated by any other path to $v$, or if it is not dominated by other paths from the source to the destination. 


\noindent
{\bf Look ahead (pruning by infeasibility).} 
We can further reduce NM's search space by omitting path candidates with endpoint $v$, if the sum of their current path distances and the residual best distances from $v$ to the destination violates any of the path constraints. 
This technique is known as ``look ahead"~\cite{exact_qos} and requires a run of Dijkstra or Bellman-Ford algorithm $p$ times to pre-compute best distances (corresponding to path constraints $p$) from all vertices to the destination. 

\junk{
For example, consider Figure$~\ref{example_n}$. We can omit path $X \rightarrow A$ from further processing as its first distance ($i.e.$, end-to-end delay) plus the best distance from $A$ to $Y$ violates the $delay$ constraint: $5+1>5$.
\noindent
In our case, we can compute best distances from the source to all other vertices and use them during backward pass. However, as we show in Section~\ref{exhaustive_sec}, look ahead helps only in a case of non lenient constraints. Moreover, for large-scale networks ($e.g.$, 1000 nodes and up) it introduces a significant path computation time overhead.
}

\noindent
{\bf Looking backwards (pruning by infeasibility).} 
A further optimization can be obtained  applying the look back technique: the best distance could be estimated from the intermediate vertex $v$ to the source (instead of the destination) for each neighborhood containing $v$ during the NM forward pass.
%
%
We then use such information during the backward pass to exclude paths from $v$ at neighborhood $N$ that violate corresponding path constraints.
%
Note that, as we coupled it with forward pass, the look back space reduction does not introduce additional overhead with respect to the look ahead optimization ($i.e.$, running Dijkstra $p+1$ times). 

\begin{figure}[t!]
\centering
\begin{subfigure}[b]{0.18\textwidth}
\centering
\includegraphics[width=1\linewidth]{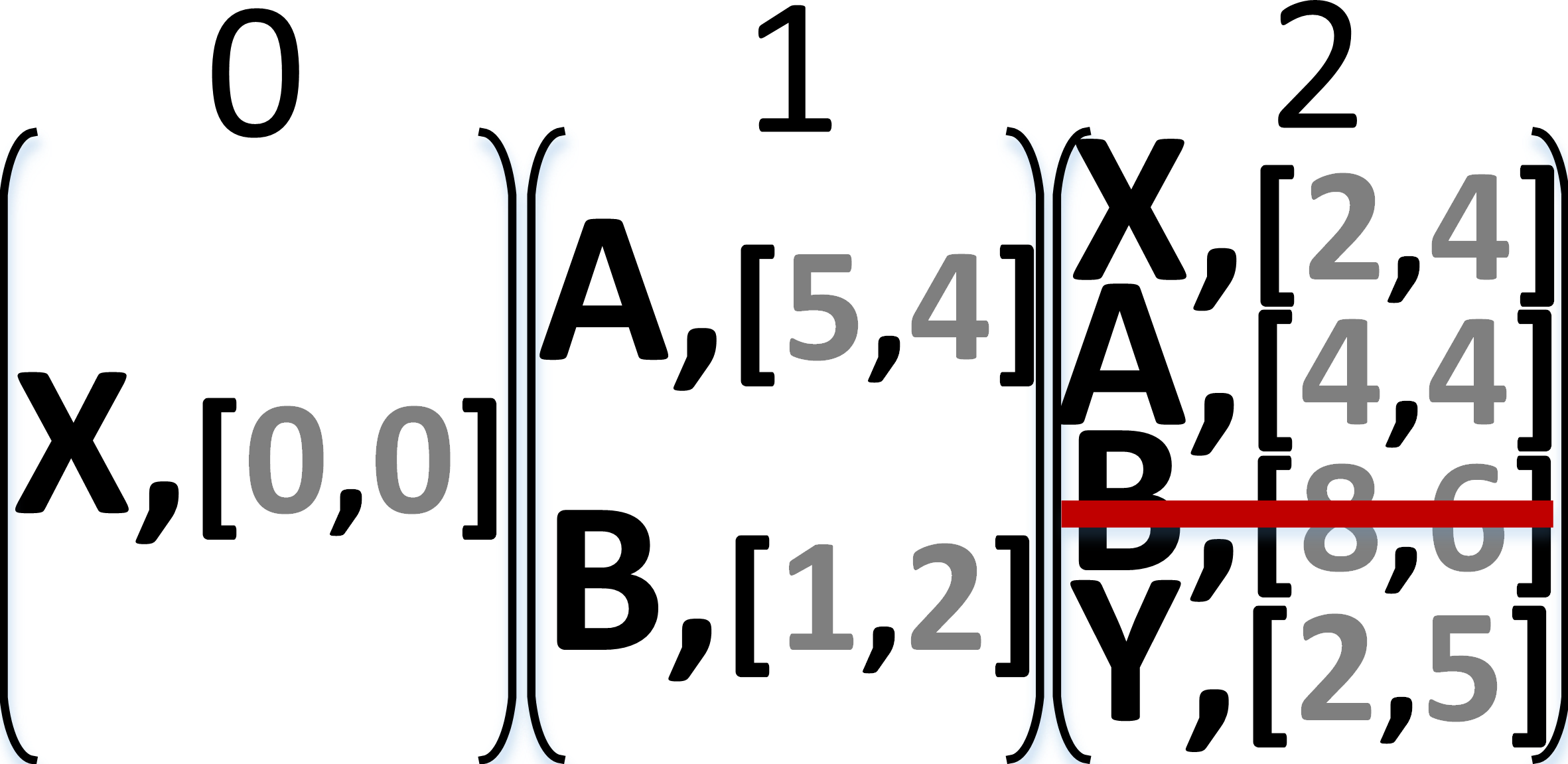}
\caption{}
\label{lb_forward}
\end{subfigure}
\qquad
\begin{subfigure}[b]{0.18\textwidth}
\centering
\includegraphics[width=1\linewidth]{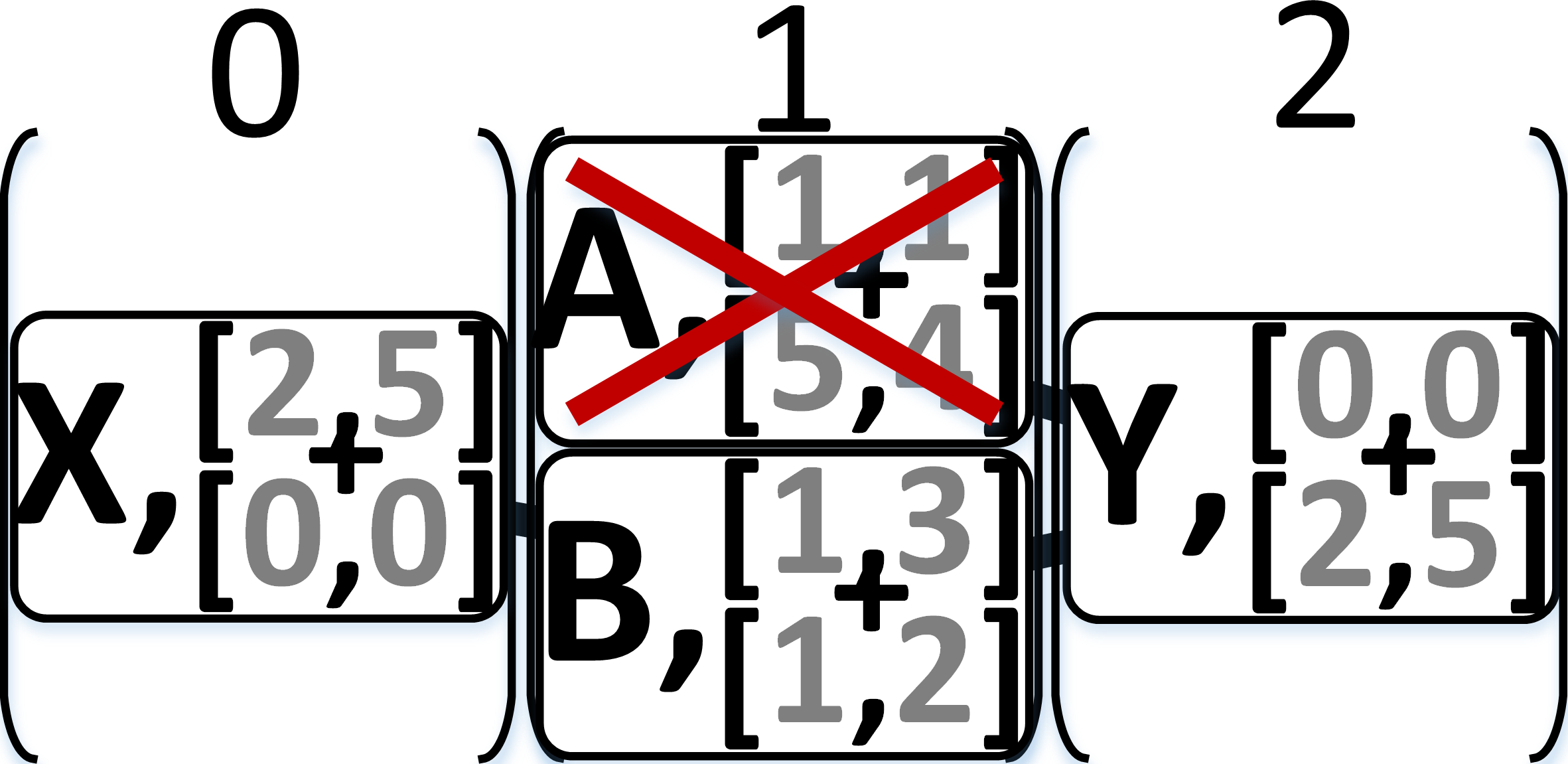}
\caption{}
\label{lb_backward}
\end{subfigure}
\vspace{-2mm}
\caption{\footnotesize{NM running in its in $l\oplus p$ general case  reduce its search space with a {\it Look Back}. (a) Forward pass: excludes vertices violating at least a path constraint; (b) Backward pass: NM looks back and removes a path candidate which sum of its current $delay$ and the best (estimated during the forward pass) residual $delay$ violates the $delay$ constraint ($i.e.$, $1+5 > 5$).}}
\label{example_lb_fig}
\vspace{-6mm}
\end{figure}

\begin{example}\label{example_lb}
In this example we revisit  Example~\ref{example0}, to describe how NM in its the general case coupled with a Look Back search may reduce the search space when finding a path from $X$ to $Y$. We assume that two path constraints $delay \leq 5$ and $cost \leq 5$ need to be satisfied. During the forward pass (Figure$~\ref{lb_forward}$) NM estimates the path length to Y, keeping track of all  constraint-satisfying distances  for each vertex and at each neighborhood. Vertices whose estimated distance violate at least one path constraint are instead removed.
During the backward pass, NM removes $A \rightarrow Y$ path candidate whose sum of the current $delay$ (which equals to 1) and the best estimated $delay$ from the source to $A$ (which equals to 5) violates $delay$ constraint $1+5>5$ (Figure$~\ref{lb_backward}$).
\end{example}

\junk{
\noindent
Applying the look back optimization brings two main advantages:
(i) it reduces the number of iterations needed for NM to converge by keeping only those nodes in neighborhoods whose best distance from the source do not violate corresponding path constraints;
and (ii) it significantly reduces number of processing paths, as we learn best distances not for each vertex ($e.g.$, as in Look Ahead case) but for each neighborhood which contains this vertex allowing more precise judging about sub path feasibility.
}

\subsection{NM with links and a single path constraint ($l/l\oplus1$ cases)}
\label{nm_l_1}

The worst case running time of NM becomes polynomial (by avoiding an exhaustive search) if either a general constrained shortest path  (see Problem~\ref{csp_def}) with only $l$ link constraints  or the resource optimal constrained path (see Problem~\ref{rocp_def}) with $l$ links and a single path constraint ($l \oplus 1$) are sought  (see Section~\ref{complexity_section}). This is because unnecessary iterations and phases (including the exhaustive search) are avoided. 
%
%
Note however that the constrained shortest path in $l\oplus1$ case is still NP-hard~\cite{edfs}. Thus, the general NM version should be used.

\begin{figure}[h!]
\vspace{-3mm}
\centering
\begin{subfigure}[b]{0.13\textwidth}
\centering
\includegraphics[width=1\linewidth]{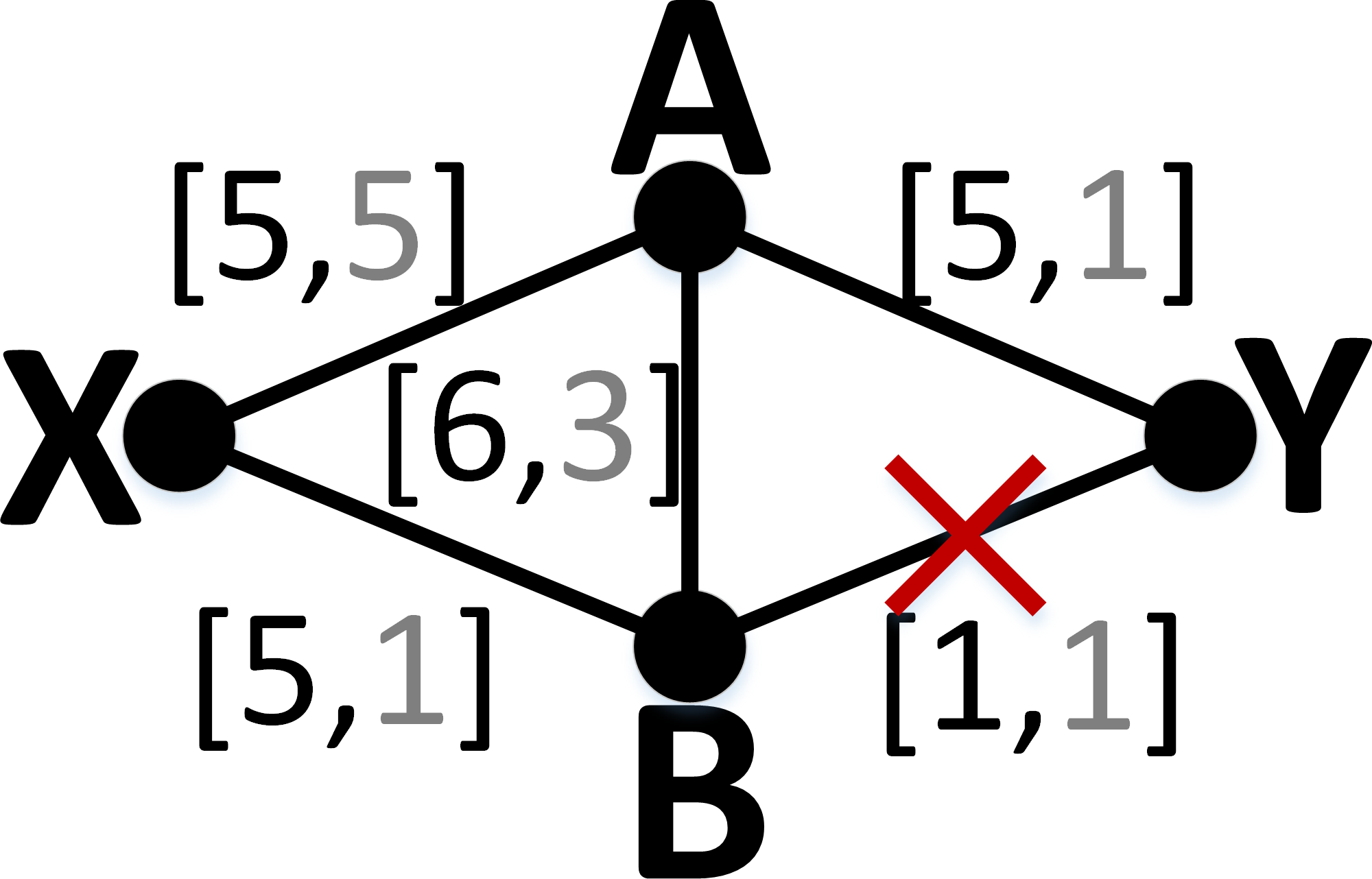}
\caption{}
\label{example_n1}
\end{subfigure}
~
\begin{subfigure}[b]{0.16\textwidth}
\centering
\includegraphics[width=1\linewidth]{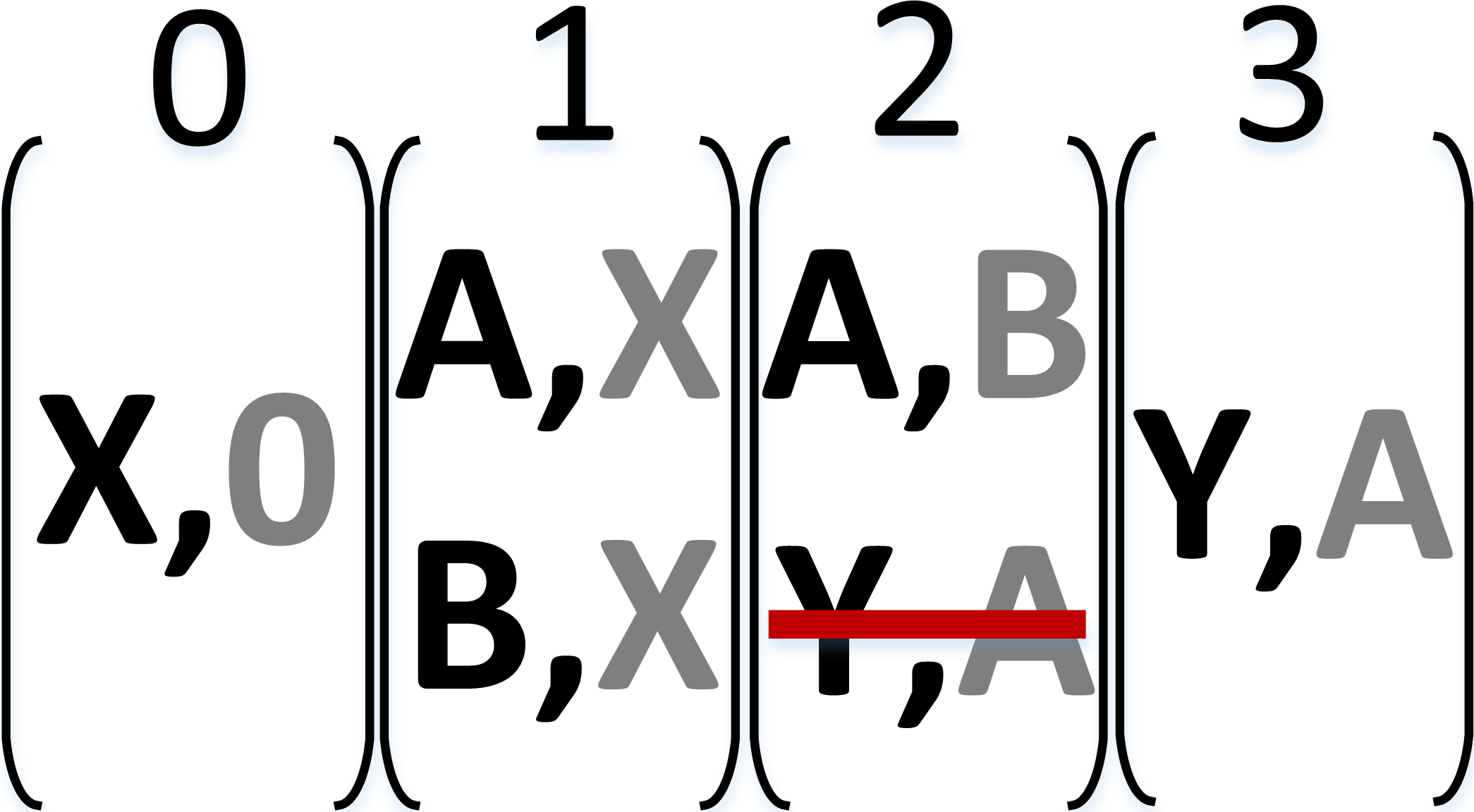}
\caption{}
\label{example_p2}
\end{subfigure}
~
\begin{subfigure}[b]{0.16\textwidth}
\centering
\includegraphics[width=1\linewidth]{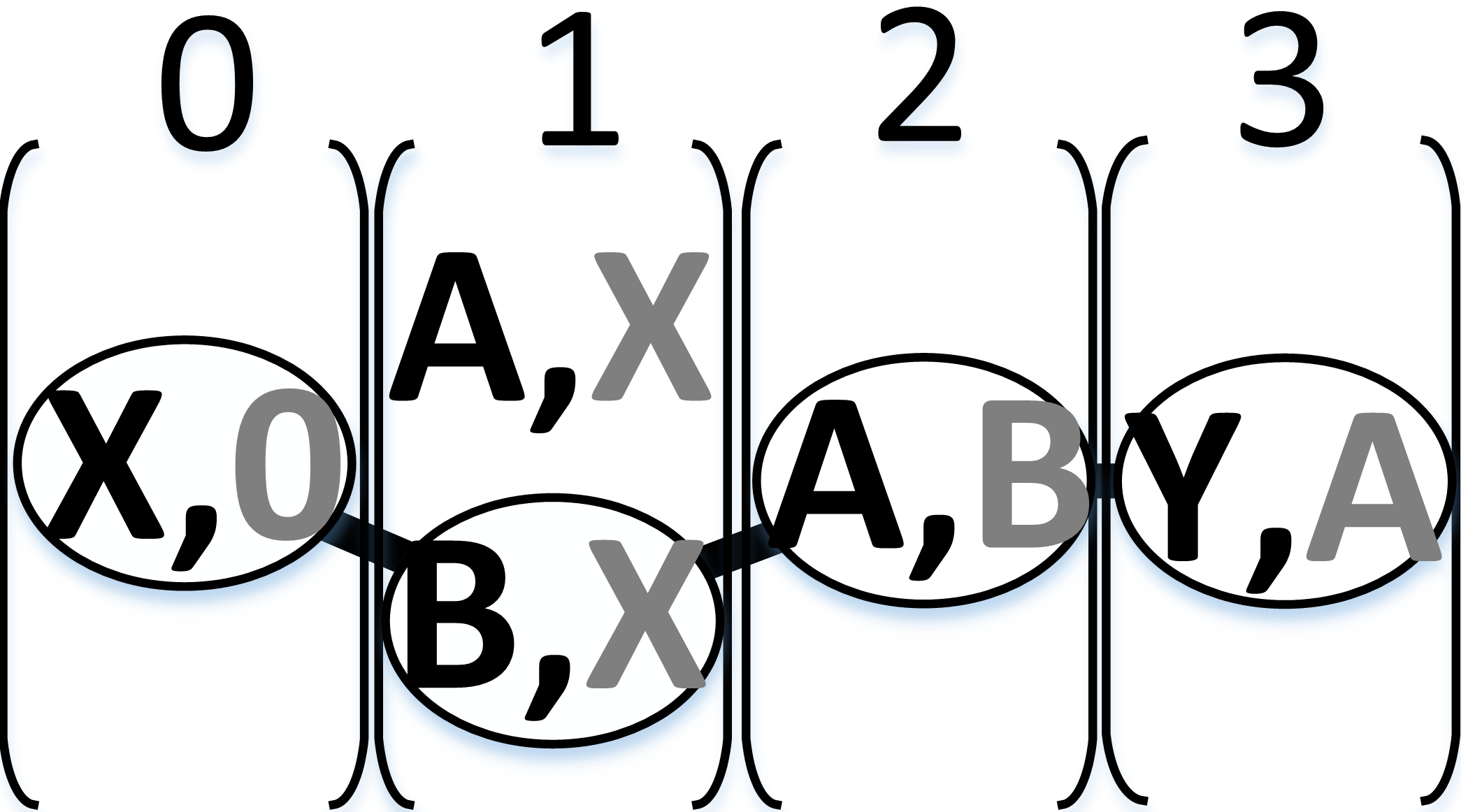}
\caption{}
\label{example_p3}
\end{subfigure}
\caption{\footnotesize{Running example of NM in $l\oplus1$ case: (a) In the \textit{pre-routing phase} NM prunes $B\rightarrow Y$ due a $bw$ violation; (b) In the \textit{forward phase} NM finds the best length to Y saving vertices' predecessors from previous neighborhoods; (c) \textit{back track phase} identifies the \textit{resource optimal constrained path} $X\rightarrow B \rightarrow A \rightarrow Y$ by recursive predecessor visits.}}
\label{example_case}
\vspace{-5mm}
\end{figure}

\begin{example}
Revisiting our example~\ref{example0}, NM finds a path from $X$ to $Y$ satisfying the two constraints $bw \geq 5$ and $delay \leq 5$ as follows: In the \textit{pre-routing phase} (see Figure$~\ref{example_n1}$) NM prunes $B\rightarrow Y$ due a $bw$ violation. In the \textit{forward phase} (see Figure$~\ref{example_p2}$) NM finds the best length to Y saving information about vertices' predecessors from previous neighborhoods, discarding candidates that violate the path constraint.
During the  \textit{back track phase} NM recursively constructs the resource optimal constrained path by recursive predecessor visits (see Figure$~\ref{example_p3}$).
\end{example}

\noindent
{\bf NM pseudocode for $l \oplus 1$}. Algorithm~\ref{buildNbWithPath} describes the NM procedure with only $l\oplus1$ constraints.
The neighborhoods data structure $<NH>$  identifies each vertex with its predecessors in previous neighborhoods; each vertex $u$ also contains a path distance $D(u)$. 
During the pre-routing phase, $D(u)$ (line $3$), the weights of each edge not satisfying all $l$ constraints (line $7$) are set to $\infty$.

During the forward phase of NM, we successively build the neighborhoods $<NH>$; for each {\it current} neighborhood $cNH$ of $u$ we exclude all neighbors that do not satisfy the path constraint or that have previous distance lower than $dist$ (line $18$). 
Note that to compute $dist$ (line $17$) we assume $D(u)$ to be a distance of $u$ after its inclusion in current neighborhood ($cNH$).
%
The forward phase ends when the destination $Y$ appears in $cNH$ (line $13$). If the number of neighborhoods $<NH>$ is equal to the number of vertices $|V|$,  the algorithm  terminates concluding that a negative weight cycle is detected (line $28$), or that a path does not exist (line $30$).
During the back-track phase, NM then recursively finds the \textit{resource optimal constrained path} 
between $X$ and $Y$. 

\begin{algorithm}[h]
\scriptsize
\KwIn{$X$:= src, $Y$:= dest, $l$ link constraints, $p$  single path constraint.}
\KwOut{The \textit{optimal} $path$ between $X$ and $Y$ (which satisfies $l\oplus1$).} 
\SetAlgoLined
\Begin{
\tcc{\textbf{pre-routing phase:}}
\ForEach{Vertex $u \in |V|$}
{
$D(u) \leftarrow \infty$\\
}
\ForEach{Edge $u\rightarrow v \in |E|$}
{
\If{$u\rightarrow v$ does not satisfy $l$ }
{
$w(u\rightarrow v)\leftarrow \infty$\\
}
}

\tcc{\textbf{forwarding phase:}}
$D(X) \leftarrow 0$\\
$cNH \longleftarrow (X,NIL)$\\
$<NH> \longleftarrow <NH> \cup cNH$\\
\While{$Y \notin cNH$} 
{
$NH \longleftarrow \emptyset$\\
\ForEach{Vertex $u \in cNH$}
{
\ForEach{Neighbor $v \in adjacent(u)$ }
{
$dist \leftarrow D(u) + w(u\rightarrow v)$\\
\If{$dist < D(v)$ and $dist\le p$} 
{
$D(v) \leftarrow dist$\\
$NH \longleftarrow NH \cup (v,u)$\\
}
}
} 
\uIf{$NH \notin \emptyset$ and $|<NH>| < |V|$}
{
$<NH> \leftarrow <NH> \cup NH$\\
$cNH \leftarrow NH$\\
}
\uElseIf{$|<NH>| == |V|$}
{Return Negative Weight Cycle is detected. }
\Else{Return $Y$ is unreachable. }
}

\tcc{\textbf{back track phase:}}
$path \leftarrow Y$\\
$k \leftarrow |<NH>|$\\
\While{$k > 0$}
{
$predecessor \leftarrow <NH>[k, path(1)]$\\
$path \leftarrow predecessor \cup path$\\
$k \leftarrow k-1$\\
}
}
\caption{NM in $l\oplus1$ case}
\label{buildNbWithPath}
\end{algorithm}

\subsection{NM with only link constraints ($l$ case)}
In presence of only $l$ link constraints and when the resource optimal constrained path is sought,  NM can be further simplified to run in linear time by omitting the path constraint checking and by using hop count as a vertex distance metric (Algorithm~\ref{buildNbWithPath}, line $18$). As a result, NM traverses each vertex exactly once, and hence all neighborhoods contain unique and distinct vertices, $i.e.$, a common vertex for two or more neighborhoods cannot exist.
%

\subsection{NM Optimal Solution}
\label{generalTh}

\begin{th_nm}
\emph{(Theorem of the Optimal Solution)}
\label{th1}
\\NM always finds \textit{the optimal path} if it exists.
\end{th_nm}
\begin{proof} 
See Appendix~\ref{NM_proof}. 
\end{proof} 
%

\section{Asymptotic Complexity Analysis}
\label{complexity_section}
In this section, we provide a complexity analysis comparison among our NM and related algorithms such as: IBF~\cite{net_utilization}, EDijkstra~\cite{wacr}, EDFS~\cite{edfs} and EBFS~\cite{pareto_bfs,widyono} --- common branch-and-bound exhaustive search approaches. 
%
%
Table~\ref{table1} summarizes the main results of this comparison, where $|V|$ is the total number of vertices, and $|E|$ is the total number of edges. 

\begin{table*}[t!]
\renewcommand{\arraystretch}{1.0}
\caption{Virtual path embedding algorithms complexity}
\label{table1}
\scriptsize
\centering
\begin{tabular}{c|c|c|c|c}
Case:&EDijkstra~\cite{wacr}&IBF~\cite{net_utilization}&EDFS~\cite{edfs}, EBFS~\cite{pareto_bfs,widyono}&NM\\
\hline
\hline
\multirow{3}{*}{$l$}
&Time: $O(|V|+|E|\cdot \it{l})$&Time: $O(|V| + |E|\cdot \it{l})$&Time: $O(|V|+|E|\cdot \it{l})$&Time: $O(|V| + |E|\cdot \it{l})$\\

&Space: $O(|V|+|E|)$&Space: $O(|V|+|E|)$&Space: $O(|V|+|E|)$&Space: $O(|V|+|E|)$\\

&\textit{resource optimal constrained path}&\textit{resource optimal constrained path}&\textit{resource optimal constrained path}&\textit{resource optimal constrained path}\\
\hline
\multirow{4}{*}{$l/l\oplus 1$}
&Time: $O(|V|log|V|+|E|\cdot \it{l})$&Time: $O(|V||E| + |E|\cdot \it{l})$&Time: $O\big(\big(\frac{|E|}{|V|}\big)^{|V|}+|E|\cdot \it{l}\big)$&Time: $O(|V||E| + |E|\cdot \it{l})$\\

&Space: $O(|V|+|E|)$&Space: $O(|V||E|)$&Space: $O\big(\big(\frac{|E|}{|V|}\big)^{|V|}\big)$&Space: $O(|V||E|)$\\
&\textbf{constrained shortest path}/&\textbf{constrained shortest path}/&\multirow{2}{*}{\textbf{constrained shortest path}}&\textbf{constrained shortest path}/\\
&constrained path only&\textit{resource optimal constrained path}&&\textit{resource optimal constrained path}\\
\hline
\multirow{3}{*}{$l\oplus p$}&\multirow{3}{*}{N/A}&\multirow{3}{*}{N/A}&Time: $O\big(\big(\frac{|E|}{|V|}\big)^{|V|}p+|E|\cdot \it{l}\big)$&Time: $O\big(\big(\frac{|E|}{|V|}\big)^\frac{|V|}{2}p+|E|\cdot \it{l}\big)$\\

&&&Space:$O\big(\big(\frac{|E|}{|V|}\big)^{|V|}p\big)$&Space: $O\big(\big(\frac{|E|}{|V|}\big)^\frac{|V|}{2}p\big)$\\

&&&\textbf{constrained shortest path}&\textbf{constrained shortest path}\\
\end{tabular}
\vspace{-6mm}
\end{table*}

\noindent
{\bf Theoretical results summary.} 
The major benefits of NM arise when we are seeking the NP-hard~\cite{edfs, pareto_bfs} constrained shortest paths in $l \oplus 1$ and $l \oplus p$ cases. 
We show how, under $l$ constraints, EDijkstra finds constrained (shortest) paths faster than IBF and NM. 
We also show how NM is an alternative for IBF to find the resource optimal constrained path when accepting $l \oplus 1$ constraints (and thus \textit{flexible}), and how EDijkstra loses any path optimality guarantees in that case. 
%
Finally, when accepting $l \oplus 1$ and $l \oplus p$ constraints, we show how time and space EBFS and EDFS complexities are quadratically higher with respect to our NM.

\subsection{Complexity Analysis for the $l$ and $l\oplus 1$ Cases}

\noindent
\textbf{EDijkstra complexity ($l$ and $l\oplus 1$ cases).}
Given the presence of merely link constraints in the $l$ case, EDijkstra can find the constrained shortest path by minimizing an arbitrary cost function and its variant the resource optimal constrained path by minimizing path hop count. 
The original Dijkstra algorithm can run in $O(|V|log|V| + |E|)$ time utilizing Fibonacci Heap~\cite{fibonacci}. Its min hop count variant can be reduced to $O(|V| + |E|)$ with Thorup's algorithm~\cite{thorup}. The corresponding space complexity for both Dijkstra variants is $O(|V| + |E|)$. 
Before applying Dijkstra, we also need to verify the $l$ constraints satisfaction by pruning all edges which violate $l$ in O($|E|\cdot \it{l}$).
Thus, the total time complexity of finding the constrained shortest path by EDijkstra is $O(|V|log|V| + |E|\cdot \it{l})$ with a space complexity of $O(|V| + |E|)$.
For the resource optimal constrained path, the total time complexity of EDijkstra is reduced to $O(|V| + |E|\cdot \it{l})$ with the same space complexity. 

Given the presence of link and a singe path constraints in the $l\oplus 1$ case, EDijkstra has to minimize distance which is a metric for the required path constraint to guarantee feasibility of the solution, $i.e.$, to find a feasible solution if it exists. 
Thus, EDijkstra loses its ability to any path length optimization. 
The total time complexity of finding a constrained path by EDijkstra in $l \oplus 1$ case is again $O(|V|log|V| + |E|\cdot \it{l})$ with a space complexity of $O(|V| + |E|)$.
Note that the constrained shortest path in $l \oplus 1$ case is NP-hard~\cite{edfs}, and thus requires an exhaustive search for the solution. 
As a result, EDijkstra cannot find any variant of the constrained shortest path with $l \oplus 1$ constraints.



\noindent
\textbf{IBF complexity ($l$ and $l \oplus 1$ cases).}
The original Bellman-Ford shortest path algorithm~\cite{bellman} runs in $O(|V||E|)$ time and similarly to EDijkstra can be extended to meet $l$ constraints in $O(|E|\cdot \it{l})$ additional time.
Thus, in $l$ case we can find the constrained shortest path by minimizing an arbitrary cost function with Bellman-Ford algorithm with $O(|V||E| + |E|\cdot \it{l})$ time and $O(|V||E|)$  space complexities.
In contrast to EDijkstra, Bellman-Ford can also iteratively find shortest paths in ascending hop count order. 
Hence,  the iterative version of Bellman-Ford (IBF) algorithm can minimize path hop count by visiting each vertex and ``relax" its adjacent edges only once, starting from the source while advancing towards the destination. 
This steps takes $O(|V|+|E|)$ allowing IBF to find the resource optimal constrained path in $l$ case with $O(|V|+|E| \cdot \it{l})$ time and $O(|V| + |E|)$ space complexities.

The same iterative property of IBF can be used to find the resource optimal constrained path in $l \oplus 1$ case. 
To this aim, at each iteration when the next best hop count candidate is found, we check its path constraint feasibility. 
Once, the feasible path is found, IBF returns the resource optimal constrained path, $i.e.$, the min hop count path which satisfy a single path constraint. 
Thus, IBF finds the resource optimal constrained path in $l \oplus 1$ case with the same $O(|V||E| + |E|\cdot \it{l})$ time and $O(|V||E|)$  space complexities as for the constrained shortest path in $l$ case. 
Due to NP-hardness of the constrained shortest path in $l \oplus 1$ case, IBF is also not applicable to find it.

\noindent
\textbf{EDFS and EBFS complexities ($l$ and $l\oplus 1$ cases).}
In  case of only $l$ link constraints both EDFS and EBFS can find the resource optimal constrained path (by minimizing hop count) without an exhaustive search. To this end, we can run the original DFS and BFS algorithms which have linear time complexity $O(|V|+|E|)$. The only difference lays again in the adjacent link checking that ensures the $l$ constraints satisfaction resulting in a time complexity of $O(|V|+|E|\cdot \it{l})$, and a space complexity of $O(|V|+|E|)$.

However, to find the constrained shortest path with $l$ constraints, or both the constrained shortest and the resource optimal constrained paths with $l\oplus 1$ constraints, EDFS and EBFS algorithms would have to build and check all possible paths from the source node (exhaustive search). For each potential path, algorithms can compute cost and path metrics and check for a path constraint satisfaction in $O(2)$.  All constraint violating edges could be pruned prior to running EDFS or EBFS. Note how, as EDFS and EBFS  algorithms use a branch-and-bound approach, they may visit each vertex more than once. The total number of path candidates can be bound as $O(b^d)$~\cite{korf}, where $b$ is the number of neighbors, and $d$ is the maximum loop-free path hop count. The average number of neighbors per vertex $b$ can be obtained from the hand-shaking lemma\footnote{Without loss of generality, we can assume undirected network graphs as in a directed graph, $b$ equals to the average outdegree: $b=\frac{|E|}{|V|}$.}: 
\begin{equation}
b=\frac{\sum_{i=v}^V{(neighbors \: of\: vertex\: v)}}{|V|}=\frac{2 |E|}{|V|}.
\label{eq1}
\end{equation}
%
Using Equation~\ref{eq1} and based on the fact that the maximum loop-free path hop count equals to $|V-1|$, the EBFS time complexity $O^{l\oplus 1}$ is:
\begin{equation}
O^{l\oplus 1}=O(2b^d+|E|\cdot \it{l})=O \Big(\Big(\frac{|E|}{|V|}\Big)^{|V|}+|E|\cdot \it{l} \Big).
\label{eq4}
\end{equation} 
Based on Equation~\ref{eq4}, the EBFS space complexity is $O\Big(\Big(\frac{|E|}{|V|}\Big)^{|V|}\Big)$.

\noindent
\textbf{NM complexity ($l$ and $l\oplus1$ cases).}
To estimate the time complexity of NM of finding the constrained shortest path with $l$ constraints or the resource optimal constrained path with $l\oplus 1$ constraints, we first obtain the average number of neighbors per vertex $b$ from the hand-shaking lemma shown in Equation~\ref{eq1}.
During the \textit{pre-routing phase}, NM visits each vertex and edge to either set initial values or prune an edge due to a constraint violation in $O(|V|+|E|\cdot \it{l})$. During the \textit{back track phase}, which runs in linear time, NM recursively visits each predecessor starting from the destination in $O(|V|)$.
During the \textit{forward phase} we have quadratic complexity as to construct a neighborhood with best path distances, we loop over all $b$ neighbors of each node of the previous neighborhood in $O(|V|b)$. The total number of neighborhoods is at most the maximum loop-free path hop count which equals to $|V-1|$.
Here we assume that a vertex look up and placement takes $O(1)$ time. The overall NM complexity is hence:
\begin{equation}
O^{l \oplus 1}=O(|V|^2b + 2|V| + |E|l)=O(|V||E| + |E|\cdot \it{l}).
\label{eq_nm_complexity}
\end{equation}
Based on Equation~\ref{eq_nm_complexity}, the NM space complexity is $O(|V||E|)$. 

As for related algorithms, the process of finding the resource optimal constrained path with $l$ constraints can be further simplified to linear complexity.
%
To minimize path hop count, all neighborhoods need to contain only unique and distinct vertices, $i.e.$, a common vertex for two or more neighborhoods  cannot exist; this means that  the maximum size of all neighborhoods is $|V|$. For each vertex in the neighborhoods set, the complexity of retrieving its neighbors is $O(b)$, assuming that a vertex look up and placement takes $O(1)$ time. Therefore, the time complexity of NM in this case is  $O(|V|b + 2|V| + |E|\cdot \it{l}) = O(|V| + |E|\cdot \it{l})$ and its space complexity is $O(|V| + |E|)$.

\subsection{Complexity Analysis for the $l\oplus p$ Case}

\noindent
\textbf{EDijkstra and IBF complexities ($l\oplus p$ case).}
For completeness, we mention that neither EDijkstra nor IBF are applicable to any variant of the constrained shortest path in the $l \oplus p$ case.


\noindent
\textbf{EDFS and EBFS complexities ($l\oplus p$ case).}
In this case, both EDFS and EBFS can find any variant of the constrained shortest path or $k$ such paths in exponential time through exhaustive search.
As in the previous case, edge pruning phase can be done prior to EDFS of EBFS runs, which build all possible paths from the source node, but now for each new path they check $p$ constraints satisfaction by calculating $p+1$ new path and cost metrics $O(2p)$. Hence, both EDFS and EBFS have the following time complexity $O^{l\oplus p}$:
\begin{equation}
O^{l\oplus p}=O(2p\cdot b^d + |E|l)=O\Big(\Big(\frac{|E|}{|V|}\Big)^{|V|}p + |E|\cdot \it{l}\Big)
\label{eq7}
\end{equation} 
The space complexity equals to $O\Big(\Big(\frac{|E|}{|V|}\Big)^{|V|}p\Big)$ due to the fact that we have to store $p$ metrics for each path.

\begin{figure}[t!]
\centering
\includegraphics[width=0.8\linewidth]{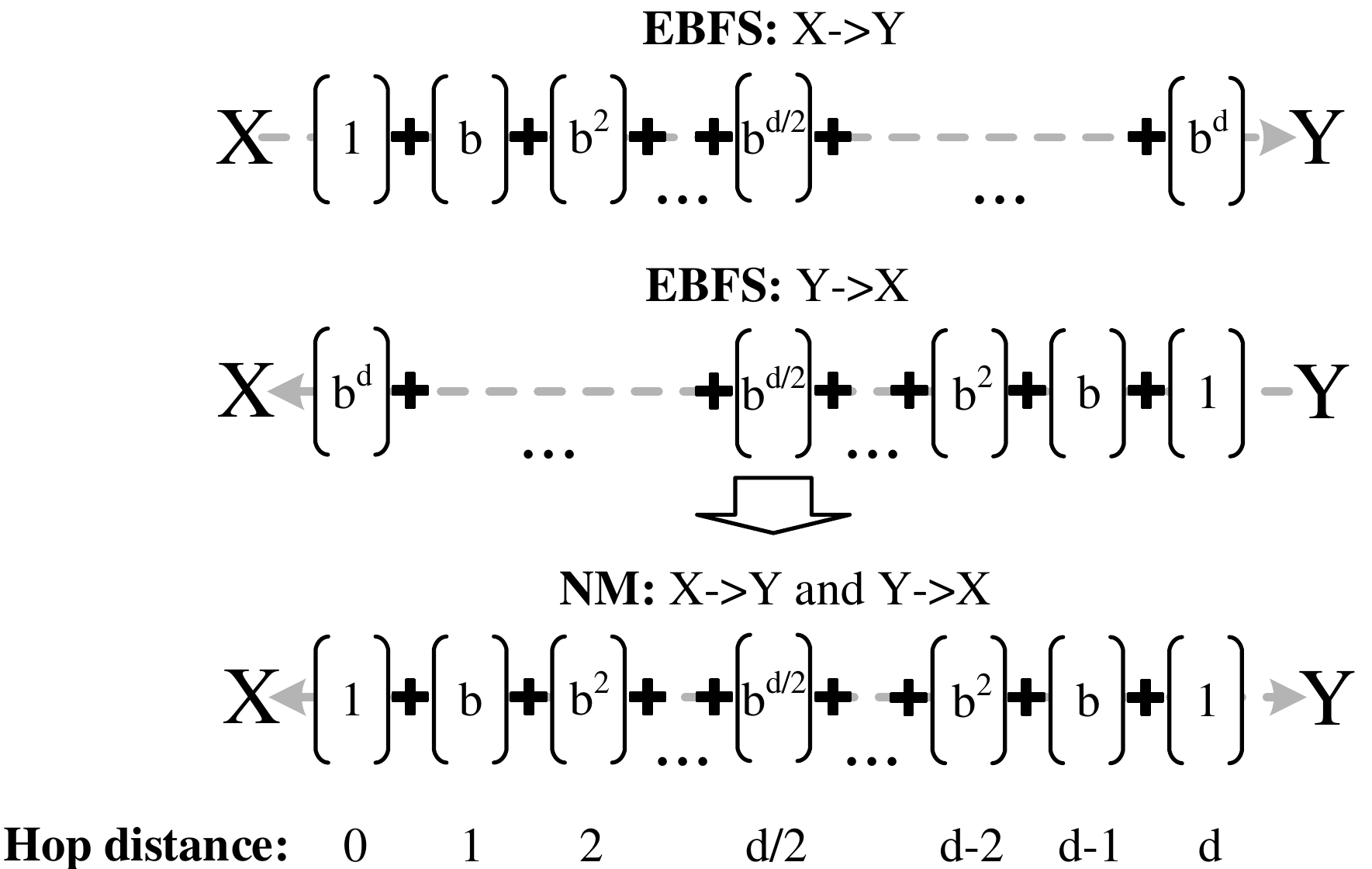}
\caption{\footnotesize{Differently than the branch-and-bound exhaustive search of EBFS, in which traversed paths grow between X and Y, NM's, reduces the search space significantly reducing the path candidates (from $O(b^d)$ to $O(b^{\frac{d}{2}})$ paths)  due to its forward and backward passes.}}
\label{complexity_fig}
\vspace{-6mm}
\end{figure}

\noindent
\textbf{NM complexity ($l\oplus p$ case).}
Similar to EDFS and EBFS, NM can find any variant of the constrained shortest path or $k$ such paths with $l \oplus p$ constraints in exponential time using exhaustive search.
As in the previous case, neighborhoods contains non-unique nodes, and the total number of their nodes can be up to $|V|^2$. For each  neighbor, NM checks if it already appears in the neighborhood $O(b)$. Taking into account the edge pruning phase, to ensure the constraints satisfaction and to reduce a search space, the time complexity of the neighborhoods building step $O_1^{l\oplus p}$ is:
\begin{equation}
O_1^{l\oplus p}=O(|V|^2b + |E|l)=O(|V||E|+ |E|\cdot \it{l})
\label{eq8}
\end{equation}
During the backward pass, NM builds all possible paths from the destination node using any exhaustive search methods such as EDFS or EBFS. \textit{However, there is a difference in that, we process only those vertex neighbors which appear in its previous neighborhood}. This allows us to significantly reduce the total number of the processing paths: 
instead of processing $1+b+b^2+...+b^d$ or $O(b^d)$ paths, due to double pass ($i.e.$, forward and backward passes) we process only: $1\cap b^d + b\cap b^{d-1} + b^2 \cap b^{d-2}+...+b^\frac{d}{2} \cap b^\frac{d}{2}+...+b^{d-2} \cap b^{2}+b^{d-1} \cap b + b^d \cap 1 \le 1 + b + b^2+...+b^\frac{d}{2}+...+b^{2}+b + 1$ or $O(b^{\frac{d}{2}})$ paths. Hence, the time complexity of the backward pass step $O_2^{l\oplus p}$ is quadratically lower than for EDFS or EBFS (see Equation~\ref{eq7}):
\begin{equation}
O_2^{l\oplus p}=O(2p\cdot b^\frac{d}{2})=O\Big(\Big(\frac{|E|}{|V|}\Big)^\frac{|V|}{2}p\Big)
\label{eq9}
\end{equation} 
The total time complexity of the NM is $O\Big(\Big(\frac{|E|}{|V|}\Big)^\frac{|V|}{2}p+|V||E| +|E|\cdot \it{l}\Big)$ or just $O\Big(\Big(\frac{|E|}{|V|}\Big)^\frac{|V|}{2}p + |E|\cdot \it{l}\Big)$. Similarly as for EDFS and EBFS, the total space complexity is $O\Big(\Big(\frac{|E|}{|V|}\Big)^\frac{|V|}{2}p\Big)$.

\section{NM Prototype Implementation}
\label{sdn}

\begin{figure}[t!]
\centering
\includegraphics[width=0.95\linewidth]{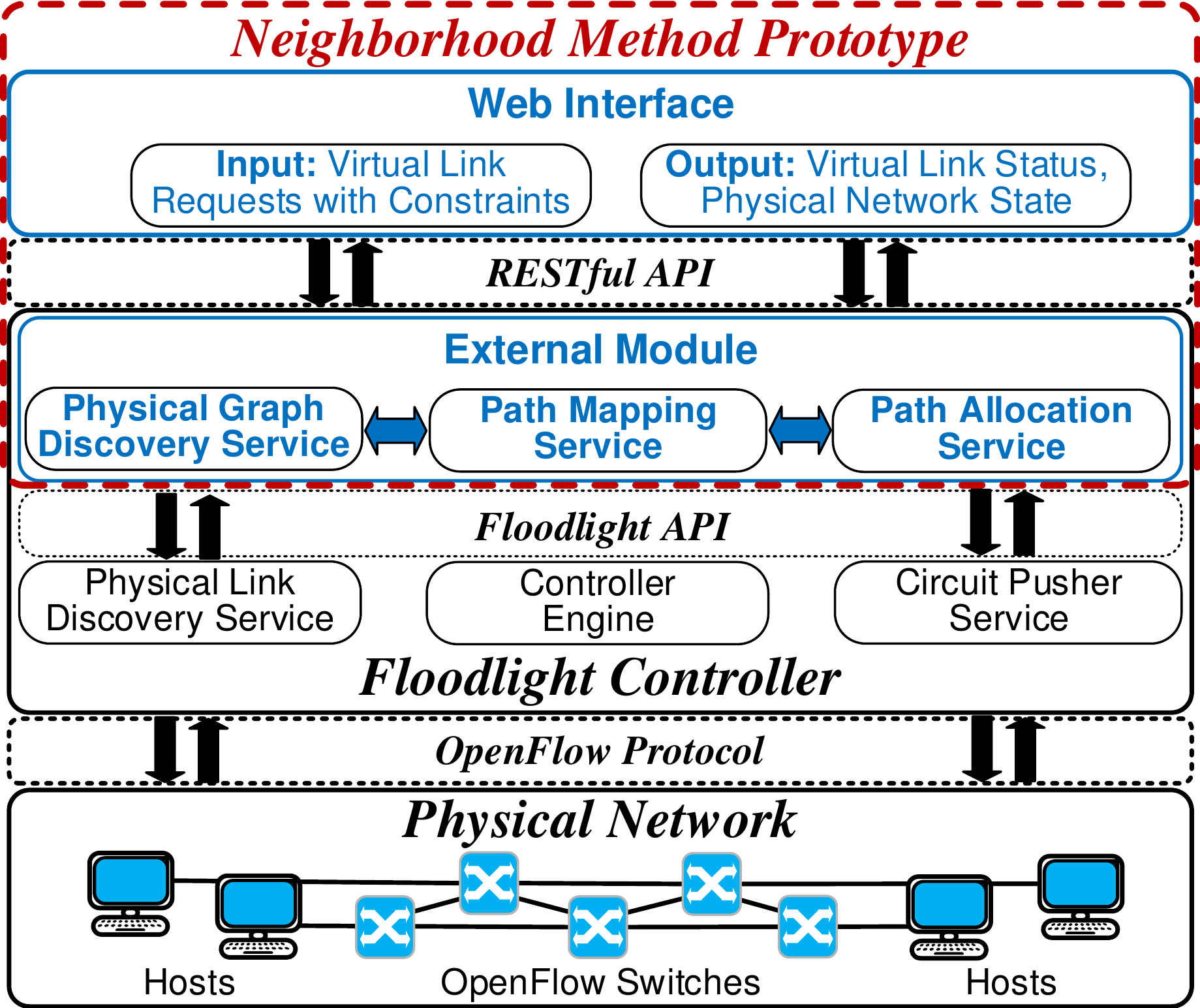}

\caption{\footnotesize{System architecture of our NM prototype (which is a module of the Floodlight OpenFlow controller~\cite{floodlight}) includes four main logical components: a \textit{physical graph discovery} service, a \textit{path mapping} service, a \textit{path allocation} service and a user \textit{web interface} that interacts with the controller. 
The prototype source code is publicly available under a GNU license at~\cite{nm_repo}.
}}
\label{prototype_architecture}
\vspace{-6mm}
\end{figure}
In this section, we establish the practicality of our approach for network virtualization with a prototype implementation over a Software Defined Networking infrastructure. Our source code is publicly available at~\cite{nm_repo}. In particular, our prototype implementation extends the Floodlight OpenFlow controller~\cite{floodlight}. Our system architecture is shown in Figure~\ref{prototype_architecture}. 
Our prototype includes four main logical components: a \textit{physical graph discovery} service, a \textit{path mapping} service, a \textit{path allocation} service and a user \textit{web interface} that interacts with the controller.
 In the rest of the section we describe with some more details each of the four components of our prototype.

\noindent
{\bf Physical graph discovery.}
Upon bootstrapping an SDN setup, all configured OpenFlow switches connect to the controller allowing a dynamic  \textit{switch-to-switch} link discovery. After this phase, the NM module tracks all unknown incoming packets to detect hosts and their corresponding \textit{host-to-switch} links.
We specify the main physical link properties, such as capacity and cost ($e.g.$, delay) through an XML configuration file. The  XML configuration file indirection allows  our NM prototype to easily interact with foreign measurement services, for example for real-time awareness of its link properties.
The information collected by the \textit{path discovery} service is then used in the \textit{path mapping} and the path \textit{allocation} steps.

\noindent
{\bf Path mapping.}
To map constrained virtual link requests, the path mapping module of our prototype uses information from the physical path discovery service and runs one of the following routing algorithm policies: NM for $l$ and $l\oplus1$ cases (default policy), NM for $l\oplus p$ cases, EDijkstra, IBF or EBFS.
In the current version of our prototype the routing policy is static and has to be set via our XML configuration file before starting the Floodlight controller.

\noindent
{\bf Path allocation.} In this last phase of the path embedding, the NM module sets all appropriate flow rules in all switches via the OpenFlow~\cite{openflow} protocol, along with the computed path obtained during the path mapping phase. Specifically, using OpenFlow protocol messages the module assigns active flows to corresponding queues, $i.e.$, applies an \textit{ENQUEUE} action, for guaranteed bandwidth provisioning. 
In our XML configuration file, users may specify also the type of timeout for each flow $i.e.$, the flow duration time can start from either the previously matched packet or from the first flow instantiation. 
To estimate the available bandwidth on each physical link, the NM module uses both the capacity information derived from the XML configuration file, and its allocated bandwidth queried from the flow stored in the OpenFlow switch.

\noindent
{\bf Web interface.}
To request virtual links and/or to monitor physical network states, we have implemented a web user interface, which uses a RESTful API to communicate with our NM prototype module. 
The user interface uses technologies such as HTML, PHP, AJAX, JavaScript, and CSS. 
%
%

\section{Performance Evaluation}
\label{implementation}

In this section, we evaluate NM's \dcR{scalability and flexibility} performance through simulations and our prototype implementation in the context of its applicability to several complementary virtual network services. 
\dcR{To assess the flexibility of NM, we compare, with and without it: 
$(i)$ the embedding performance of several online VNE and real-time NFV-SC mechanisms within the management plane; $(ii)$ several traffic engineering solutions within the data plane. 
Our goal is to show how our NM can be used to improve the overall network utilization (allocation ratio or total flow throughput), optimality (load balancing or fairness of flows), as well as energy consumption within both planes;
$(iii)$ To assess NM scalability, we then compare it} with related solutions under different network scales and service requests/topology scenarios when accepting multiple link and multiple path constraints; \dcR{$(iv)$} Finally, we confirm our main simulation results with our NM prototype running over the GENI testbed~\cite{geni}. 

\noindent
{\bf Evaluation settings.} In our simulations,  we used a machine with  
an \textit{Intel Core i5} processor with dual core CPU of $2.7$ GHz and $8$GB RAM. 
We use the BRITE~\cite{brite} topology generator to create our physical and virtual networks. Our results are consistent across physical networks that follow Waxman and Barabasi-Albert models~\cite{waxman}, that are known to approximate well subsets of Internet topologies~\cite{internet}. For lack of space we only show results relative to Waxman connectivities.
%
In our NM prototype evaluation instead, we use a physical network obtained with the GENI testbed~\cite{geni}.
All our results show $95\%$ confidence interval, and our randomness lays in both the service request ($i.e.$, in its type and constraints to accept) 
and in the physical network topology.  In most of our physical topologies, the average node degree equals to $4$, a known common value within Internet topologies~\cite{internet}.

\noindent
{\bf Results summary.} 
{\it Efficiency and Provider's revenue:} During the virtual network formation (management plane), we found that using NM within recent VNE or NFV-SC algorithms increases their allocation ratio while \dc{improving the network utilization by better load balancing (close to the optimal)}, which in turn decreases energy consumption.
{\it Network Utilization:} Our results evaluating NM within the data plane instead show how utilizing a set of paths found with NM is beneficial for TE in terms of minimum path hop count, network utilization and in some cases even energy consumption.
{\it Time to Solution (Convergence Time):} 
When we attempted to allocate flows (virtual links) with multiple link and multiple path constraints over large scale physical networks using NM with the proposed search space reduction techniques, we found a path computation speed-up of almost an order of magnitude w.r.t. common exhaustive search algorithms. Moreover, we also found almost 3 orders of magnitude running time improvement w.r.t. the same integer programming problßem solved with CPLEX~\cite{cplex}. 
{\it Prototype Evaluation:} Finally, using our NM prototype over GENI, we were able to reproduce our main results. Moreover, our measurements show how the constrained shortest path computation (virtual link or flow mapping phase in NM) is up to an order of magnitude faster than the path allocation phase ($i.e.$, setting up appropriate flow rules within all switches along the found path) \dc{on small scale networks}. 
%
\dc{This along with our scalability results confirm how at large scale the time needed for a path computation with NM will have the same order of magnitude as the time needed for the path allocation introducing no significant bottleneck for the end-to-end virtual link embedding.}
%

\subsection{Management Plane Evaluation}
\label{management_evaluation}
\noindent
{\bf Simulation settings.}
To assess the impact of NM on the virtual network embedding, we include results obtained with simulated physical networks of $20$ nodes (as in similar earlier studies~\cite{vne_path}), following Waxman connectivity model, where each physical node and each physical link have uniformly distributed CPU and bandwidth from 0 to 100 units, respectively. 
\dc{Note that} we use fairly small scale physical networks due to complexity of the integer programming.
We attempt to embed a pool of $40$ VN \dc{(service chain)} requests with $6$ virtual nodes 
and random \dc{(linear)} virtual topologies. 
Each virtual node and each virtual edge have uniformly distributed CPU and bandwidth demand from 1 to 10 units, respectively.
In the \dc{virtual network embedding} case, we vary the virtual node degree from $1$ (the VN has linear topology) to $5$ (VN is a fully connected topology).  \dc{Moreover, we also assume that each virtual edge has uniformly distributed latency constraint from 1 to 4  of a propagation delay stretch, defined as  the propagation delay encountered  traversing  the diameter of the physical network area.}
In the \dc{real-time service chaining} case, we vary latency constraint of virtual links (service-to-service communication) from $\infty$  ($i.e.$, SC is not real-time sensitive) to 1/4 ($i.e.$, SC is highly real-time sensitive) of the propagation delay stretch.

\noindent
{\bf Evaluation metrics.}  
TTo demonstrate the advantages of using NM within the virtual network embedding (VNE) and real-time service chaining (SC) mechanisms, we compared four representative VNE algorithms. To compare them, we replace their (original) Dijkstra-based shortest path with our NM.
We have chosen to compare against~\cite{vne_path, vne_opt} as the optimal VNE scheme formulated as  integer programming multi-commodity flow is the best to our knowledge solution for online VNE (yet intractable for large scale networks). We refer to this solution as Optimal and we denote it as $Opt$.
\dcR{Note however, that $Opt$ solution can be still suboptimal with respect to the optimal solution of the offline VNE problem, where all requests are known in advance.}
Opt in fact attempts to minimize the ratio between the provider's costs of embedding a VN request and the available substrate resources provided for this request, with the aim of balancing the network load. 
%
%
We also compare Opt against its version where a path (or column) generation approach is used to make Opt more scalable~\cite{vne_path}. 
We refer to this scheme as PathGen and, even in this case,  substitute its original shortest path algorithm (used to find new paths within the multi-commodity flow) with our path management solution (see details in Section~\ref{csp_vne_sc_sec}).
%
\dc{Note that we used the one-shot VNE approximation algorithm proposed  in~\cite{lika} as an initial solution for the column generation approach to avoid two stage VNE limitations when physical network is initially unbalanced~\cite{vne_path}.}
%
%
Finally, we compare against a Consensus-based Auction mechanism (CAD)~\cite{catena,nodeembedding}, the first policy-based distributed VNE approximation algorithm with convergence and optimality guarantees.  \dcR{Note that a version of CAD can be also used to solve the NFV-SC problem~\cite{catena}.} The link embedding of CAD is a policy that  runs a shortest path algorithm to either pre-compute the k-shortest paths or to find these paths dynamically. For fairness of comparison, we assume that the latter holds and as in the PathGen case, we substitute the currently used shortest path algorithm, Dijkstra, with our NM constrained shortest path finder solution.

In this simulation scenario we have tested the potential revenue loss by specifying the fraction of VN request accepted over the VN requested (allocation ratio), to what extent physical links were utilized (link utilization) and how many paths of the particular length in total were used per VN pool.
Finally, we used the idle energy model proposed previously~\cite{energy_model} to access the energy consumption of the network:
\begin{equation}
\mbox{Energy Consumption}=\underset{e \in E}{\sum} (M-E_0)U_e + E_0
\label{energy_eq}
\end{equation} 
where $E$ is a set of physical edges, $U_e$ is an edge $e$ utilization, and $M$ and $E_0$ are numerical values taken from~\cite{energy_model} \dc{that correspond to the maximum and idle energy consumption of the switch interface, respectively. We use $M=2$ and $E_0=1.7$ maximum and idle energy consumptions (measured in Watts) assuming gigabit channel communications}. In our results, we show an energy consumption increase relative to the idle network state (in \%). 

\begin{figure*}[t!]
\centering

\begin{subfigure}[b]{0.95\textwidth}
\centering
\includegraphics[width=1\linewidth]{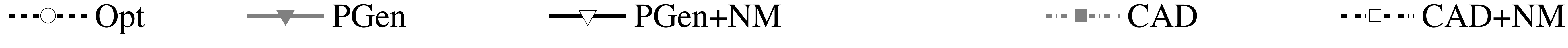}
\end{subfigure}

\vspace{2mm}

\begin{subfigure}[b]{0.2425\textwidth}
\centering
\includegraphics[width=1\linewidth]{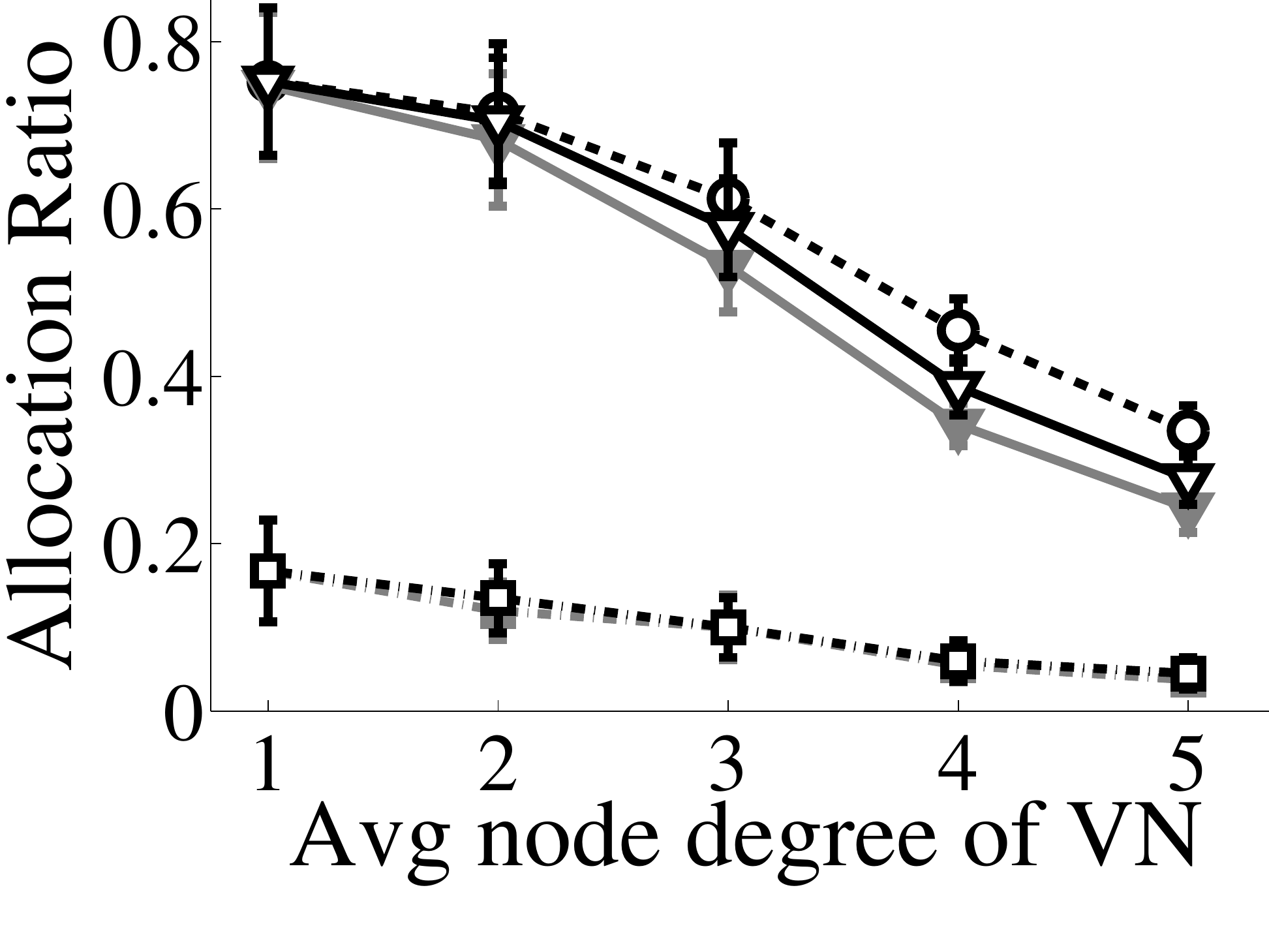}
\caption{}
\label{vne_alloc}
\end{subfigure}
%
\begin{subfigure}[b]{0.2425\textwidth}
\centering
\includegraphics[width=1\linewidth]{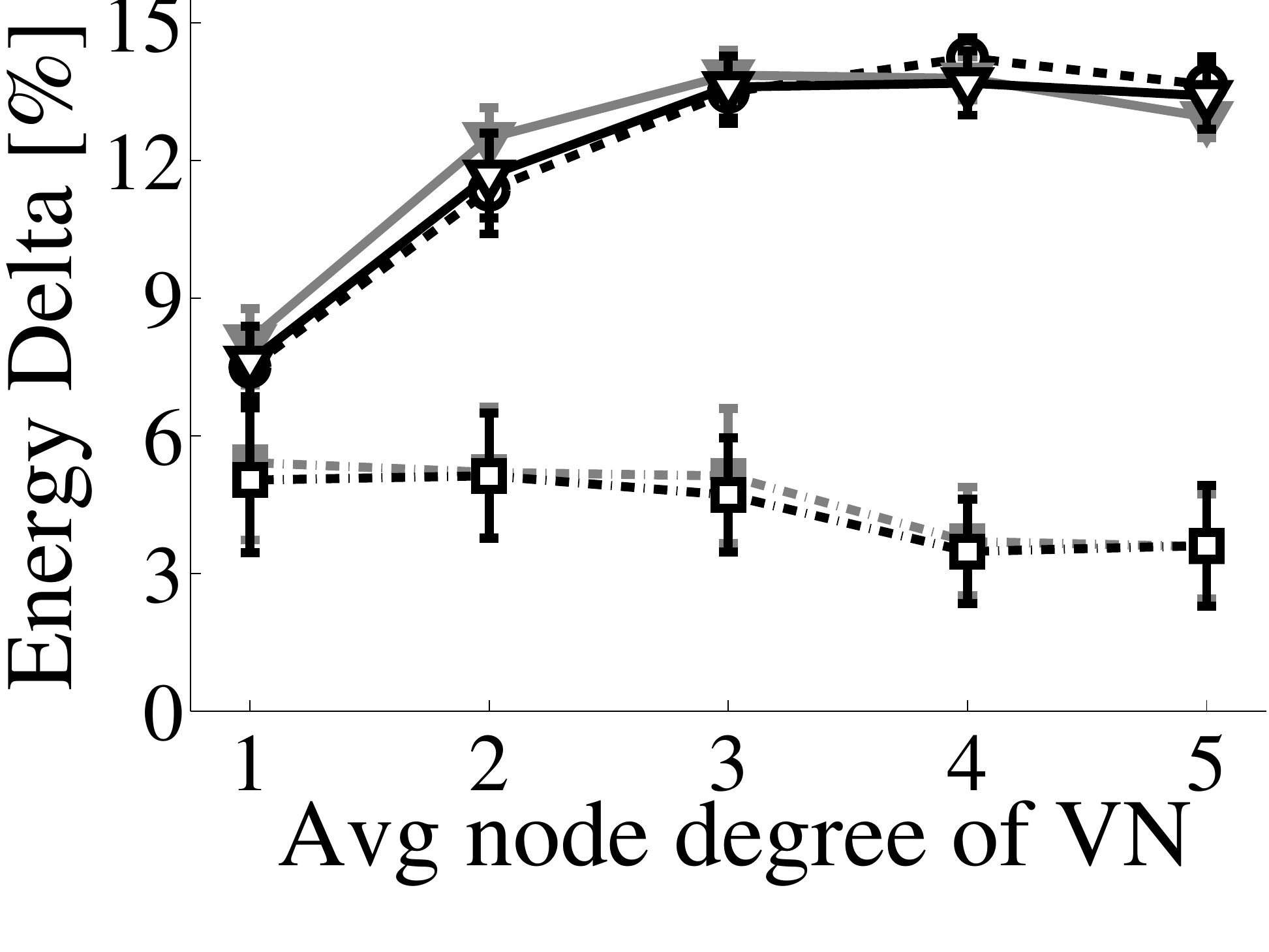}
\caption{}
\label{vne_energy}
\end{subfigure}
\begin{subfigure}[b]{0.2425\textwidth}
\centering
\includegraphics[width=1\linewidth]{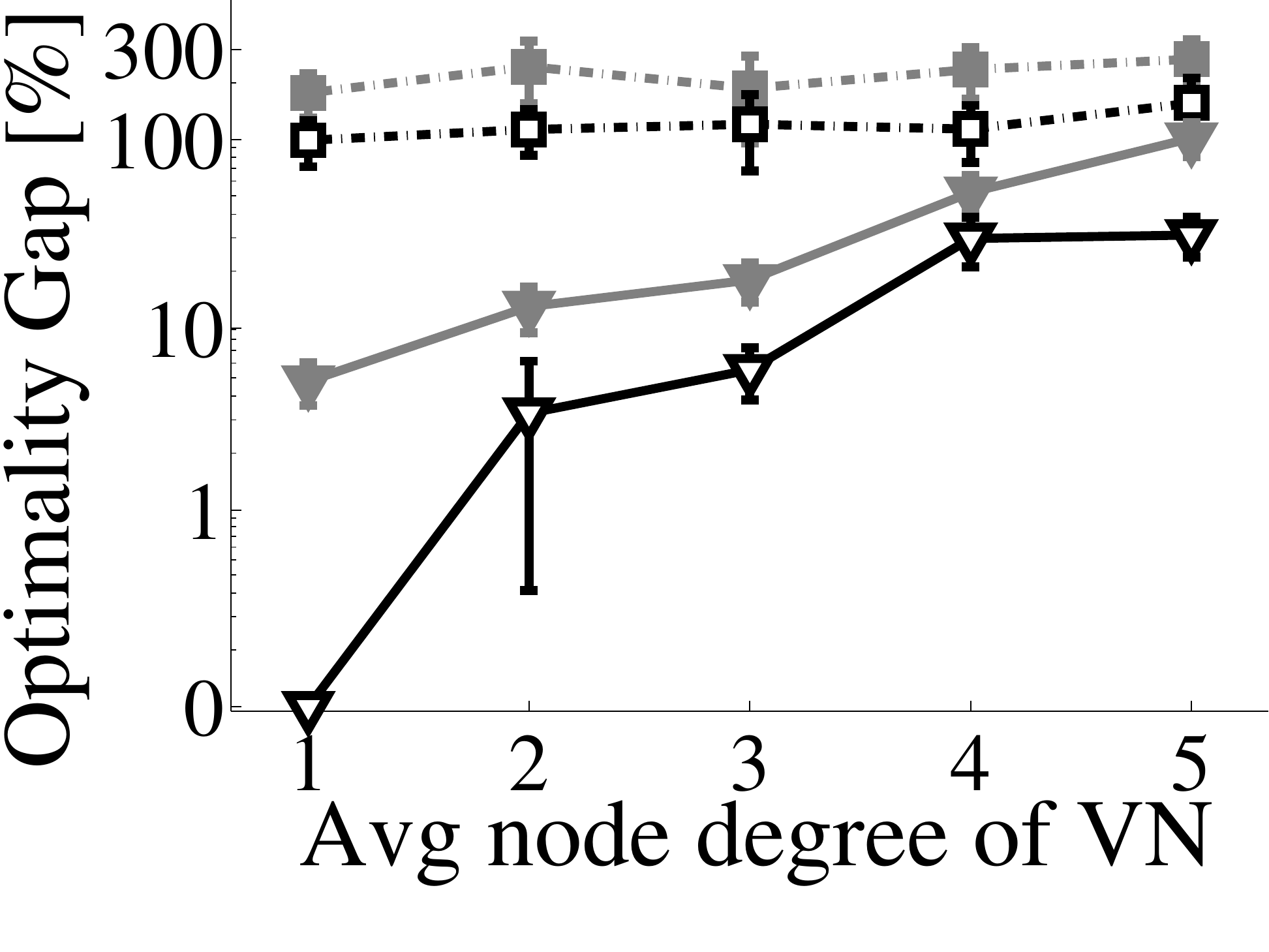}
\caption{}
\label{link_utilization_full}
\end{subfigure}
\begin{subfigure}[b]{0.2425\textwidth}
\centering
\includegraphics[width=1\linewidth]{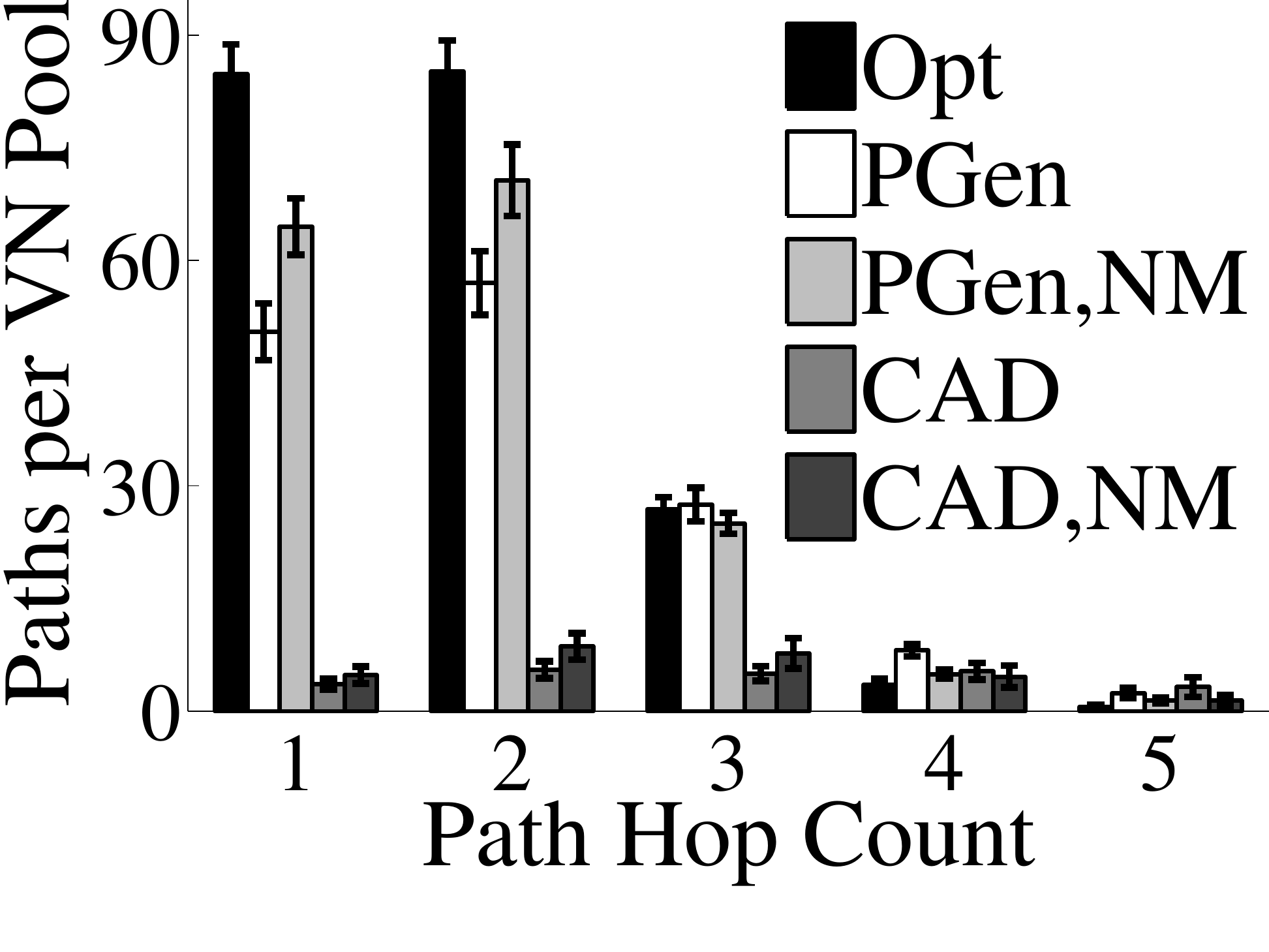}
\caption{}
\label{vne_length}
\end{subfigure}

\vspace{2mm}


\begin{subfigure}[b]{0.2425\textwidth}
\centering
\includegraphics[width=1\linewidth]{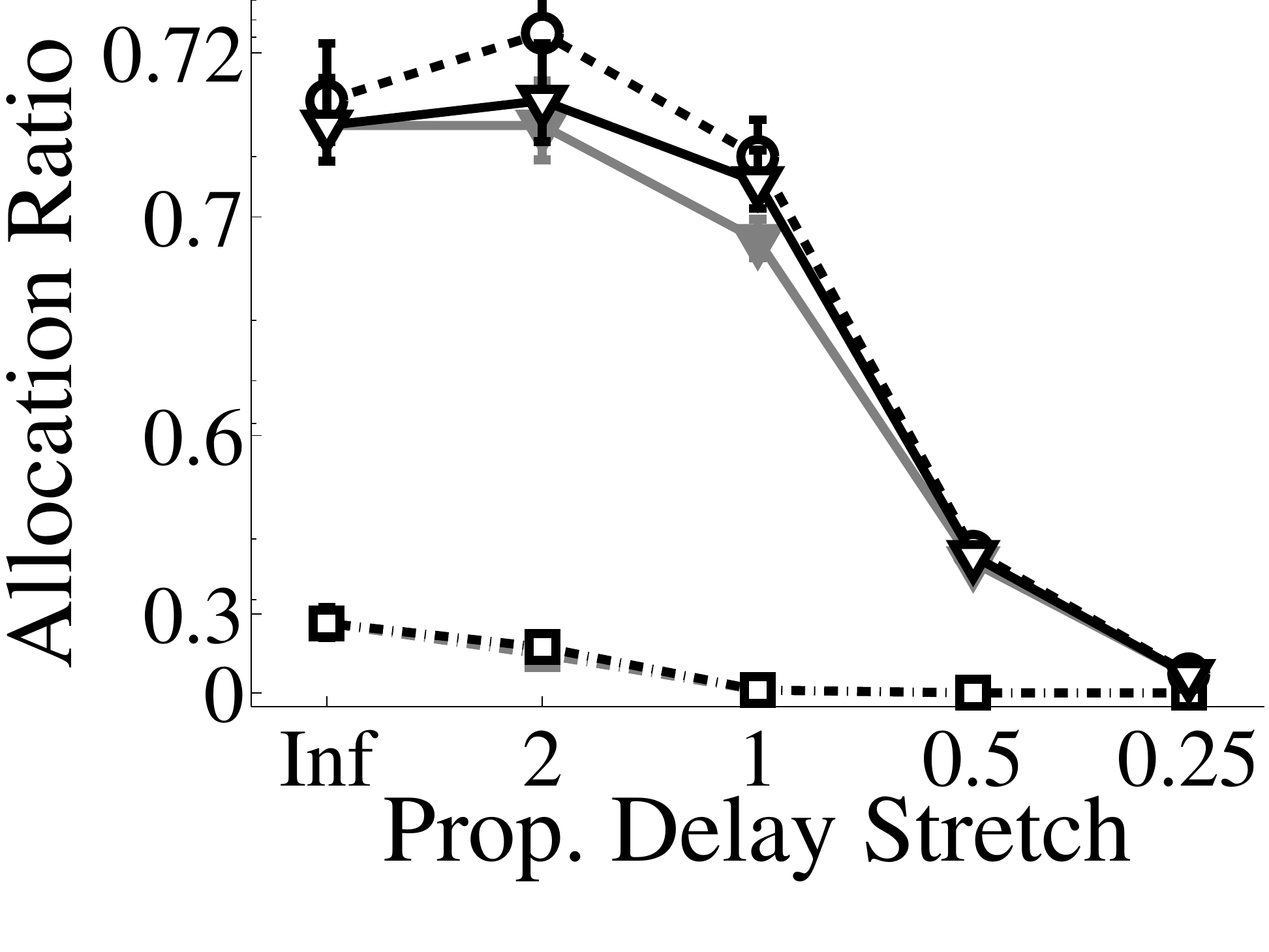}
\caption{}
\label{sc_alloc}
\end{subfigure}
\begin{subfigure}[b]{0.2425\textwidth}
\centering
\includegraphics[width=1\linewidth]{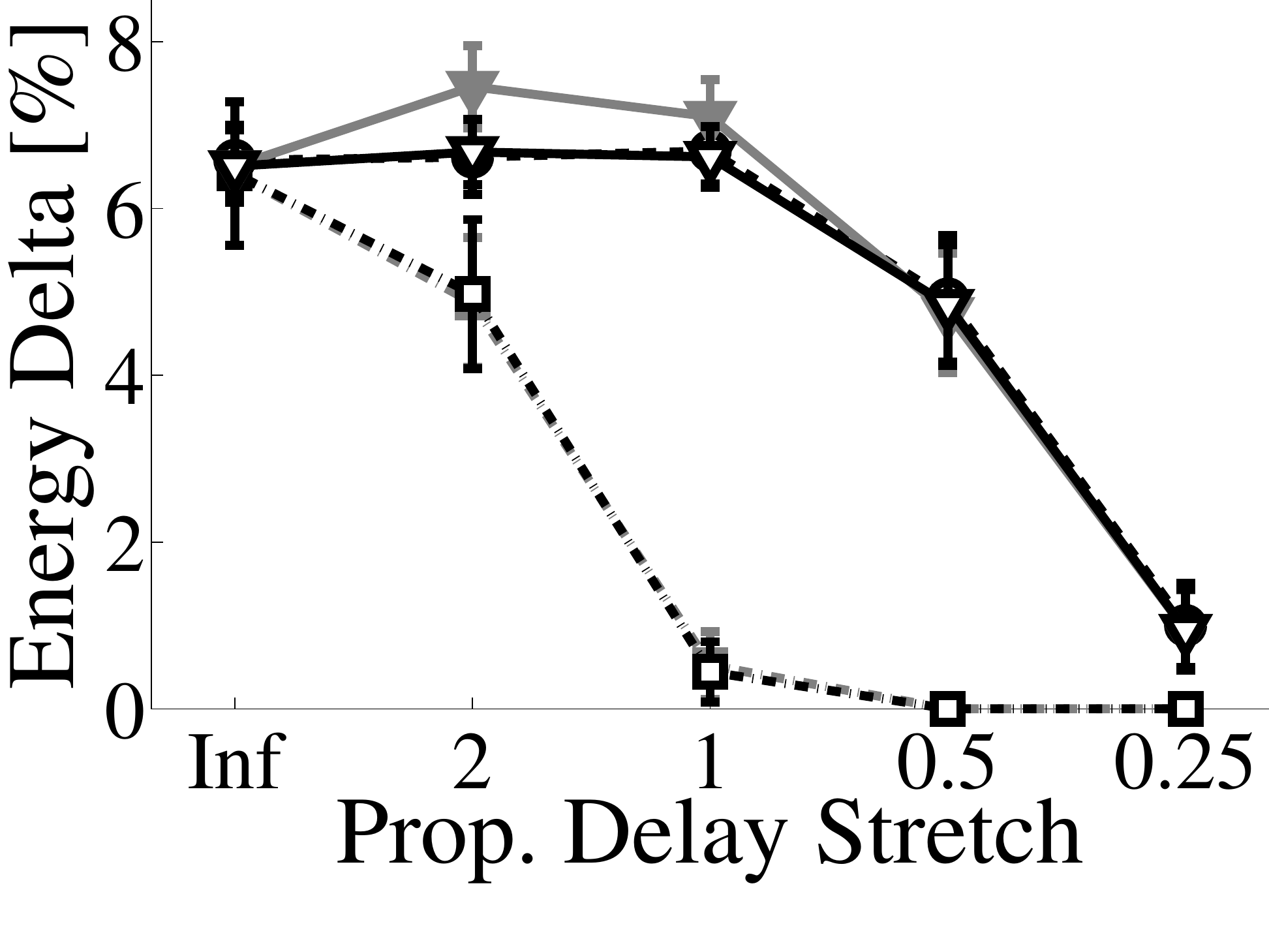}
\caption{}
\label{sc_energy}
\end{subfigure}
\begin{subfigure}[b]{0.2425\textwidth}
\centering
\includegraphics[width=1\linewidth]{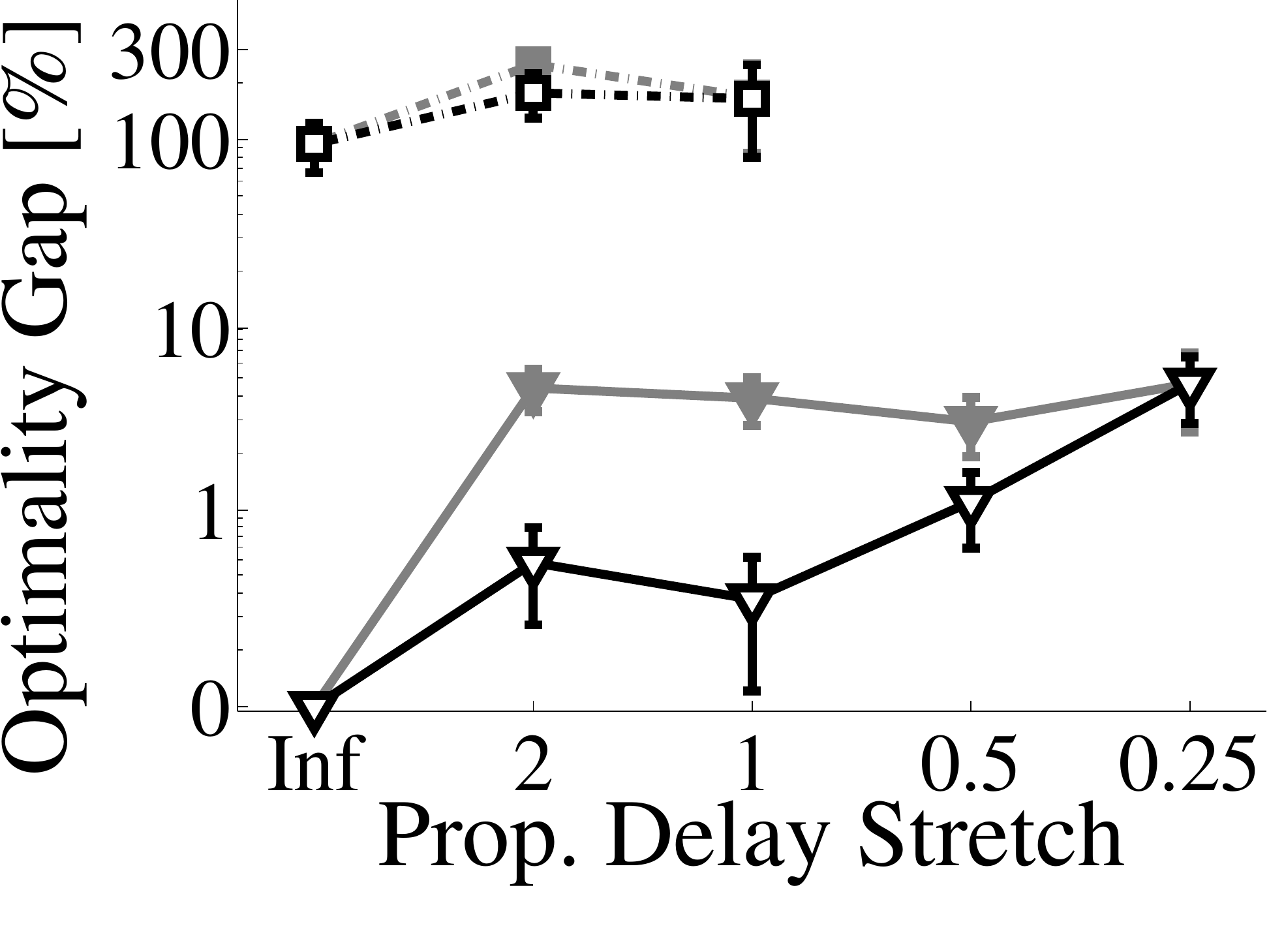}
\caption{}
\label{sc_util_high}
\end{subfigure}
\begin{subfigure}[b]{0.2425\textwidth}
\centering
\includegraphics[width=1\linewidth]{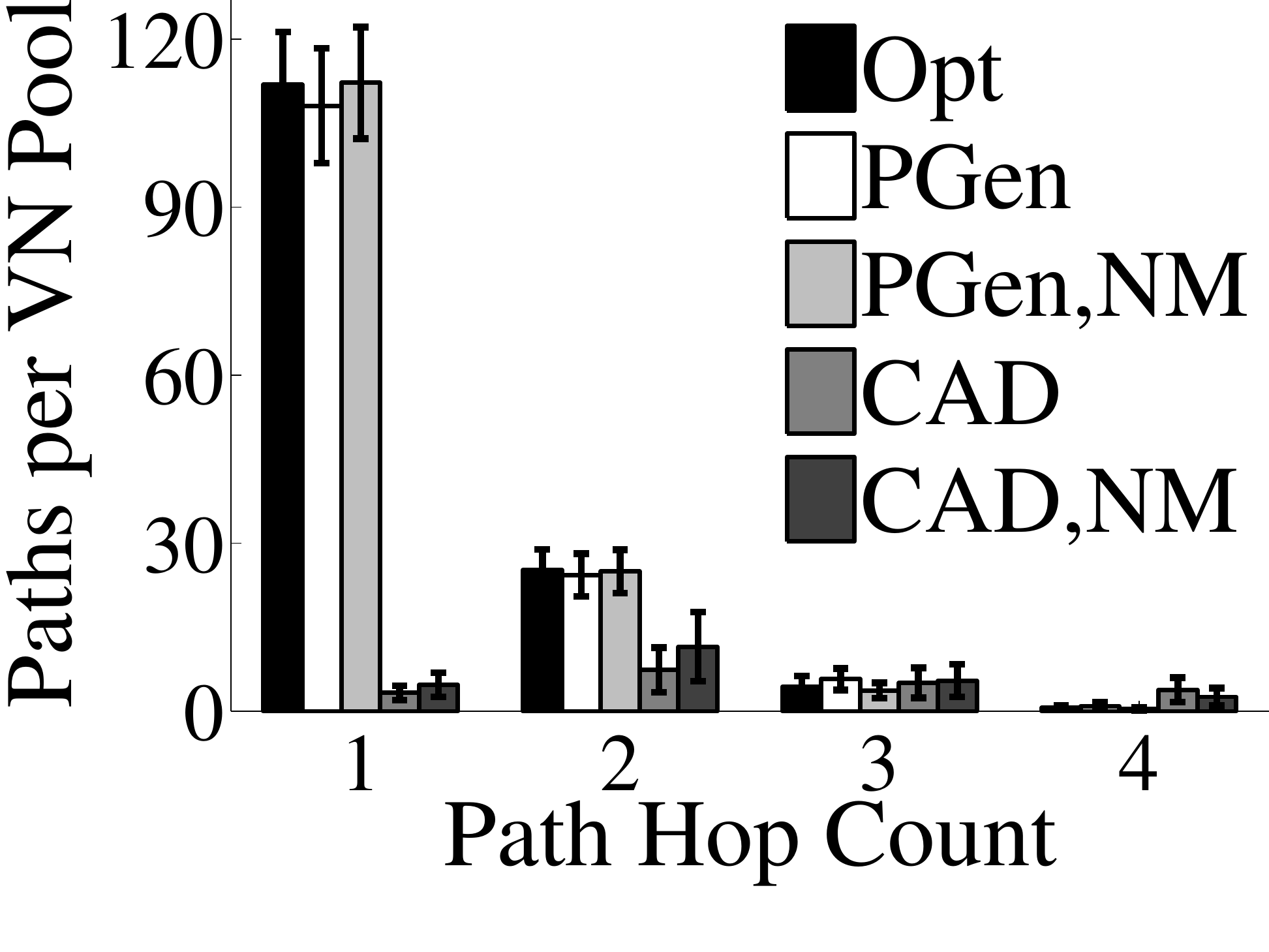}
\caption{}
\label{sc_len_high}
\end{subfigure}

\caption{\footnotesize{
Virtual network (VN) embedding and real-time NFV service chaining (SC) results obtained with physical networks of $20$ nodes following Waxman connectivity model: by addressing the constrained shortest path problem with NM versus using commonly adopted shortest path algorithms ($e.g.$, Disjktra), the allocation ratio of VNE (a) and real-time NFV-SC (e) can be 
improved when the virtual node degree increase or when requests are not highly sensitive to delays. 
\dc{Optimality gap of VNE (c) and NFV-SC (g) can be also improved (resulting in our case into a better load balancing) which leads} to a lower energy consumption (b) and (f) relative to the network idle state~\cite{energy_model}, respectively. 
NM's benefits are due to \dc{its ability of finding more paths \dcR{with a lower hop count} that can satisfy latency demands and simultaneously improve objective value for full-mesh} VNs (d) and \dc{moderate} real-time sensitive SC (h) pools.
}}
\label{cad_gen}
\vspace{-6mm}
\end{figure*}

\noindent
{\bf NM improves VNE/NFV-SC allocation ratio and energy efficiency.} 
%
Figures$~\ref{vne_alloc}$ and$~\ref{sc_alloc}$ show how including constrained shortest paths ($e.g.$, found with NM) during column generation of  the PathGen approach can improve overall VNE and NFV-SC acceptance ratio.  Particularly, the highest acceptance ($i.e.$, allocation) ratio gains arise in the case ofdense (e.g., full-mesh) VNs or under \dc{moderate} real-time sensitivity of NFV-SCs. 
%
Moreover, we can see how in some cases, utilizing NM with\dc{in} PathGen leads to a lower energy consumption \dc{(see Figures$~\ref{vne_energy}$ and$~\ref{sc_energy}$). This is due to an improved network utilization because of a better (closer to the optimal) load balancing (see Figures$~\ref{link_utilization_full}$ and$~\ref{sc_util_high}$).}
%
%
%

\dc{These optimality} gains \dc{in turn} arise due to NM's ability to \dc{find} more paths \dc{that can satisfy all virtual link constraints ($e.g.$, bandiwdth and latency) and simultaneously improve the objective value} (see Figures$~\ref{vne_length}$ and$~\ref{sc_len_high}$). These results confirm our expectations in Section~\ref{csp_vne_sc_sec}.
%
%
\dc{At the same time, optimality improvements with NM demonstrate no significant benefits (in terms of allocation ratio or energy efficiency) for the CAD over standard shortest path management. This is due to the fact that in our settings, separate node and link embedding approach demonstrates the worst performance caused by significantly limited feasible space for the virtual link mapping. Such limitations are due to randomized capacities of physical nodes and edges ($i.e.$, due to initially unbalanced physical network) further exacerbated by randomized virtual link capacity and latency constraints.}

\begin{figure*}[t]
\centering
\begin{subfigure}[b]{0.99\textwidth}
\centering
\includegraphics[width=1\linewidth]{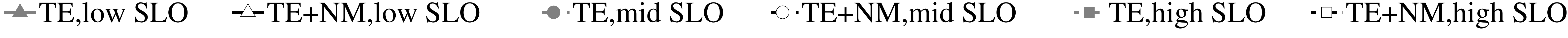}
\end{subfigure}

\begin{subfigure}[b]{0.2425\textwidth}
\centering
\includegraphics[width=1\linewidth]{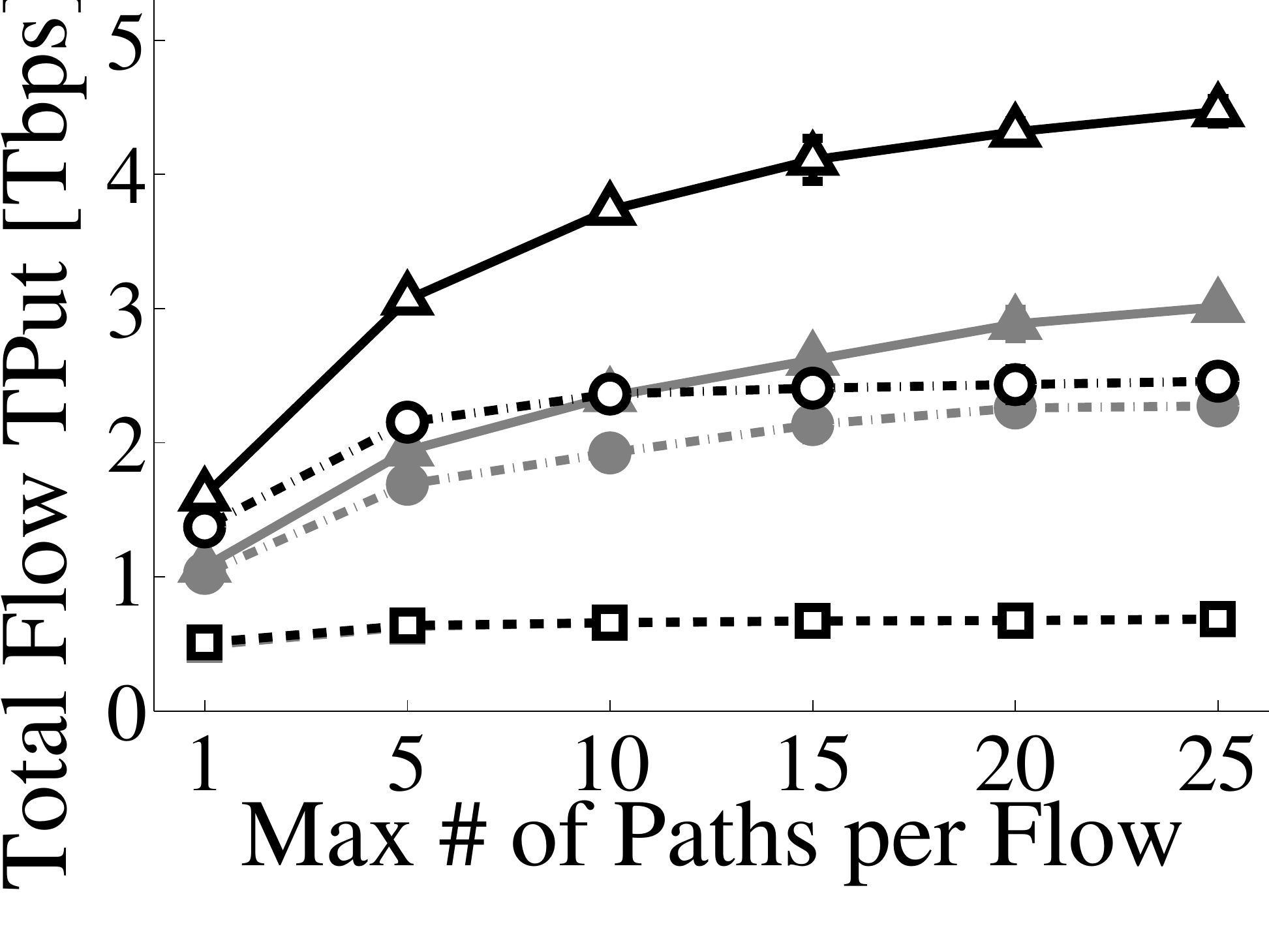}
\caption{}
\label{lp_thr}
\end{subfigure}
\begin{subfigure}[b]{0.2425\textwidth}
\centering
\includegraphics[ width=1\linewidth]{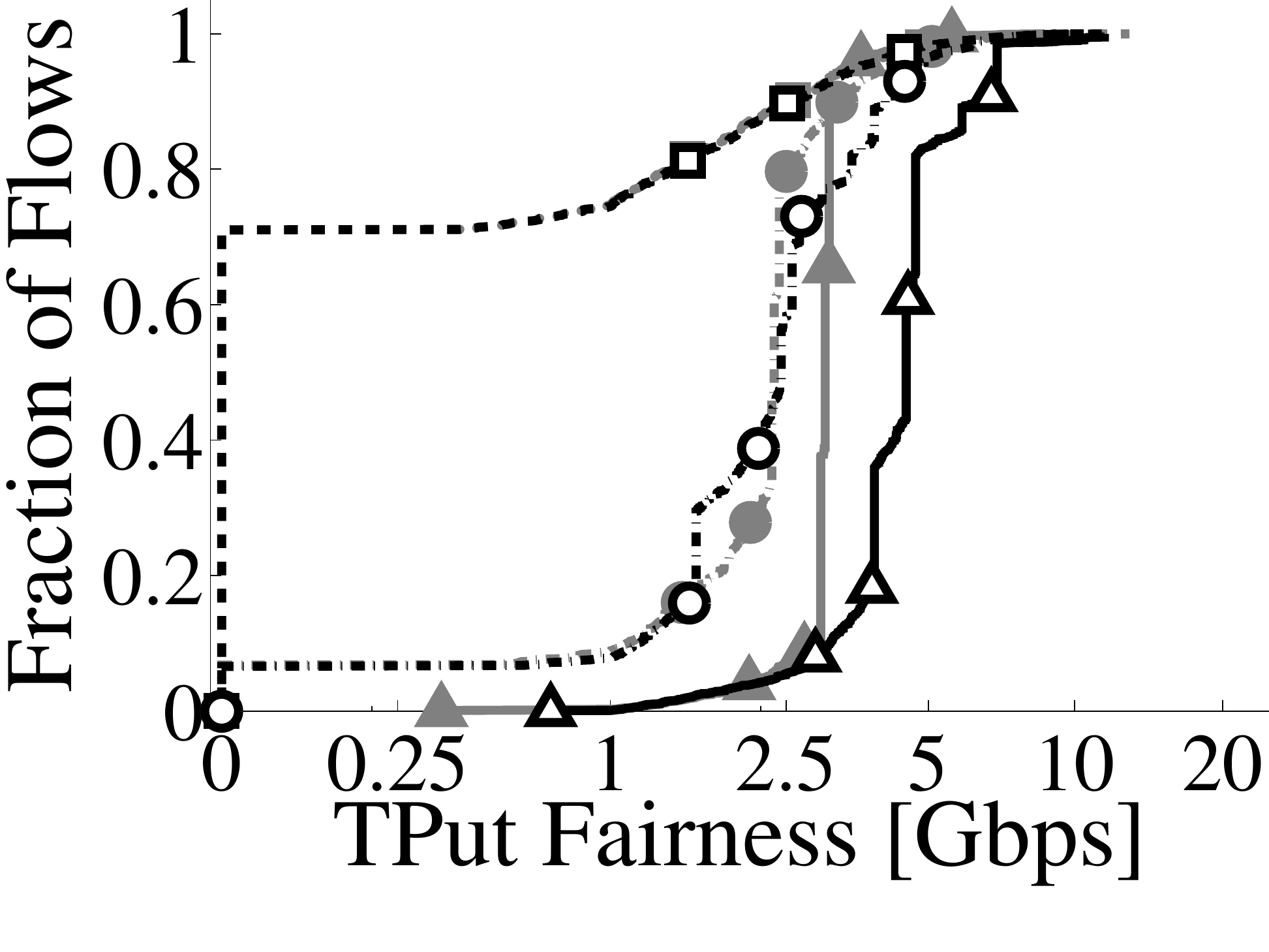}
\caption{}
\label{lp_thr_cdf}
\end{subfigure}
\begin{subfigure}[b]{0.2425\textwidth}
\centering
\includegraphics[width=1\linewidth]{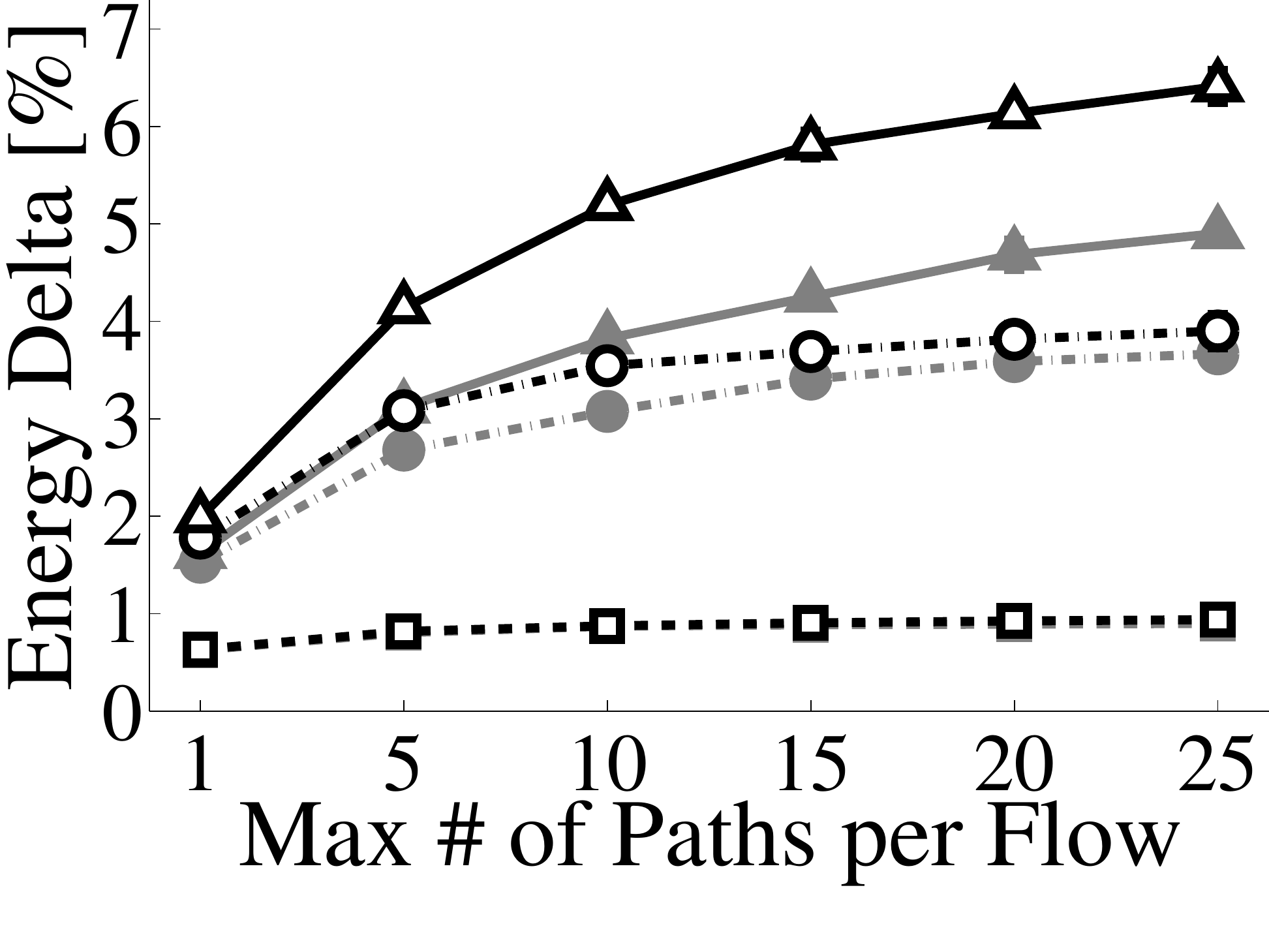}
\caption{}
\label{lp_energy}
\end{subfigure}
\begin{subfigure}[b]{0.2425\textwidth}
\centering
\includegraphics[width=1\linewidth]{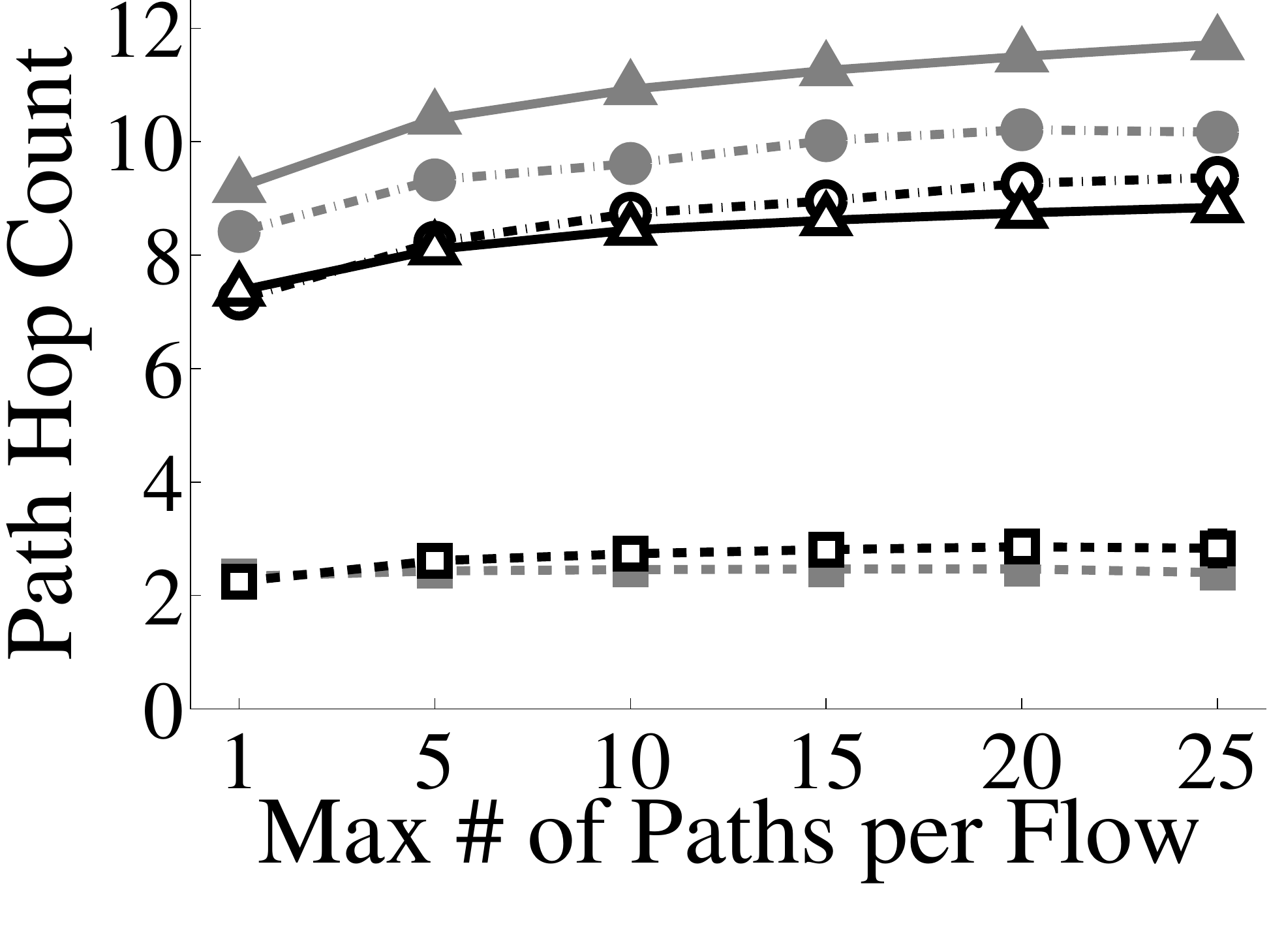}
\caption{}
\label{lp_len}
\end{subfigure}

\begin{subfigure}[b]{0.2425\textwidth}
\centering
\includegraphics[width=1\linewidth]{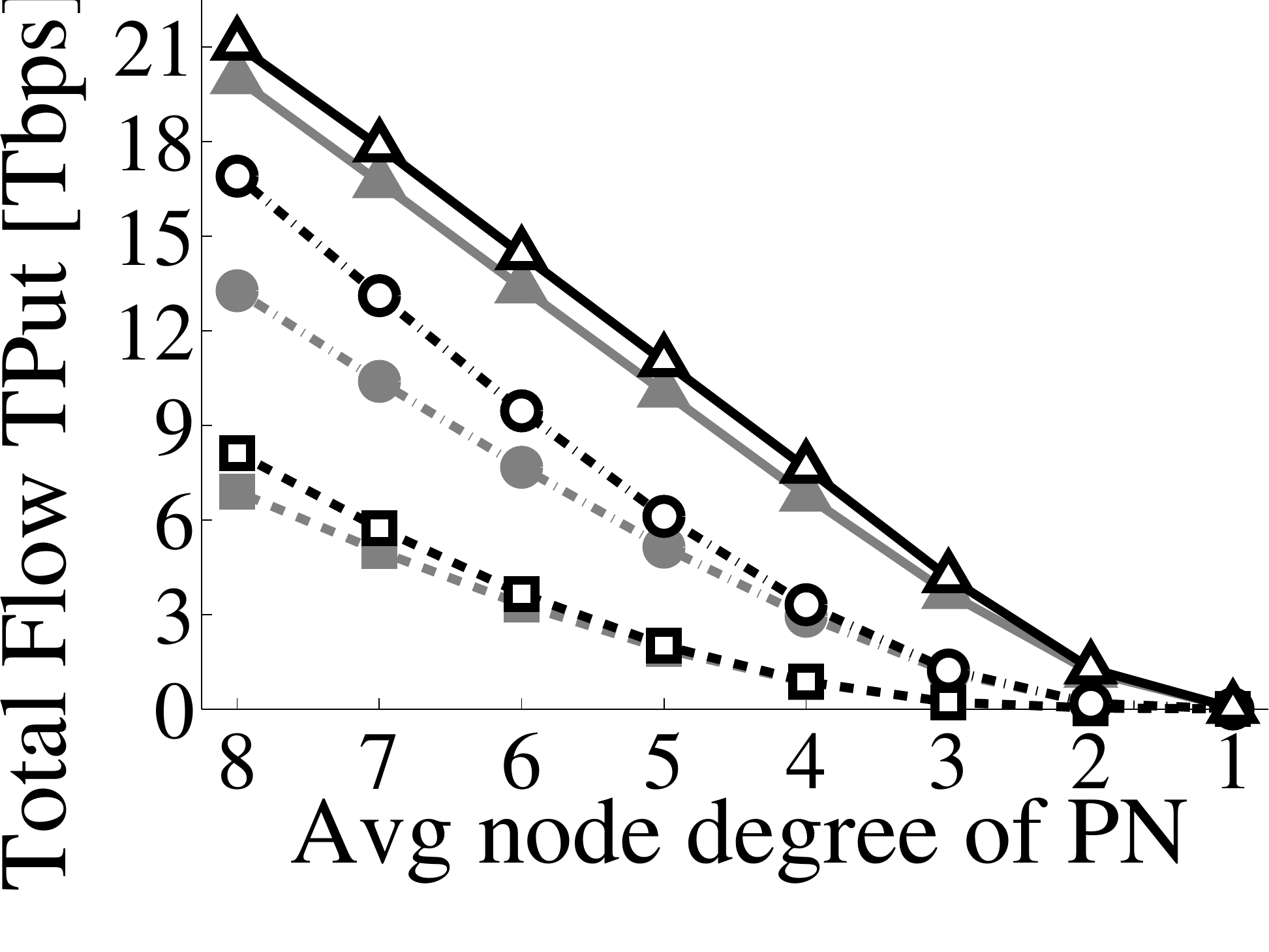}
\caption{}
\label{b4_thr}
\end{subfigure}
\begin{subfigure}[b]{0.2425\textwidth}
\centering
\includegraphics[ width=1\linewidth]{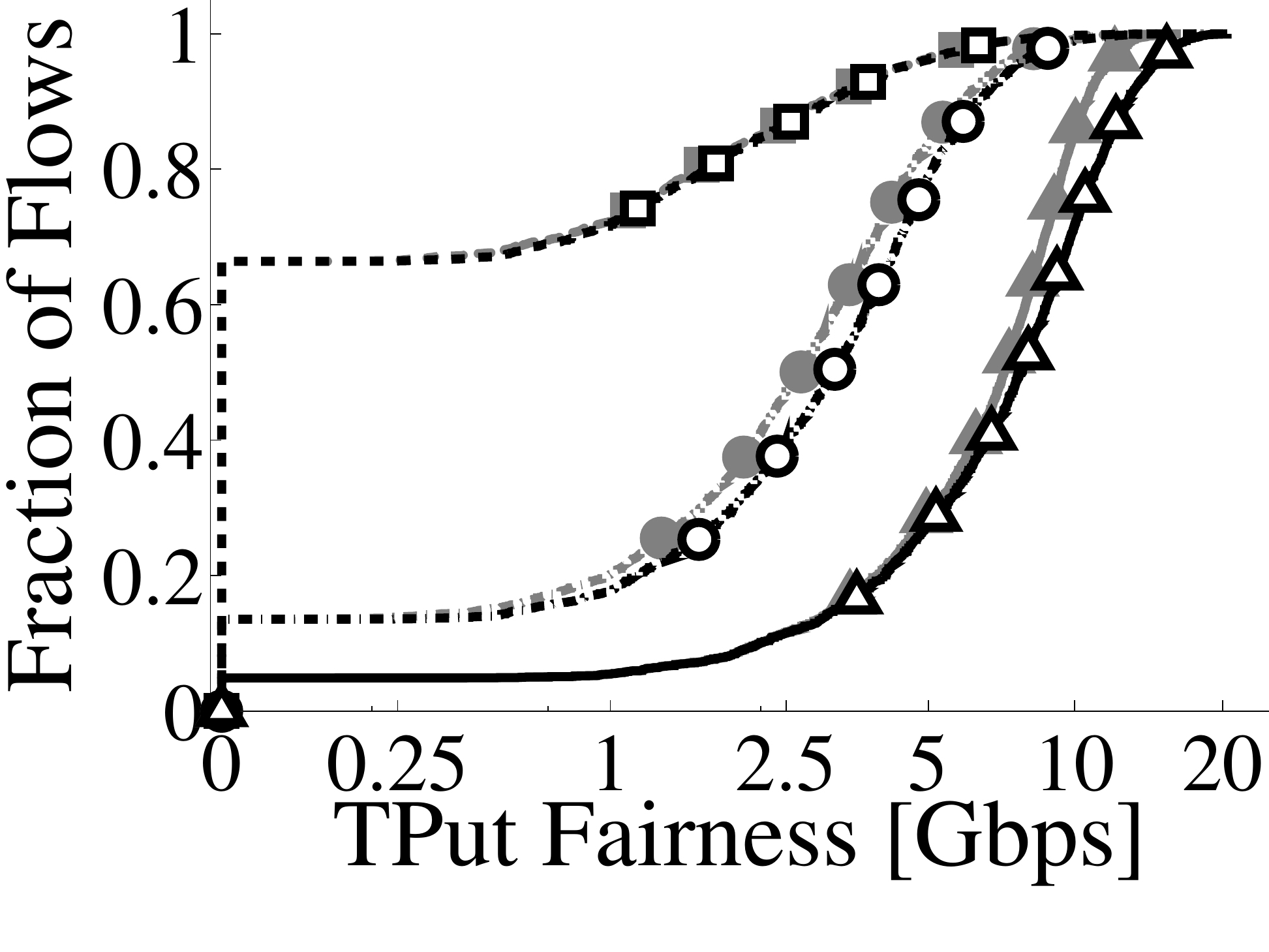}
\caption{}
\label{b4_thr_cdf}
\end{subfigure}
\begin{subfigure}[b]{0.2425\textwidth}
\centering
\includegraphics[width=1\linewidth]{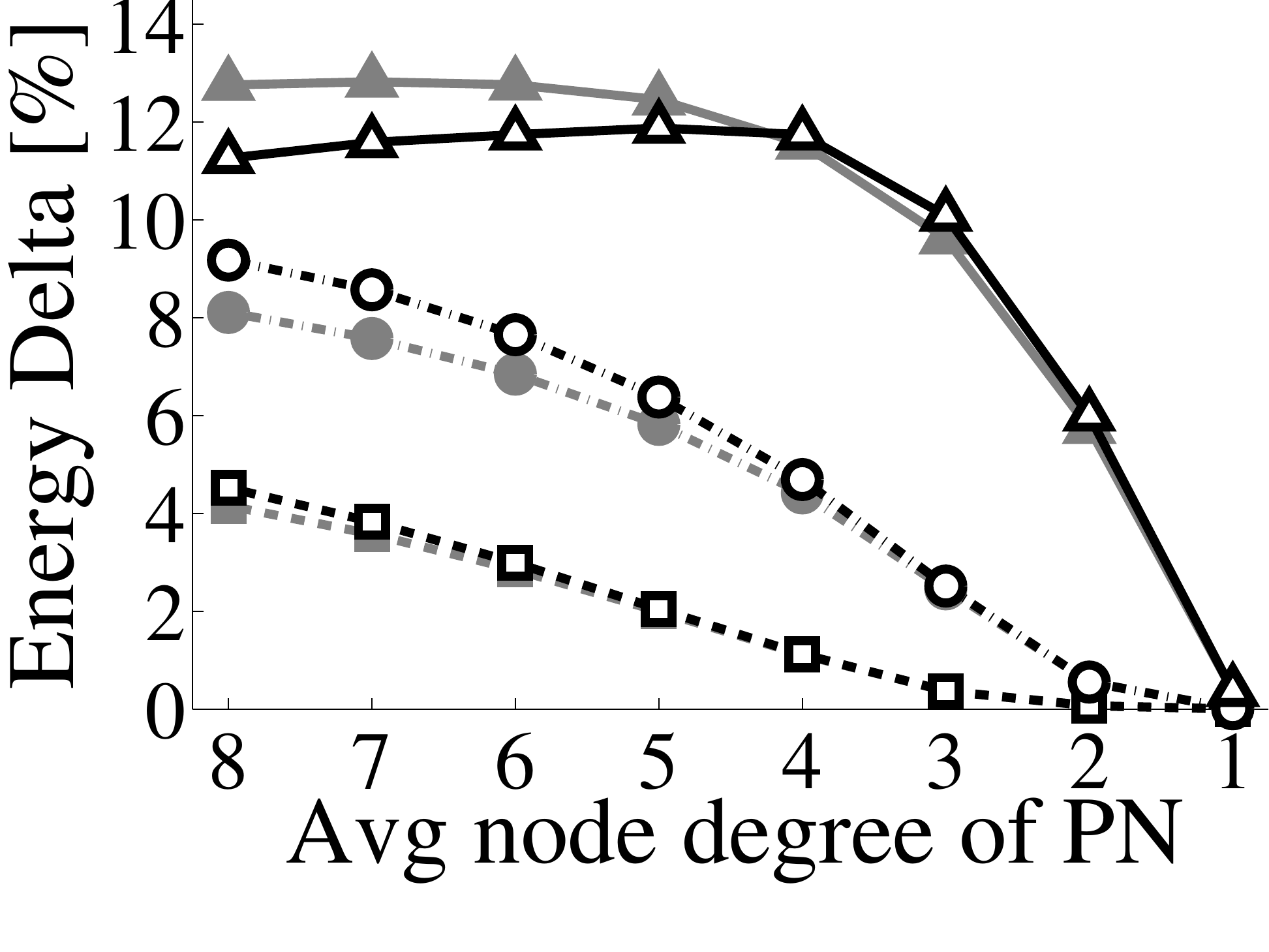}
\caption{}
\label{b4_energy}
\end{subfigure}
\begin{subfigure}[b]{0.2425\textwidth}
\centering
\includegraphics[width=1\linewidth]{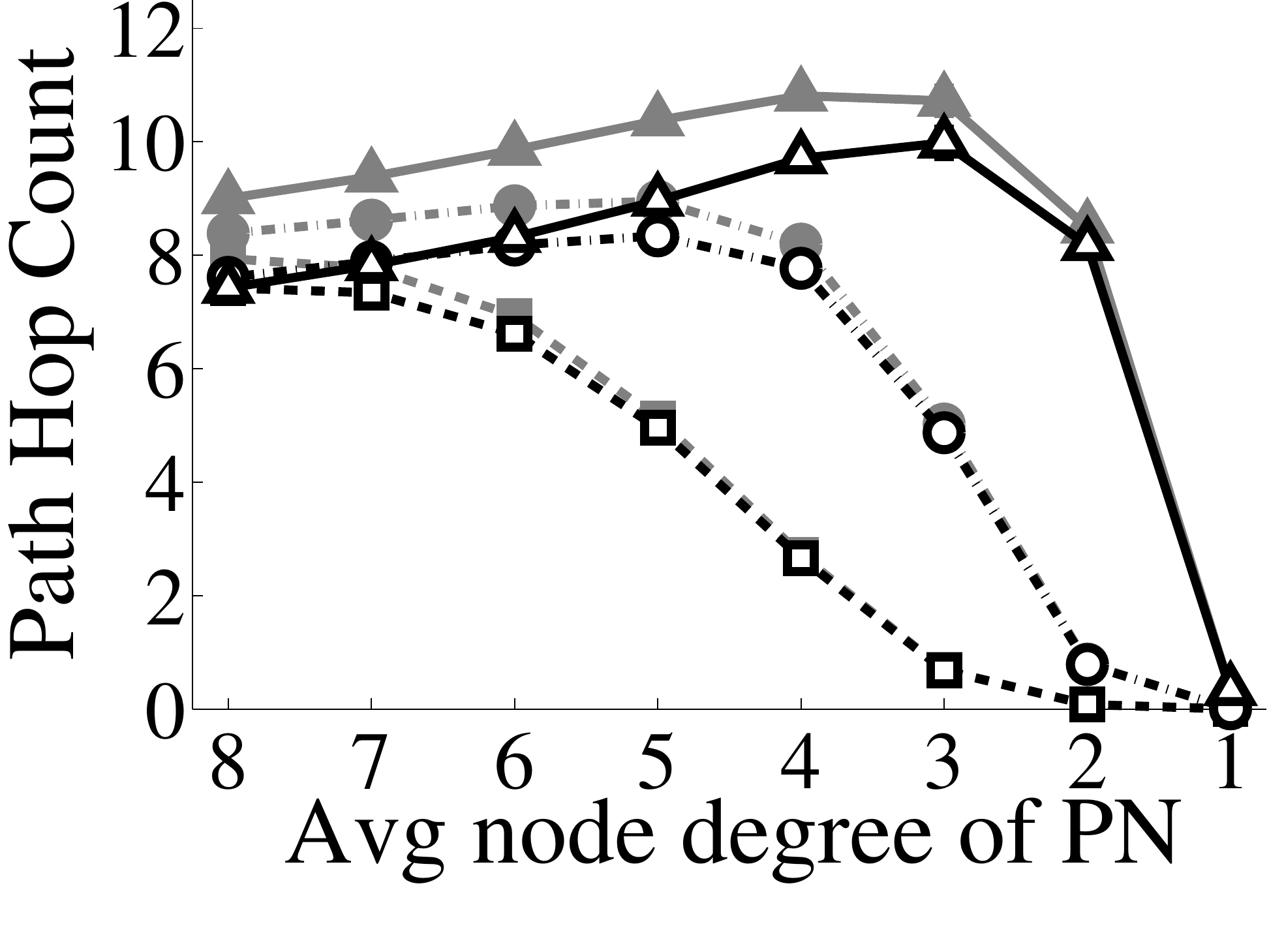}
\caption{}
\label{b4_len}
\end{subfigure}
\vspace{-1mm}
\caption{\footnotesize{
Performance analyses of LP-based (\textit{Top}) and greedy (\textit{Bottom}) max-min fairness Traffic Engineering (TE) algorithms~\cite{swan,b4} utilizing \dc{the constrained shortest path management with} NM versus \dc{their original shortest path management with} Dijkstra on Waxman topologies in terms of: (a) and (e) total gained flow throughput; (b) and (f) cumulative distribution of flow throughput for 25 paths per flow and for average node degree 4 (common for Internet~\cite{internet}), respectively; (c) and (g) energy consumption increase relative to the network idle state~\cite{energy_model}; and (d) and (h) number of average path hops per flow.
}} 

\label{topo_waxman}
\vspace{-5mm}
\end{figure*}

\subsection{Data Plane Evaluation}
\label{te_evaluation}
%
In the next set of results we analyze the benefits of using our solution within the data plane by evaluating NM performance within standard Traffic Engineering schemes~\cite{swan, b4}, under different physical network topologies and under different severity of service level objectives.

\noindent
{\bf Simulation settings.}
To evaluate the impact of NM (used to compute constrained shortest paths) within Traffic Engineering~\cite{swan, b4}, we use a physical network topology of $10,000$ nodes, where each physical link has bandwidth uniformly distributed  between $1$ and $10$ Gbps. We attempt to allocate flows for $1000$ random  source destination pairs by solving max-min fairness problem shown in Equation~\ref{te_objective_eq} with the fixed latency SLO demands.
To evaluate the maximum possible gains, we assume infinite bandwidth demands of the flows and we omit any constraints imposed by hardware granularity due to rule count limits or flow quantization limitations~\cite{swan,b4}.
For clarity, we also assume that all flows have the same priority. Thus, the fairness of the flow is its total allocated throughput.
We denote 
with low, medium and high delay SLO constraints, $4$, $1.5$, and $1$ times of a propagation delay stretch defined in Section~\ref{management_evaluation}, respectively.
%

In the first simulation scenario, we use LP-based solution (that is costly to address in practice~\cite{swan, b4}) with fixed average physical node degree equal to 4 (common for the Internet~\cite{internet}), where we vary the maximum number of paths available for each flow allocation.
%
In the second scenario, we use instead scalable greedy solution proposed in~\cite{b4} with unrestricted number of paths per flow. To this end, once the best currently available path (or tunnel) gets fully saturated, we find the next best path dynamically. 
In this scenario, we vary average physical node degree from $8$ to $1$.

\noindent
{\bf Evaluation metrics.}
To evaluate the performance of NM for data plane TE solutions, we compare NM with 
(extended) Dijkstra algorithm within the greedy-based TE ($i.e.$, in the second scenario), and their corresponding $k$-shortest path~\cite{eppstein} and $k$-constrained shortest path (that uses general version of NM coupled with \textit{Look Back} technique) algorithms within LP-based TE ($i.e.$, in the first scenario).
%
We remark that NM's superior performance w.r.t. Dijkstra\dc{-based TE} is expected due to its ability of minimizing provider's associated cost for flow allocation $e.g.$, minimizing provisioned physical bandwidth (see Problem~\ref{rocp_def}) under an arbitrary set of ($e.g.$, SLO) constraints. 
This difference is expected to degrade in the first scenario (where LP-based formulation is used) with either number of maximum paths per flow or SLO severity increase, as in this case the $k$-shortest path set converge to the $k$-constrained shortest path set resulting in the equal LP formulation. 
However, in the second scenario, as we allocate bandwidth for flows on their most preferable tunnels (best paths) first, that superior performance is expected to be preserved and can vary with different node degree or SLO severity.
For comparison we use the following four metrics: total gained throughput of all flows, cumulative distribution of flows' throughput which corresponds to their fairness, energy consumption relative to the network idle state (see Equation~\ref{energy_eq}) and path hop count. 
%

\noindent
{\bf Path hop savings lead to network utilization and flow fairness gains.}
%
In Figure$~\ref{lp_thr}$, we show how the correlation between the total gained throughput across all flows and the maximum number of paths available per each flow is a logarithmic function --- increasing number of paths linearly brings a logarithmic growth to the total gained throughput.
Figure$~\ref{b4_thr}$ shows how the total gained throughput of all allocated flows changes when multiple physical links become available (the average physical node degree increases).
This dependence is an affine function: the maximum possible total gained throughput increases linearly with the available physical links. 

In both scenarios, we can see how the total flow throughput and the resulting flow fairness ($e.g.$, the particular flow throughput) are higher for NM than for Dijkstra\dc{-based TE} (see Figures$~\ref{lp_thr}$,$~\ref{lp_thr_cdf}$,$~\ref{b4_thr}$ and$~\ref{b4_thr_cdf}$).
%
%
These results demonstrate how minimizing the total physical bandwidth provisioned for a single flow with NM can significantly benefit \st{even} traffic engineering solutions.
In particular, due to the path hop count optimization under SLO constraints within the data plane, NM gains up to 50\%  of total flow throughput under low SLO in the first (LP-based) scenario, and up to 20\% of total flow throughput under mid SLO in the second (greedy-based) scenario w.r.t. Dijkstra\dc{-based TE} (see Figures$~\ref{lp_thr}$ and$~\ref{b4_thr}$). 
%
%
Such gains allow, in turn also to improve flow fairness (see Figures$~\ref{lp_thr_cdf}$ and$~\ref{b4_thr_cdf}$).
Note that such large throughput gains in the first scenario ($i.e.$, up to 50\%) are partly due to the $k$-shortest path~\cite{eppstein} and the general version of NM (that finds $k$-constrained shortest paths) algorithms difference, $i.e.$, the former has a higher probability of finding paths with more shared edges then the later.
As expected, NM gains decrease with the node connectivity, as less physical path choices are available to map virtual links. Also, for LP-based scenario these gains decrease with increase of the maximum number of paths or SLO severity, as both shortest and constrained shortest path sets converge to each other resulting in equal LP formulations. 

\noindent
{\bf  Average path length and energy consumption tradeoff.}
We further investigate the reasons why we observed such gains in total throughput of all flows w.r.t. Dijksra\dc{-based TE}.
In particular, observing Figures$~\ref{lp_len}$ and$~\ref{b4_len}$ we note that there are $\approx2-3$ hops difference in the average path length between NM and Dijkstra\dc{-based TE} when allocating low SLO flows.
At the same time, for the medium SLO constraints this difference is reduced to circa one hop. Finally, when the SLO constraints are high, there is no significant physical path length difference.
%
%
To understand why the average path length changes with the constraint severity, note how the longer is an end-to-end physical path, the lower is the probability that
the entire path satisfies the SLO constraints. On the other hand, the higher the number of hops, the higher is the number of candidates paths, and so the higher is the probability of finding one which satisfies these constraints. This explains the trade-offs in average path length behavior observed in Figure$~\ref{b4_len}$.

The hop count savings minimize the physical bandwidth provisioned for a single path, allowing the provider to accept more flows or allocate more bandwidth for a single flow. 
As a result, the overall link utilization increases leading to a higher energy consumption (see Figures$~\ref{lp_energy}$ and$~\ref{b4_energy}$).
We observe one exception when throughput gains are low and path hop count savings are high (as observed in the management plane scenario in Section~\ref{management_evaluation}). 
An example of such situation can be also observed in the second scenario for dense physical networks (with average node degree $\ge 5$) when allocating flows with low SLO demands. In that case, we can see a small reduction (of $\approx$ 2\%) in the energy consumption simultaneously with low throughput gains (of $\approx$ 5\%).

\subsection{Scalability Results}
\label{exhaustive_sec}
In the next set of results we test the scalability performance of NM when accepting multiple link and multiple path constraints ($l \oplus p$ case). 
We remark that in this case only exponential exact solutions exist for the constrained shortest path problem due to its NP-hardness~\cite{edfs}.

\noindent
{\bf Scalability simulation settings.}
To \dc{assess} NM scalability, we simulate on-line requests for allocating constrained virtual links (or traffic flows).  
In particular, we generate physical network topologies of $10$, $100$, $1K$ and $10K$ nodes, where each physical link has bandwidth uniformly distributed  between $1$ and $9$ Gbps. In addition, we set each physical link with a cost uniformly distributed between $1$ and $10$. We attempt to find the constrained shortest path variant --- the resource optimal constrained path for $10\%$ of physical network nodes random source-destination pairs, where for each pair we allocate as many virtual links as possible with the fixed demands.
%
We denote with low and medium bandwidth constraints, $1$ and $4$ Gbps, respectively; these values represent approximately $10\%$ and $45\%$ of the maximum physical link capacity. Similarly, we denote with low and medium (propagation) delay SLO constraints, $4$ and $2.5$ times of a propagation delay stretch defined in Section~\ref{management_evaluation}, respectively. 
In addition, we denote with low and medium cost constraints, $100$ and $50$ that represent $10$ and $5$ times of the maximum physical link cost.
 %

\noindent
{\bf Scalability evaluation metrics.}
To evaluate the NM scalability, we compare the general version of NM with the EBFS (common branch-and-bound exhaustive search) algorithm and with the CPLEX~\cite{cplex} performance (that uses 4 parallel threads) of solving the common arc-based constrained shortest path formulation. Both NM and EBFS are coupled with dominant paths search space reduction techniques. Moreover, we couple EBFS with a Look Ahead (EBFS+LA) search space reduction technique~\cite{exact_qos} and NM with a Look Back (NM+LB) search space reduction technique - a variant of Look Ahead without complexity overhead (see Section~\ref{space_reduction_sec}). 

\noindent
We compare NM with EBFS and CPLEX across two metrics: the number of traversed paths required to find the constrained shortest path and the average path computation time. Note that in case of CPLEX, the number of traversed paths corresponds to the total number of iterations.

\vspace{2mm}
%

\begin{figure}[t!]
\centering

\hspace{-3mm}
\begin{subfigure}[b]{0.47\textwidth}
\centering
\includegraphics[width=1\linewidth]{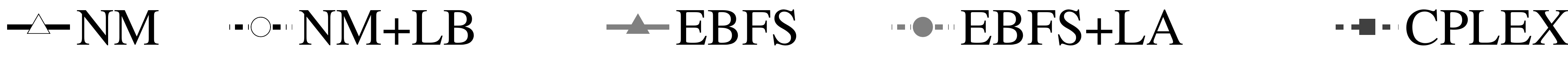}
\end{subfigure}

\begin{subfigure}[b]{0.235\textwidth}
\centering
\includegraphics[width=1\linewidth]{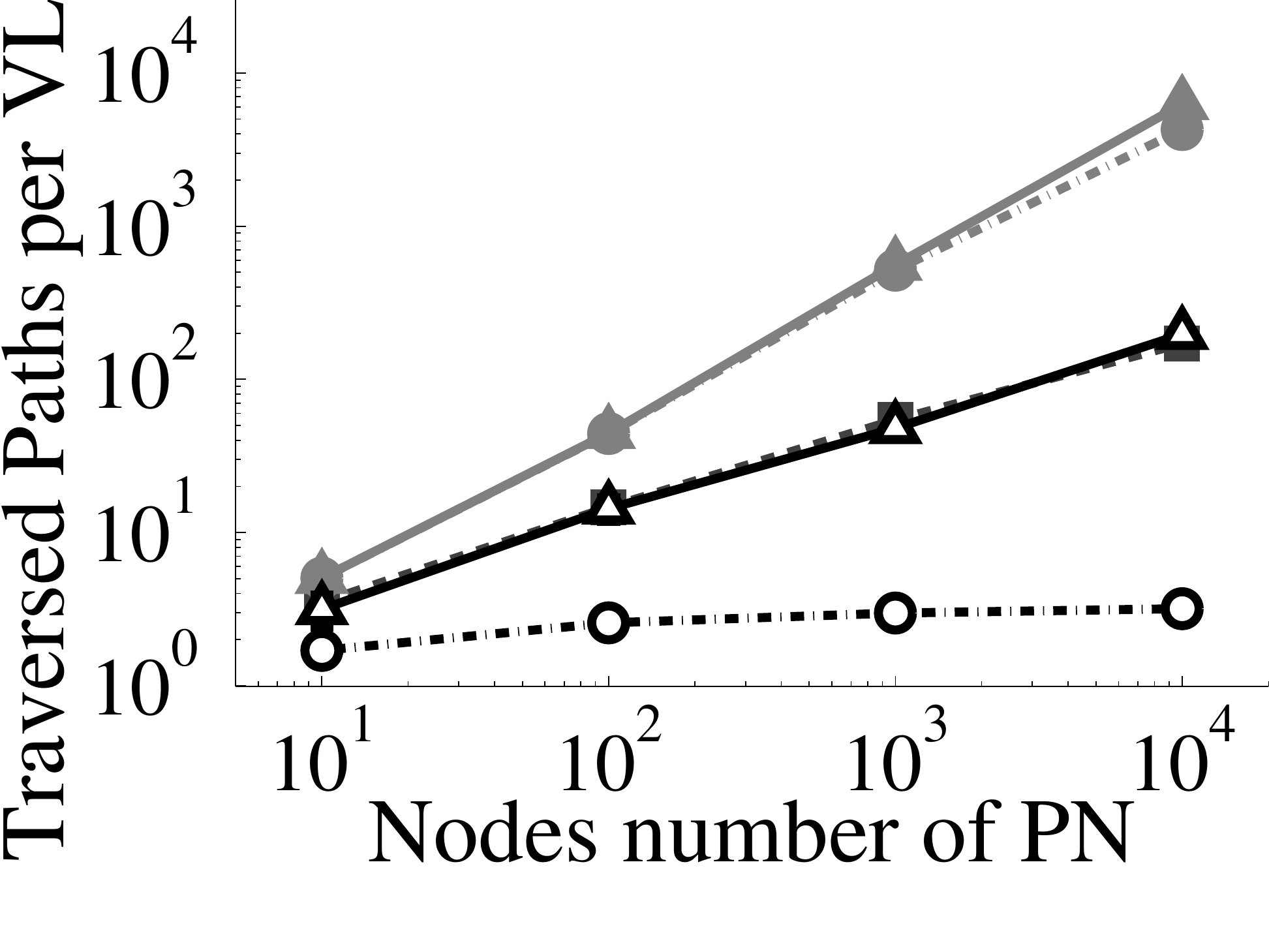}
\caption{}
\label{ex_paths}
\end{subfigure}
%
\begin{subfigure}[b]{0.235\textwidth}
\centering
\includegraphics[ width=1\linewidth]{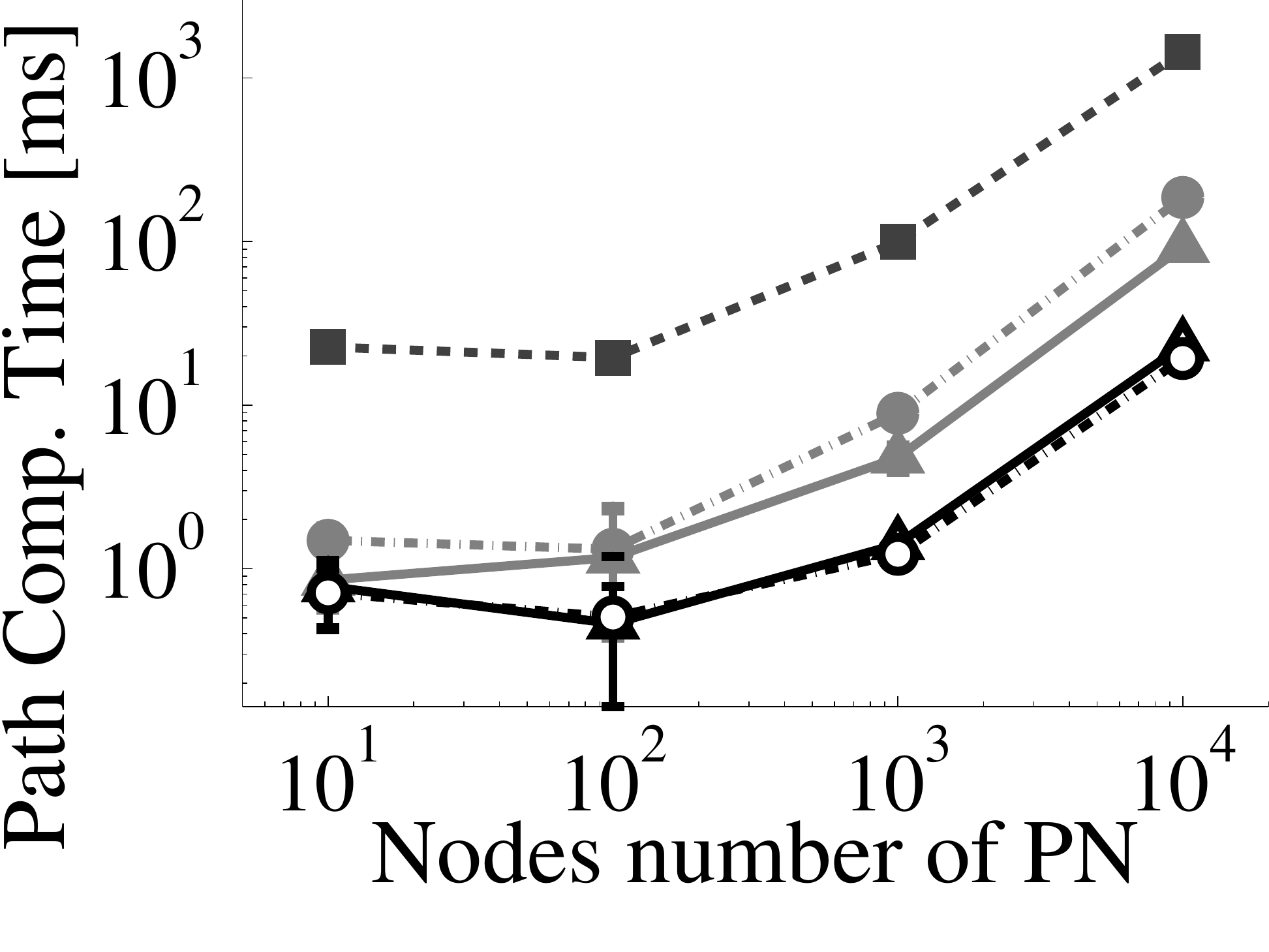}
\caption{}
\label{ex_time}
\end{subfigure}

\vspace{-0.5mm}

\begin{subfigure}[b]{0.235\textwidth}
\centering
\includegraphics[width=1\linewidth]{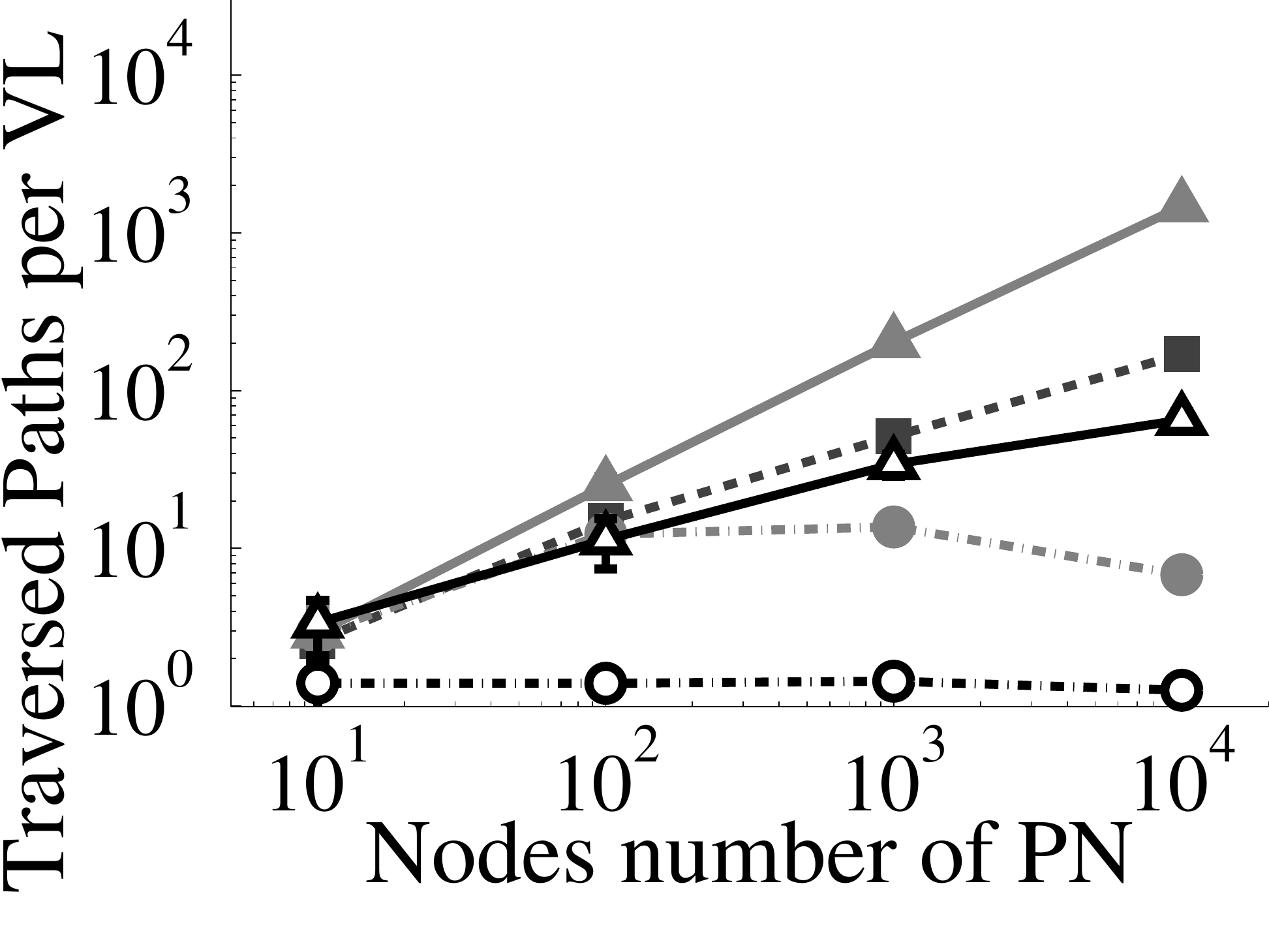}
\caption{}
\label{ex_paths_m}
\end{subfigure}
%
\begin{subfigure}[b]{0.235\textwidth}
\centering
\includegraphics[ width=1\linewidth]{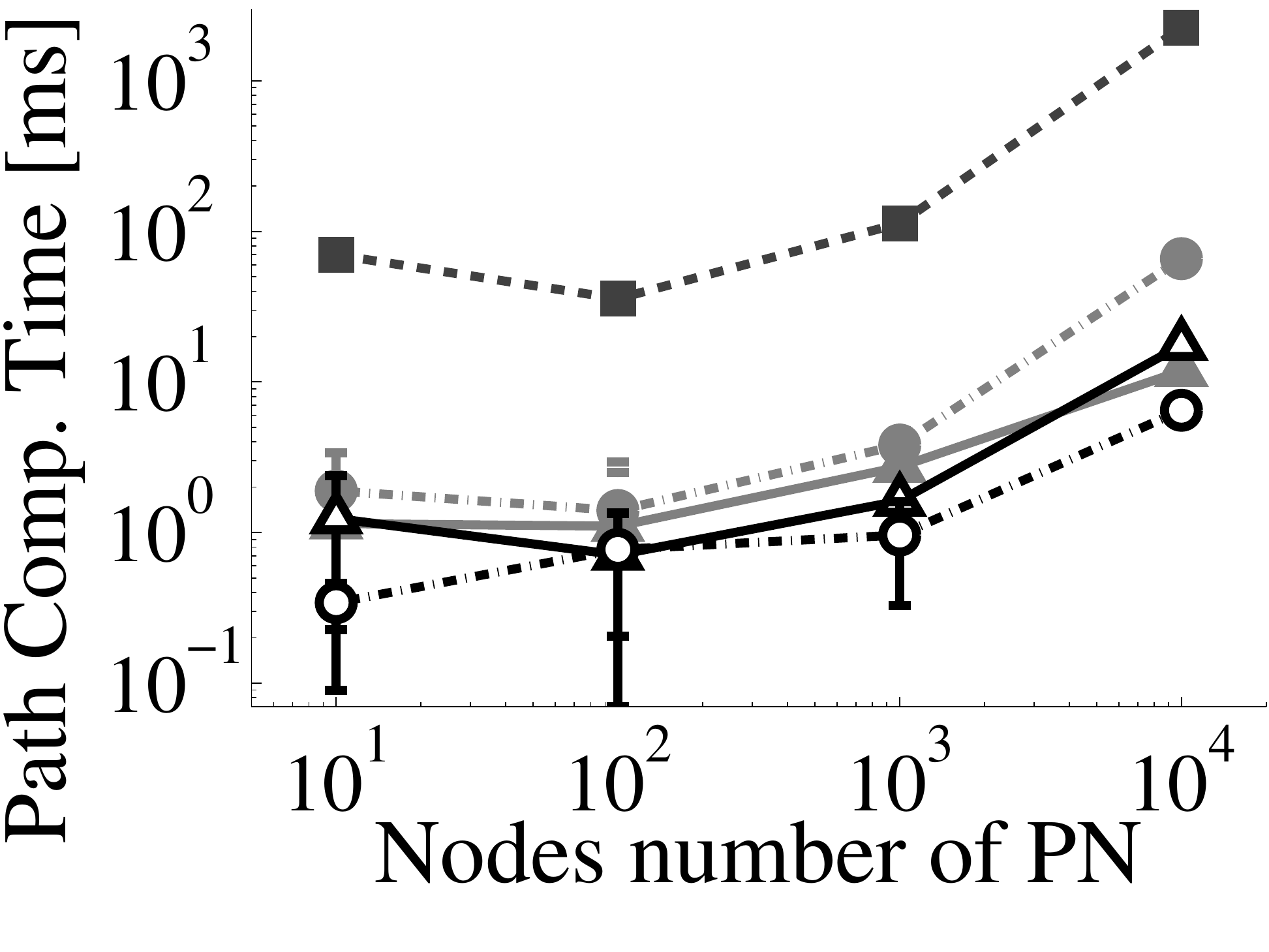}
\caption{}
\label{ex_time_m}
\end{subfigure}

\vspace{-2mm}
\caption{\footnotesize{
Performance analyses of general NM versus EBFS with dominant paths, look-back for NM (NM+LB) and look-ahead for EBFS (EBFS+LA) search space reduction techniques, and versus IBM CPLEX solver for the  constrained shortest path formulation in Problem~\ref{rocp_def} for low (top row) and medium (bottom row) SLO constraints in terms of: (a,c) number of traversed paths to find the \textit{optimal} virtual path; (b,d) the virtual path computation time.} 
}
\label{ex_waxman}
\vspace{-6mm}
\end{figure}

\noindent
{\bf Dominant paths prevent intractabilities.}
Figures$~\ref{ex_paths}$ and$~\ref{ex_paths_m}$ show that dominant paths technique reduces the number of traversed paths per virtual link to a linear function of a physical network size for both EBFS and NM.
Moreover, due to its ``double pass" technique, NM traverses up to two orders of magnitude paths less then EBFS.
However, we can see how EBFS works slightly faster than NM for large scale physical networks with medium link and path constraints (see Figure$~\ref{ex_time_m}$). That can be explained with the more expensive forward pass of NM for larger physical networks ($\ge10$K nodes). The NM's unnecessary iterations can be however reduced by the proposed Look Back technique.
%


\noindent
{\bf NM scalability with backward pass and look back.} Our experiments show that when NM backward phase is coupled with a Look Back technique, the number of traversed paths by NM  further reduces. This reduction is almost independent from the size of the  physical network (see Figures$~\ref{ex_paths}$ and$~\ref{ex_paths_m}$).  Using the Look Back search space optimization does not introduce any significant path computation overhead, while the same cannot be said for the Look Ahead search space reduction technique (see Figures$~\ref{ex_time}$ and$~\ref{ex_time_m}$).

Even though  CPLEX uses 4 parallel threads (instead of a single thread for NM and EBFS) and traverses moderate number of path (similar to NM without Look Back technique), it shows the worst performance in all cases. 
That is due to the fact, that finding constrained shortest paths with the commonly utilized arc-based integer programming formulation 
is NP-hard and no existing techniques can reduce that complexity to pseudo-polynomial.
On the contrary, NM and EBFS complexities can be reduced to pseudo-polynomial by applying the dominant paths search space reduction technique~\cite{exact_qos}. Proposed novel \textit{double pass} and \textit{Look Back} search space reduction techniques further reduce the practical complexity of finding constrained shortest paths. Thus, NM is almost an order of magnitude faster in comparison with EBFS and is almost 3 orders of magnitude faster than CPLEX for large-scale physical networks, and hence \textit{scales} better. 
%
We remark that such \textit{scalability} improvements over existing constrained shortest path algorithms are essential for the virtual network service management at large scale. \dcR{For example, the column generation approach can generate tens of thousands paths per a single VN request at large scale. Thus, it will take $\text{100 ms x }10K \approx 17$ minutes when EBFS is used. On the contrary, we need only  $\text{10 ms x }10K \approx 100$ seconds with NM which significantly reduces VN request blocking probability.}
\dcR{Note that additional scalability results comparison of our NM with respect to the EDijkstra shortest path scheme can be found in our prior work~\cite{nm}.}

%


\subsection{Prototype Evaluation}\label{sec:floodlight-experiments} 


In this final set of results, we use our NM prototype to estimate the impact of the on-demand constrained shortest path computation on the end-to-end virtual link embedding performance. We also confirm our main simulation results.

\noindent
{\bf Experiment settings.}
Our setup for the performance experiments includes $15$ virtual machines (VMs) from the GENI testbed~\cite{geni}: Ten of these VMs are OpenFlow Virtual Switches (OVS)~\cite{ovs},  and others are hosts. Each \textit{host-to-switch} physical link has $10$ Mbps bandwidth and a $0$ arbitrary cost, and each \textit{switch-to-switch} physical link has both bandwidth (measured in Mbps) and an arbitrary cost uniformly distributed between 1 and 10. 
Note, that our arbitrary cost is an additive metric and therefore can represent any path metric, $e.g.$, delay, losses, jitter, etc.
%
%
We request virtual links with low SLO constraints, $i.e.$, $\ge 1$ Mbps bandwidth and $\le 50$ arbitrary cost  ($5$ times greater than the maximum physical link cost), between $5$ random $<src,dst>$ pairs of hosts, where for each pair of endpoints we allocate as many virtual links as possible. 

\noindent
{\bf Experiment metrics.}
For each virtual link request we again measure the total gained throughput and the virtual link path hop count. In addition, we measure the time required to compute a path (virtual link mapping), and the time to allocate the computed path ($i.e.$, set appropriate flow rules within OpenFlow switches along the computed path). Note that the overall time for end-to-end virtual link embedding includes both virtual link mapping and its allocation. 
Our experiment goals are twofold: first, we want confirm our simulation results in real settings; secondly, we want to estimate an overhead of addressing constrained shortest path problem in real settings.

\begin{figure}[t!]
\centering
\begin{subfigure}[b]{0.41\textwidth}
\centering
\includegraphics[width=1\linewidth]{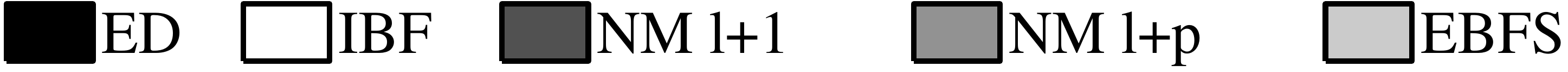}
\end{subfigure}

\vspace{2mm}
\vspace{-1mm}
\begin{subfigure}[b]{0.152\textwidth}
\centering
\includegraphics[width=1\linewidth]{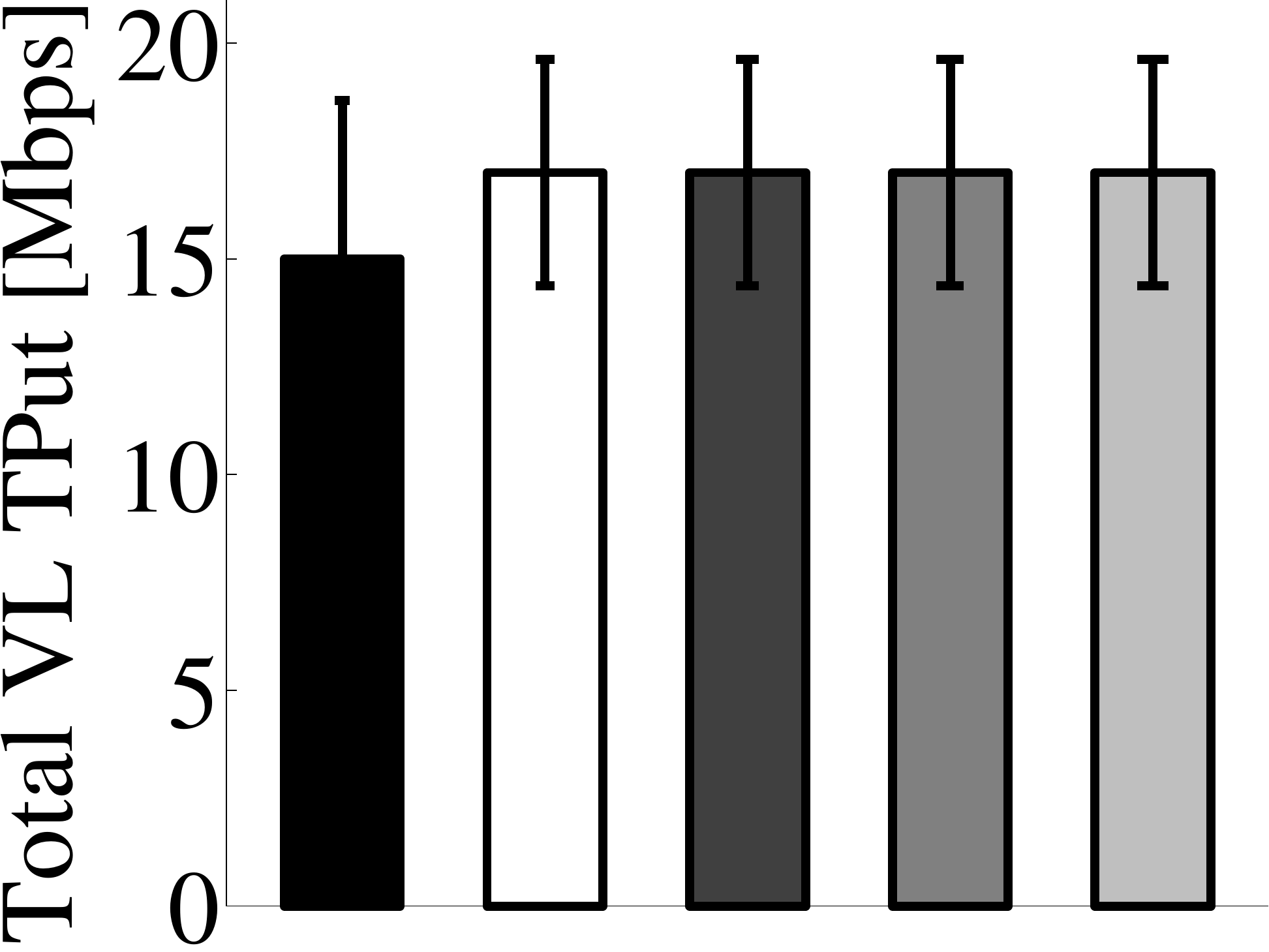}
\caption{}
\label{sdn_total}
\end{subfigure}
\begin{subfigure}[b]{0.152\textwidth}
\centering
\includegraphics[width=1\linewidth]{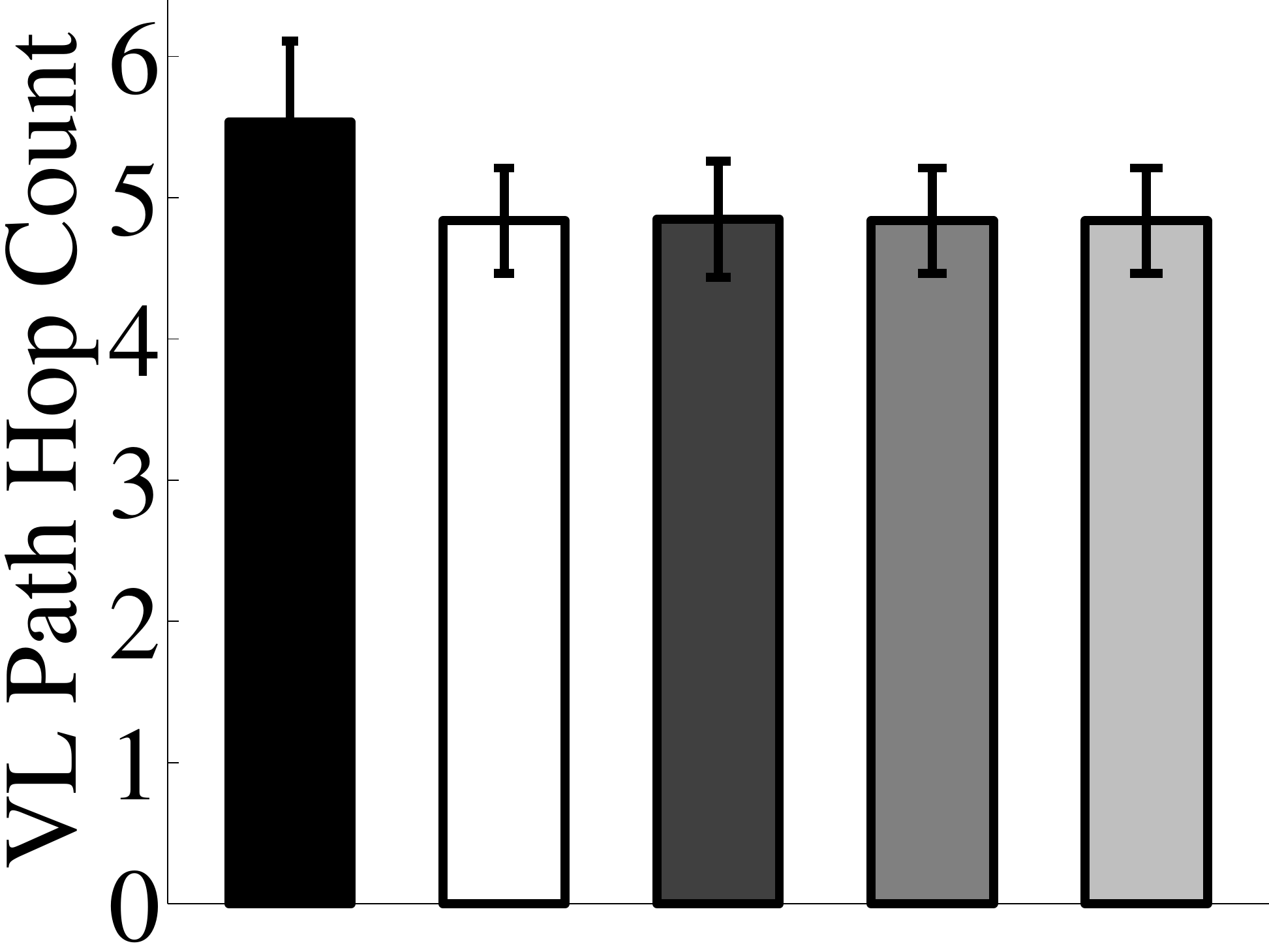}
\caption{}
\label{sdn_length}
\end{subfigure}
\begin{subfigure}[b]{0.152\textwidth}
\centering
\includegraphics[width=1\linewidth]{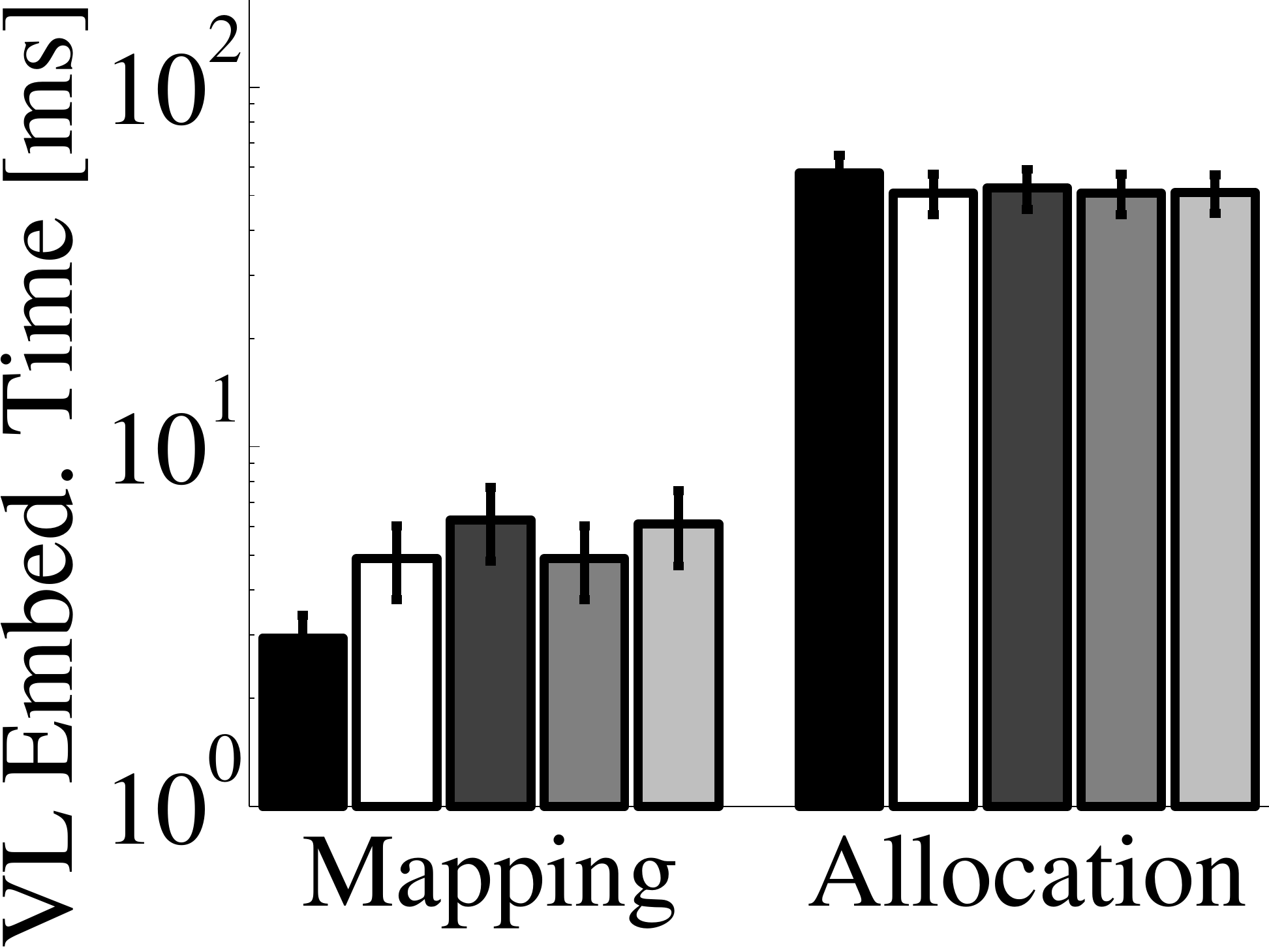}
\caption{}
\label{sdn_time}
\end{subfigure}
\vspace{-3mm}
\caption{\footnotesize{Performance analysis of the shortest path algorithm such as the extended version of Dijkstra (ED) versus constrained shortest path algorithms such as NM (in both $l\oplus1$ and $l\oplus p$ cases), EBFS and IBF on a reserved in GENI small SDN testbed in terms of: (a) total gained throughput; 
(b) number of path hops per VL; and (c) average time per VL embedding, i.e., VL path computation (mapping) and its consequent allocation.}} 

\label{sdn_data}
\vspace{-6mm}
\end{figure}

\noindent
{\bf NM gains are confirmed experimentally.}
Using real-world settings, we were able to confirm constrained shortest path algorithms ($i.e.$, IBF, NM and EBFS)  \dcR{for the online TE} produce superior performance \dcR{even on a small physical network scale. This is similar to superior results of the offline TE which utilizes the constrained shortest path algorithms (see Section~\ref{te_evaluation})}. 
%
Specifically, IBF, NM and EBFS show gains of up to $12\%$ in total VL throughput (network utilization) 
and find almost 1 hop shorter VL path in average, w.r.t. \dc{extended Dijkstra (ED) shortest path scheme} as shown in Figures~\ref{sdn_total} 
and~\ref{sdn_length}.
Note however that IBF is applicable only for in $l$ and $l\oplus1$ cases (see Table~\ref{table1}). 

\noindent
{\bf NM running time scales well with physical network size.}
Figure~\ref{sdn_time} shows how VL mapping  is an order of magnitude faster for small scale physical networks (of $\approx10^1$ nodes) than its allocation for all routing schemes that have been implemented.
This is because the path computation  is a local (in-memory) operation but the virtual link allocation requires setting up of flow rules within all switches along the loop-free underlying physical path found. Hence, its speed depends on the Round Trip Time between switches and the OpenFlow controller.
%
\dcR{In reference to Figures~\ref{ex_time} and~\ref{ex_time_m}, we can see how for large scale networks ($\ge 10K$ nodes), 
the running time of classical constrained shortest path algorithms such as EBFS can become prohibitive (up to two orders of magnitudes larger than in a case of a single path computation for small scale networks). As a result, the VL mapping time becomes a bottleneck. On the contrary, NM is just an order of magnitude slower at large-scale than at small scale. 
%
Thus, NM does not bottleneck the end-to-end VL embedding at large scale.}

%
 
\section{Conclusion}
\label{conclusion}

In this paper, we motivated the problem of achieving a \textit{flexible} and \textit{scalable} constrained shortest path management approach for virtual network services deployed across multiple data centers. 
%
%
To cope with constrained shortest paths NP-hardness (that limits both its scalability and flexibility), we first introduced a novel algorithm viz., `Neighborhoods Method' (NM), which utilizes a double pass (a synergy of dynamic programming and a branch-and-bound exhaustive search) and ``Look Back" search space reduction techniques.
%
Our computational complexity analysis indicates NM's quadratically lower complexity upper-bound than for recent branch-and-bound exhaustive search methods, and our scalability evaluation results 
show how NM is faster by an order of magnitude than these methods. 
Thus, NM \textit{scales} better for large scale networks of $\ge10,000$ nodes.
Via 
numerical simulations of diverse network topology scenarios, we were able to show how the constrained shortest path management improves the network utilization (gains were seen up to 50\%) and in some cases even in the energy efficiency of the recent management plane algorithms for virtual network embedding or network function virtualization service chaining. Additionally, we found improvements in the recent data plane algorithms for traffic engineering. Thus, we demonstrated that our proposed NM is also \textit{flexible} and can be applied to diverse virtual network service scenarios.
Finally, we were able to reproduce our main simulation results in a real-world GENI testbed with an NM implementation, whose source code is publicly available under a GNU license at~\cite{nm_repo}.  

As part of future work, we aim to develop new VNE/NFV-SC
algorithms that can better utilize our proposed constrained
shortest path scheme at larger network scales. To obtain a good
lower bound solution within large network scale simulations, we can relax
integrality constraints of the optimal VNE/NFV-SC integer programs to $e.g.$, use efficient linear programming.


\section*{Acknowledgements}
This work has been partially supported by the National Science Foundation
awards CNS-1647084, CNS-1647182, the Coulter Foundation Translational Partnership
Program and by RFBR according to the research project 16-07-00218a and
the public tasks of the Ministry of Education and Science of the Russian
Federation (2.974.2017/4.6). Any opinions, findings or conclusions expressed
in this publication are those of the author(s) and do not necessarily reflect the
views of the funding agencies.

\appendix
\subsection{Proof of NM Optimality Theorem}
\label{NM_proof}
\begin{proof} 
To prove the NM optimality we need to prove: firstly, that the forward pass of the general NM estimates all possible hop count distances of simple (loop-free) paths between the source and the destination vertices in ascending order; secondly, that the backward pass of NM can return any path of a given length $N$; and finally that NM returns the \textit{constrained shortest path}.

The first thesis can be proved by contradiction: assume that there is the 
{\it optimal path} with ($i$) the minimum or ($ii$) the maximum hop count distance to destination, and this distance has not been estimated by NM. Firstly, the forward pass step starts building neighborhoods from the first neighborhood, meaning that NM estimates all possible 1 hop paths first. If there is a path with length lower than $1$, we have the special case in which the source node is also the destination assumed to be satisfied, which in turn contradicts with assumption ($i$). 
Further, the forward pass step continues to build neighborhoods (estimate possible path hop count distances in ascending order) until their number is equal to the number of vertices, or the destination vertex appears in the last neighborhood. In the first case, the forward pass ends by estimating the maximum possible distance of a simple (loop-free) solution, and there will never be a case in which a complex path (with loops) can be provided due to a contradiction with the \textit{constrained shortest path} definition (see Definition~\ref{csp_def}). In the second case, the forward pass ends by finding the hop count distance to the destination. In both cases we contradict assumption ($ii$). 

The second thesis can be also proved by contradiction: assume that NM has not found a path with the $N$ hops length between the source and the destination nodes. This is possible only if at least one of the path's nodes has not appeared in the neighborhoods $<NH>$. In this case, it suggests that this node is not accessible from the source node within $N$ hops, i.e., if it is unreachable, or path does not have the $N$ hops length, which contradicts our assumption.

The third thesis is easy to prove: NM successively checks paths for constraints satisfaction in ascending hop count order, where NM continues to iterate until either one of the path will satisfy all constraint or the solution length will be equal to the number of vertices. In the first case, NM finds the \textit{min hop count} path which satisfies all constraints and by definition is the \textit{resource optimal constrained} path (see Definition~\ref{rocp_def}).
In the second case, this length is the maximum possible length of a simple path, and there will never be a case in which a complex path can be provided due to a contradiction with the \textit{constrained shortest path} definition (see Definition~\ref{csp_def}). 

In summary, we proved that the forward pass of NM estimates all possible hop count distances of simple paths between the source and the destination vertices in ascending order. At the same time, the backward pass can return any path of a given hop count length which provided by the forward pass to check for constraints satisfaction. 
Consequently, we conclude that - if  the \textit{constrained shortest path} exists, then NM will find it.
\end{proof}

\vspace{-20mm}

\begin{IEEEbiography}[{\includegraphics[width=1in,height=1.25in,clip,keepaspectratio]{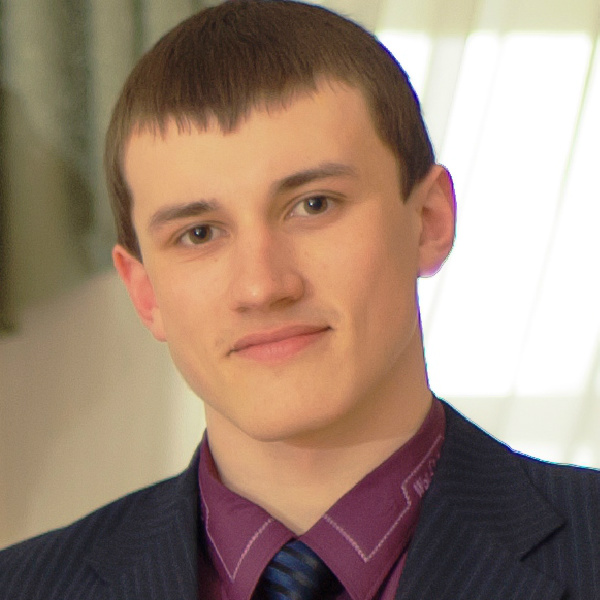}}]{Dmitrii Chemodanov} %
received his MS degree from the Department of Computer Science at Samara State Aerospace University, Russia in 2014. He is currently a PhD student in the Department of Computer Science at University of Missouri-Columbia. His current research interests include distributed and cloud computing, network and service management, and peer-to-peer networks.
\end{IEEEbiography}

\vskip -2\baselineskip plus -1fil

\begin{IEEEbiography}[{\includegraphics[width=1in,height=1.25in,clip,keepaspectratio]{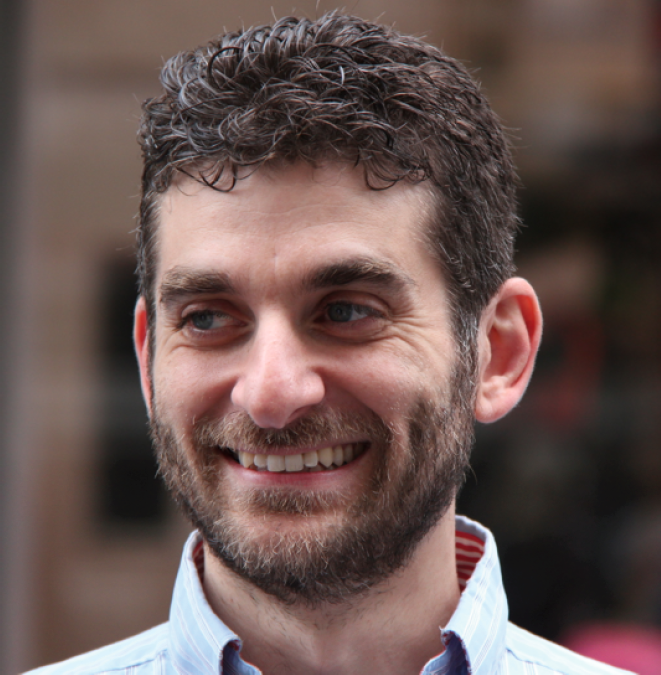}}]{Flavio Esposito} %
is an Assistant Professor in the Computer Science Department at SLU and a Visiting Research Assistant Professor in the CS Dept. at University of Missouri, Columbia. He received his Ph.D. in CS at Boston University in 2013, and his MS in Telecommunication Engineering from University of Florence, Italy.  His research interests include network management, network virtualization and distributed systems.  Prior to joining SLU, Flavio worked at Exegy, St.Louis, MO, at Alcatel-Lucent, Italy, at Bell Laboratories, NJ, at Raytheon BBN Technologies, MA, and at EURECOM, France. He was also a visiting researcher at the Center for Wireless Communications, Finland.
\end{IEEEbiography}

\vskip -2\baselineskip plus -1fil

\begin{IEEEbiography}[{\includegraphics[width=1in,height=1.25in,clip,keepaspectratio]{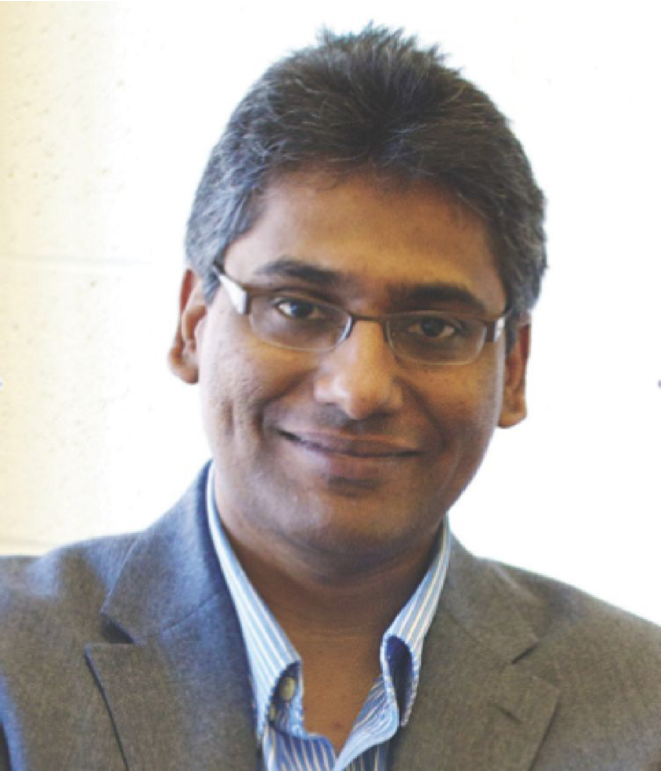}}]{Prasad Calyam} %
received his MS and PhD degrees
from the Department of Electrical and Computer
Engineering at The Ohio State University in 2002
and 2007, respectively. He is currently an Associate
Professor in the Department of Computer Science
at University of Missouri-Columbia. His current
research interests include distributed and cloud computing,
computer networking, and cyber security. He
is a Senior Member of IEEE.
\end{IEEEbiography}

\vskip -2\baselineskip plus -1fil

\begin{IEEEbiography}[{\includegraphics[width=1in,height=1.25in,clip,keepaspectratio]{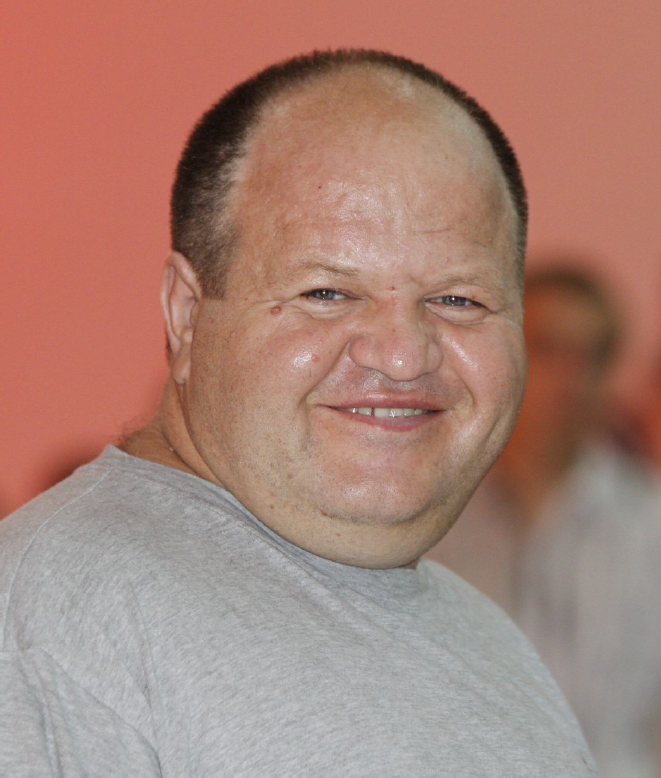}}]{Andrei Sukhov} %
is a Professor of Samara National Research University, Russia and was awarded a PhD in Moscow, in Physics and Mathematics in 1993. In 2007 he received Dr.Sc. degree in computer networking at Moscow State University of Electronics and Mathematics (MIEM HSE). Over the last 20 years he has been involved in acting as an investigator for more then 15 telecommunication projects supported by the Russian government, RFBR, INTAS, NATO, ESA, etc. His research area is the computer networks and security.
\end{IEEEbiography}

\vspace{-10mm}
\end{document}